\documentclass[12pt,journal,final,onecolumn]{IEEEtran}

\usepackage[dvips]{graphicx}
\usepackage{epsfig}
\usepackage[cmex10]{amsmath}
\usepackage{amssymb}
\usepackage{amsthm}
\usepackage{amsfonts}
\usepackage{bm}
\usepackage{dsfont}
\usepackage[mathscr]{eucal}
\usepackage[noadjust]{cite}
\usepackage{pstricks,pst-node,pst-plot,pstricks-add,pst-tree}
\usepackage[tight,footnotesize]{subfigure}

\graphicspath{{figs/}}

\input{niesen.def}

\newcommand{\SNR}{{\sf SNR}}
\newcommand{\sC}{\textup{\tsf{C}}}
\newcommand{\sP}{\textup{\tsf{P}}}

\begin{document}

\title{Interference Alignment: From Degrees-of-Freedom \\
to Constant-Gap Capacity Approximations}

\author{Urs Niesen and Mohammad A. Maddah-Ali%
\thanks{This paper was presented in part at the International Symposium
on Information Theory, July 2012.}%
\thanks{The authors are with Bell Labs, Alcatel-Lucent. Emails:
urs.niesen@alcatel-lucent.com, mohammadali.maddah-ali@alcatel-lucent.com}%
}

\maketitle

\begin{abstract}
    Interference alignment is a key technique for communication
    scenarios with multiple interfering links. In several such
    scenarios, interference alignment was used to characterize the
    degrees-of-freedom of the channel. However, these degrees-of-freedom
    capacity approximations are often too weak to make accurate
    predictions about the behavior of channel capacity at finite
    signal-to-noise ratios ($\SNR$s).  The aim of this paper is to
    significantly strengthen these results by showing that interference
    alignment can be used to characterize capacity to within a constant
    gap. We focus on real, time-invariant, frequency-flat X-channels.  The
    only known solutions achieving the degrees-of-freedom of this
    channel are either based on real interference alignment or on
    layer-selection schemes. Neither of these solutions seems sufficient
    for a constant-gap capacity approximation.

    In this paper, we propose a new communication scheme and show that
    it achieves the capacity of the Gaussian X-channel to within a
    constant gap. To aid in this process, we develop a novel
    deterministic channel model. This deterministic model depends on the
    $\tfrac{1}{2}\log(\SNR)$ most-significant bits of the channel
    coefficients rather than only the single most-significant bit used
    in conventional deterministic models. The proposed deterministic
    model admits a wider range of achievable schemes that can be
    translated to the Gaussian channel. For this deterministic model, we
    find an approximately optimal communication scheme. We then
    translate this scheme for the deterministic channel to the original
    Gaussian X-channel and show that it achieves capacity to within a
    constant gap. This is the first constant-gap result for a general,
    fully-connected network requiring interference alignment.
\end{abstract}

\section{Introduction}
\label{sec:intro}

Interference alignment has been used to achieve optimal
degrees-of-freedom (capacity pre-log factor) in several common wireless
network configurations such as
X-channels~\cite{maddah-ali08,jafar08,cadambe09,cadambe10}, interference
channels~\cite{cadambe08,motahari09}, interfering multiple-access and
broadcast channels~\cite{suh08}, multi-user systems with delayed
feedback~\cite{maddah-ali10,maleki11,abdoli11}, and distributed
computation~\cite{niesen11}, among others. The main idea of interference
alignment is to force all interfering signals at the receivers to be
aligned, thereby maximizing the number of interference-free signaling
dimensions.

\subsection{Background}
\label{sec:intro_background}

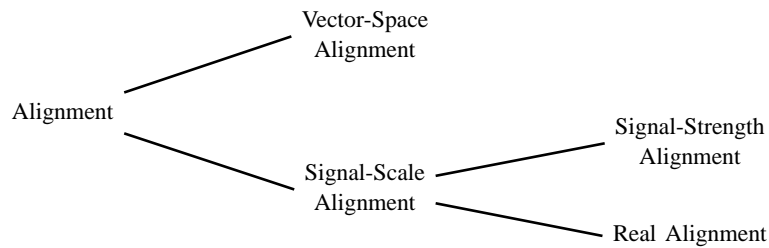
\begin{figure*}[!htbp]
    \centering
    %\documentclass[letterpaper,dvips]{article}
%\usepackage[dvips]{graphicx}
%\usepackage{epsfig}
%\usepackage[cmex10]{amsmath}
%\usepackage{amssymb}
%\usepackage{amsthm}
%\usepackage{amsfonts}
%\usepackage{bm}
%\usepackage[svgnames]{xcolor} % load before pstricks
%\usepackage{pstricks,pst-node,pst-plot,pstricks-add}
%\usepackage{pst-eps}

%\begin{document}
%\pagestyle{empty}

%\begin{TeXtoEPS}
    \begin{minipage}[b]{8.65cm} % This is necessary because \psmatrix seems to mess up the width of the picture
        \begin{pspicture}(0cm,-1.8cm)(8.65cm,1.4cm)
            %\psgrid
            \footnotesize % 8pt fontsize
            \psset{linewidth=1pt} 
            \psset{shortput=nab}
            \psset{unit=1cm} % Scale figure without scaling font size, default is 1cm

            % Draw maximal figure size for one/two column figures
            %\psframe[linecolor=gray](0in,-0.2in)(3.5in,3.3in) % One column wide
            %\psframe[linecolor=gray](0in,-0.2in)(7.25in,3.3in) % Two columns wide

            % Draw tree
            \def\psedge#1#2{\ncdiagg[nodesepA=4pt,nodesepB=4pt,angleA=180,armA=0]{#2}{#1}}
            \pstree[treemode=R,radius=3pt,levelsep=*2.5cm,treefit=loose]{\Tr{Alignment}}{%
            \Tr{\shortstack{Vector-Space\\Alignment}}%
            \pstree{\Tr{\shortstack{Signal-Scale\\Alignment}}}{%
            \def\psedge#1#2{\ncdiagg[nodesepA=4pt,nodesepB=4pt,angleA=180,armA=0]{#2}{#1}}
            \Tr{\shortstack{Signal-Strength\\Alignment}}%
            \Tr{Real Alignment}%
            }%
            }
        \end{pspicture}
    \end{minipage}
%\end{TeXtoEPS}

%\end{document}
    \caption{Different alignment approaches and their relation.}
    \label{fig:background}
\end{figure*}

Alignment approaches can be divided into two broad categories (see
Fig.~\ref{fig:background}).

\begin{enumerate}
    \item \emph{Vector-space alignment (\cite{maddah-ali08,cadambe08}
        among others)}: In this approach, conventional communication
        dimensions, such as time, frequency, and transmit/receive
        antennas, are used to align interference. At the transmitters,
        precoding matrices are designed over multiple of these
        dimensions such that the interference at the receivers is aligned in
        a small subspace. If the channel coefficients have enough
        variation across the utilized time/frequency slots or antennas,
        then such precoding matrices can be found. 
    \item \emph{Signal-scale alignment (\cite{bresler10,motahari09}
        among others)}:
        If the transmitters and receivers have only a single antenna and the
        channel coefficients are time invariant and frequency flat, the
        vector-space alignment method fails. Instead, one can make use
        of another resource, namely the signal scale. Using lattice
        codes, the transmitted and received signals are split into
        several superimposed layers. The transmitted signals are chosen
        such that all interfering signals are observed within the same
        layers at the receivers. Thus, alignment is now achieved in signal
        scale.  
\end{enumerate}

Signal-scale interference alignment can be further subdivided into two
different, and seemingly completely unrelated, approaches: alignment
schemes motivated by \emph{signal-strength deterministic models}
\cite{bresler10,huang08} and \emph{real interference alignment}
\cite{motahari09}.

For the \emph{signal-strength deterministic approach}, the channel is
first approximated by a deterministic noise-free channel. In this
deterministic model, all channel inputs and outputs are binary vectors,
representing the binary expansion of the real valued signals in the
Gaussian case.  The actions of the channel are modeled by shifting these
vectors up or down, depending on the most-significant bit of the channel
gains, and by bitwise addition of interfering vectors. The signal layers
are represented by the different bits in the binary expansion of the
signals. In the second step, the signaling schemes and the outer bounds
developed for this simpler deterministic model are used to guide the
design of efficient signaling schemes for the original Gaussian problem.

This deterministic approach has proved instrumental in deriving
constant-gap capacity approximations for several challenging multi-user
communication scenarios such as single-multicast relay
networks~\cite{avestimehr11}, two-user interference channels with
feedback~\cite{suh11} or with transmit/receive
cooperation~\cite{wang11b,wang11a}, and lossy distributed source
coding~\cite{maddah-ali10b}. In all these communication scenarios,
interference alignment is not required. For communication scenarios in
which interference alignment is required, the deterministic approach has
been less helpful. In fact, it has only been successfully used to obtain
constant-gap capacity approximations for the fairly restrictive
many-to-one interference channel, in which only one of the receivers
experiences interference while all others are interference
free~\cite{bresler10}. Even for the X-channel, one of the simplest
Gaussian networks in which interference alignment is required, only
weaker (generalized) degrees-of-freedom capacity approximations were
derived using the deterministic approach~\cite{huang08}. The resulting
communication scheme for the Gaussian X-channel is quite complicated and
cannot be used to derive a constant-gap capacity approximation.

For the \emph{real interference-alignment approach}, each transmitter
modulates its signal using a scaled integer lattice such that at each
receiver all interfering lattices coincide, while the desired lattice is
disjoint. Each receiver recovers the desired signal using a
minimum-distance decoder. A number-theoretic result concerning the
approximability of real numbers by rationals, called Groshev's theorem,
is used to analyze the minimum constellation distance at the receivers.
For almost all channel gains, this scheme is shown to achieve the full
degrees-of-freedom of the Gaussian X-channel and the Gaussian
interference channel \cite{motahari09}. While this scheme is
asymptotically optimal for almost all channel gains, there are
infinitely many channel gains for which the scheme fails, for example
when the channel gains are rational.  Moreover, this approach can again
not be used to derive stronger constant-gap capacity approximations.

At first glance, real interference alignment appears to rely on the
irrationally of the channel coefficients, preventing the desired integer
input signals from mixing with the undesired integer interference
signals.  This raises the concern that the scheme might
be severely affected by the presence of measurement errors or
quantization of the channel coefficients. In addition, arbitrarily close
to any irrational channel realization is a rational channel realization.
How are we then to engineer a communication device based on this scheme?
Quoting from Slepian's 1974 Shannon Lecture~\cite{slepian76}: ``Most of
us would treat with great suspicion a model that predicts stable flight
for an airplane if some parameter is irrational but predicts disaster if
that parameter is a nearby rational number. Few of us would board a
plane designed from such a model.''

Some of these concerns follow from the fact that real interference
alignment is somehow isolated from other known signaling schemes and
only poorly understood. Unlike the vector-space and the deterministic
approaches, no vector-space interpretation is known for real
interference alignment, making it harder to obtain intuition. On the
other hand, it is known that the degrees-of-freedom of the interference
channel are discontinuous at all rational channel
coefficients~\cite{etkin09}. It should therefore not be surprising that
the rates achieved by real interference alignment share this
characteristic. Rather, it appears that it is the degrees-of-freedom
capacity approximation that is too weak to allow accurate predictions
about the behavior of channel capacity at finite $\SNR$s, and that the
discontinuity of the degrees-of-freedom in the channel coefficients are
mainly caused by taking a limit as $\SNR$ approaches infinity. Thus, a
stronger capacity approximation is needed.

\subsection{Summary of Results}
\label{sec:intro_summary}

The main contributions of this paper are as follows.

\subsubsection{New Deterministic Channel Model}

We develop a novel deterministic channel model, in which each channel
gain is modeled by a lower-triangular, binary Toeplitz matrix. The
entries in this matrix consist of the first $\tfrac{1}{2}\log(\SNR)$
bits in the binary expansion of the channel gain in the corresponding
Gaussian model. This contrasts with the traditional signal-strength
deterministic model, which is based only on the \emph{single}
most-significant nonzero bit. The proposed lower-triangular
deterministic model is rich enough to explain the real
interference-alignment approach. Thus, it unites the so far disparate
deterministic and real interference-alignment approaches mentioned above
(see Fig.~\ref{fig:background}). Moreover, as our proposed deterministic
model is based on a vector space, it enables an intuitive interpretation
of real interference alignment. 

\subsubsection{New Mathematical Tools} 

The solution for the proposed lower-triangular deterministic model can
be translated to the Gaussian setting.  To analyze the resulting scheme
for the Gaussian setting, we develop new tools. In particular, to prove
achievability for the Gaussian case, we extend Groshev's theorem to
handle finite $\SNR$s as well as channel gains of different magnitudes,
and we prove a strengthening of Fano's inequality.

\subsubsection{New Notion of Capacity Approximation} 

We introduce the new notion of a constant-gap capacity approximation up
to an outage set.  Specifically, the aim is to provide a constant-gap
capacity approximation uniform in the $\SNR$ and the channel gains as
long as these channel gains are outside a computable outage set of
arbitrarily small measure.  This new notion of a constant-gap
approximation up to an outage set can lead to a more concise capacity
characterization as we will see next.

\subsubsection{Constant-Gap Result for the Gaussian X-Channel}

We apply these ideas to the Gaussian X-channel by deriving a
constant-gap capacity approximation up to outage for this channel. This
is the first constant-gap result for a general, fully-connected network
requiring interference alignment.  To simplify the exposition, we focus
in this paper on the most relevant situation, in which the direct links
of the X-channel are stronger than the cross links---the tools and
techniques developed here apply to the other settings as well.

To develop this result, we first consider the lower-triangular
deterministic version of the X-channel and design a signaling scheme
that achieves its capacity up to a constant gap, as long as the binary
channel matrices satisfy certain rank conditions (see
Theorems~\ref{thm:symmetric_det} and~\ref{thm:arbitrary_det} in
Section~\ref{sec:main}).  We then show that the translated version of
the solution for the deterministic model achieves the capacity of the
Gaussian X-channel to within a constant gap up to the aforementioned
outage set (see Theorems~\ref{thm:symmetric_gaussian}
and~\ref{thm:arbitrary_gaussian} in Section~\ref{sec:main}).  In
addition, we show that, similar to the MIMO broadcast
channel~\cite{jindal06}, capacity is not sensitive to channel
quantization and measurement errors smaller than $\SNR^{-1/2}$. 

One implication of these results is that the complicated solution
achieving the degrees-of-freedom of the Gaussian X-channel
in~\cite{huang08} is a result of oversimplification in the
signal-strength deterministic model rather than the properties of the
original Gaussian channel itself. Moreover, the results in this paper
imply that the discontinuity of the degrees-of-freedom of the Gaussian
X-channel with respect to the channel coefficients is due to the large
$\SNR$ limit and is not present at finite $\SNR$s.

\subsection{Organization}
\label{sec:intro_organization}

The remainder of this paper is organized as follows.
Section~\ref{sec:deterministic} introduces the new deterministic channel
model. Section~\ref{sec:model} formalizes the Gaussian network model and
the problem statement. Section~\ref{sec:main} presents the main results
of the paper---Sections~\ref{sec:proofs_det}
and~\ref{sec:proofs_gaussian} contain the corresponding proofs.
Section~\ref{sec:foundations} contains the mathematical foundations for
the analysis of the decoding algorithms. Section~\ref{sec:conclusion}
concludes the paper.

\section{Deterministic Channel Models}
\label{sec:deterministic}

Developing capacity-achieving communication schemes for multi-user
communication networks is often challenging. Indeed, even for the
relatively simple two-user interference channel, finding capacity is a
long-standing open problem.  For the Gaussian network, the difficulty is
due to the interaction between the various components of these
networks, such as broadcast, multiple access, and additive noise. For
example, the two-user interference channel mentioned before has two
broadcast links, two multiple-access links, and two additive noise
components. 

The problem of characterizing capacity can be substantially simplified
if these noise components are eliminated, so that the output at the
receivers becomes a deterministic function of the channel inputs at the
transmitters~\cite{elgamal82,avestimehr11}. Such networks are called
deterministic networks. This observation motivates the investigation of
noisy networks by approximating them with deterministic
networks~\cite{avestimehr11,bresler08,anand11}. 

This approximation has two potential advantages. First, the capacity of
the deterministic network may directly approximate the capacity of the
original Gaussian network. Second and more important, the deterministic
model may reveal the essential ingredients of an efficient signaling
scheme for the noisy network. In other words, the capacity achieving
signaling scheme for the deterministic network may be used as a road map
to design signaling schemes for the Gaussian network. If the
deterministic approximation is well chosen, then the resulting signaling
scheme for the Gaussian network is close to capacity achieving.

The first critical step in this approach is thus to find an appropriate
deterministic channel approximating the Gaussian one. This deterministic
channel model should satisfy two criteria: \emph{simplicity} and
\emph{richness}. These two requirements are conflicting. Indeed,
oversimplification of the Gaussian model can sacrifice the richness of
the deterministic model. Conversely, keeping too many of the features of
the Gaussian model can result in a deterministic model that is rich but
too difficult to analyze. Striking the right balance between these two
requirements is the key to developing a useful deterministic network
approximation. 

One of the approaches that achieves this goal is the signal-strength
deterministic model proposed by Avestimehr et al.~\cite{avestimehr11}.
We review this deterministic model in
Section~\ref{sec:deterministic_salman}. We introduce our new
lower-triangular deterministic model in
Section~\ref{sec:deterministic_triangular}.
Section~\ref{sec:deterministic_comparison} compares the two
deterministic models, explaining the shortcomings of the former and the
need for the latter.

\subsection{Signal-Strength Deterministic Model~\cite{avestimehr11}}
\label{sec:deterministic_salman}

We start with the real point-to-point Gaussian channel
\begin{equation}
    \label{eq:point-to-point}
    y[t] \triangleq 2^n h x[t] +z[t],
\end{equation}
with additive white Gaussian noise $z[t]\sim\mc{N}(0,1)$ and unit
average power constraint at the transmitter.  Here, $n$ is a nonnegative
integer, and $h \in [1, 2)$. Observe that all channel gains (and hence
$\SNR$s) greater than or equal to one can be expressed in the form
$2^nh$ for $n$ and $h$ satisfying these conditions. Since for a
constant-gap approximation the other cases are not relevant,
\eqref{eq:point-to-point} is essentially the general case.\footnote{If
the magnitude of the channel gains is less than one, then capacity is
less than one bit per channel use and hence not relevant for capacity
approximation up to a constant gap. Moreover, since capacity is only a
function of the magnitude of the channel gains, negative channel gains
are not relevant either.}

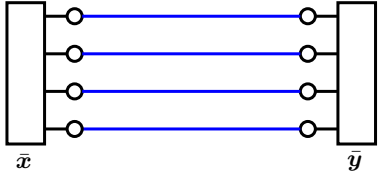
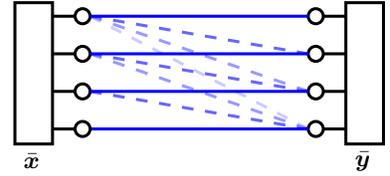
\begin{figure*}[!t]
    \centering
    \subfigure[Signal-strength deterministic model]
    {%
    % Generated with LaTeXDraw 2.0.8
% Tue Feb 12 11:18:40 EST 2013
% \usepackage[usenames,dvipsnames]{pstricks}
% \usepackage{epsfig}
% \usepackage{pst-grad} % For gradients
% \usepackage{pst-plot} % For axes
\scalebox{1} % Change this value to rescale the drawing.
{
\begin{pspicture}(0,-1.1845312)(6.4071875,1.1445312)
\psframe[linewidth=0.04,dimen=outer](1.263125,1.1445312)(0.723125,-0.77546877)
\usefont{T1}{ptm}{m}{n}
\rput(0.96671873,-0.98046875){\footnotesize $\bar{\bm{x}}$}
\usefont{T1}{ptm}{m}{n}
\rput(5.366719,-0.98046875){\footnotesize $\bar{\bm{y}}$}
\psline[linewidth=0.04cm](1.243125,0.94453126)(1.543125,0.94453126)
\pscircle[linewidth=0.04,dimen=outer](1.643125,0.94453126){0.12}
\psline[linewidth=0.04cm](1.243125,0.44453126)(1.543125,0.44453126)
\pscircle[linewidth=0.04,dimen=outer](1.643125,0.44453126){0.12}
\psline[linewidth=0.04cm](1.243125,-0.05546875)(1.543125,-0.05546875)
\pscircle[linewidth=0.04,dimen=outer](1.643125,-0.05546875){0.12}
\psframe[linewidth=0.04,dimen=outer](5.663125,1.1445312)(5.123125,-0.77546877)
\psline[linewidth=0.04cm](4.843125,0.94453126)(5.143125,0.94453126)
\pscircle[linewidth=0.04,dimen=outer](4.743125,0.94453126){0.12}
\psline[linewidth=0.04cm](4.843125,0.44453126)(5.143125,0.44453126)
\pscircle[linewidth=0.04,dimen=outer](4.743125,0.44453126){0.12}
\psline[linewidth=0.04cm](4.843125,-0.05546875)(5.143125,-0.05546875)
\pscircle[linewidth=0.04,dimen=outer](4.743125,-0.05546875){0.12}
\psline[linewidth=0.04cm](4.843125,-0.55546874)(5.143125,-0.55546874)
\pscircle[linewidth=0.04,dimen=outer](4.743125,-0.55546874){0.12}
\psline[linewidth=0.04cm,linecolor=blue](1.743125,0.44453126)(4.643125,0.44453126)
\psline[linewidth=0.04cm,linecolor=blue](1.743125,0.94453126)(4.643125,0.94453126)
\psline[linewidth=0.04cm,linecolor=blue](1.743125,-0.05546875)(4.643125,-0.05546875)
\psline[linewidth=0.04cm,linecolor=blue](1.743125,-0.55546874)(4.643125,-0.55546874)
\psline[linewidth=0.04cm](1.243125,-0.55546874)(1.543125,-0.55546874)
\pscircle[linewidth=0.04,dimen=outer](1.643125,-0.55546874){0.12}
\end{pspicture} 
}%
    \label{fig:Deter_Salman}%
    }
    \hfill
    \subfigure[Lower-triangular deterministic model]
    {%
    % Generated with LaTeXDraw 2.0.8
% Tue Feb 12 11:14:12 EST 2013
% \usepackage[usenames,dvipsnames]{pstricks}
% \usepackage{epsfig}
% \usepackage{pst-grad} % For gradients
% \usepackage{pst-plot} % For axes
\scalebox{1} % Change this value to rescale the drawing.
{
\begin{pspicture}(0,-1.1845312)(6.4071875,1.1445312)
\definecolor{color1887}{rgb}{0.8,0.8,1.0}
\definecolor{color1888}{rgb}{0.6,0.6,1.0}
\definecolor{color1890}{rgb}{0.4,0.4,1.0}
\psline[linewidth=0.04cm,linecolor=color1887,linestyle=dashed,dash=0.16cm 0.16cm](1.743125,0.94453126)(4.643125,-0.55546874)
\psline[linewidth=0.04cm,linecolor=color1888,linestyle=dashed,dash=0.16cm 0.16cm](1.743125,0.44453126)(4.643125,-0.55546874)
\psline[linewidth=0.04cm,linecolor=color1888,linestyle=dashed,dash=0.16cm 0.16cm](1.743125,0.94453126)(4.643125,-0.05546875)
\psline[linewidth=0.04cm,linecolor=color1890,linestyle=dashed,dash=0.16cm 0.16cm](1.743125,-0.05546875)(4.643125,-0.55546874)
\psline[linewidth=0.04cm,linecolor=color1890,linestyle=dashed,dash=0.16cm 0.16cm](1.743125,0.44453126)(4.643125,-0.05546875)
\psline[linewidth=0.04cm,linecolor=color1890,linestyle=dashed,dash=0.16cm 0.16cm](1.743125,0.94453126)(4.643125,0.44453126)
\psframe[linewidth=0.04,dimen=outer](1.263125,1.1445312)(0.723125,-0.77546877)
\usefont{T1}{ptm}{m}{n}
\rput(0.96671873,-0.98046875){\footnotesize $\bar{\bm{x}}$}
\usefont{T1}{ptm}{m}{n}
\rput(5.366719,-0.98046875){\footnotesize $\bar{\bm{y}}$}
\psline[linewidth=0.04cm](1.243125,0.94453126)(1.543125,0.94453126)
\pscircle[linewidth=0.04,dimen=outer](1.643125,0.94453126){0.12}
\psline[linewidth=0.04cm](1.243125,0.44453126)(1.543125,0.44453126)
\pscircle[linewidth=0.04,dimen=outer](1.643125,0.44453126){0.12}
\psline[linewidth=0.04cm](1.243125,-0.05546875)(1.543125,-0.05546875)
\pscircle[linewidth=0.04,dimen=outer](1.643125,-0.05546875){0.12}
\psframe[linewidth=0.04,dimen=outer](5.663125,1.1445312)(5.123125,-0.77546877)
\psline[linewidth=0.04cm](4.843125,0.94453126)(5.143125,0.94453126)
\pscircle[linewidth=0.04,dimen=outer](4.743125,0.94453126){0.12}
\psline[linewidth=0.04cm](4.843125,0.44453126)(5.143125,0.44453126)
\pscircle[linewidth=0.04,dimen=outer](4.743125,0.44453126){0.12}
\psline[linewidth=0.04cm](4.843125,-0.05546875)(5.143125,-0.05546875)
\pscircle[linewidth=0.04,dimen=outer](4.743125,-0.05546875){0.12}
\psline[linewidth=0.04cm](4.843125,-0.55546874)(5.143125,-0.55546874)
\pscircle[linewidth=0.04,dimen=outer](4.743125,-0.55546874){0.12}
\psline[linewidth=0.04cm,linecolor=blue](1.743125,0.44453126)(4.643125,0.44453126)
\psline[linewidth=0.04cm,linecolor=blue](1.743125,0.94453126)(4.643125,0.94453126)
\psline[linewidth=0.04cm,linecolor=blue](1.743125,-0.05546875)(4.643125,-0.05546875)
\psline[linewidth=0.04cm,linecolor=blue](1.743125,-0.55546874)(4.643125,-0.55546874)
\psline[linewidth=0.04cm](1.243125,-0.55546874)(1.543125,-0.55546874)
\pscircle[linewidth=0.04,dimen=outer](1.643125,-0.55546874){0.12}
\end{pspicture} 
}%
    \label{fig:Deter_LT}%
    }
    \caption{Comparison of the signal-strength deterministic
    model~\cite{avestimehr11}, and the lower-triangular deterministic
    model proposed in this paper. In the figure, solid lines depict
    noiseless binary links of capacity one bit per second. Dashed lines
    depict noiseless links of either capacity one or zero bits per
    channel use (depending on whether the corresponding entry in the channel
    matrix $\bar{\bm{H}}$ is one or zero). Links with the same
    color/shade have the same capacity.}
    \label{fig:Deterministic}
\end{figure*}

To develop the deterministic model and for simplicity, we assume that
$x[t]$ and  $z[t]$ are positive and upper bounded by one. We can then
write $x$ and $z$ in terms of their binary expansions as 
\begin{subequations}
    \label{eq:BE}
    \begin{align}
        \label{eq:XBE}    
        x & = \sum_{i=1}^{\infty }[x]_i 2^{-i}=0. [x]_{1} [x]_{2} [x]_{3} \ldots,\\
        \label{eq:ZBE}  
        z & = \sum_{i=1}^{\infty }[z]_i 2^{-i} =  0. [z]_{1} [z]_{2} [z]_{3} \ldots.
    \end{align}
\end{subequations}
The Gaussian point-to-point channel~\eqref{eq:point-to-point} can then
be approximated as
\begin{align*}
    y 
    & = \sum_{j=-\infty}^{\infty}\!\![y]_j 2^{-j} \\
    & \approx 2^n x +z  \\
    & = \sum_{j=1}^{n }[x]_{j} 2^{n-j}
    +\sum_{j=1}^{\infty}([x]_{j+n} + [z]_{j})2^{-j}\\
    & \approx \sum_{j=1}^{n }[x]_{j} 2^{n-j},
\end{align*}
or, more succinctly,
\begin{equation*}
    [y]_{j-n} 
    \approx [x]_{j}, \text{ for } 1 \leq j \leq n,
\end{equation*}
see Fig.~\ref{fig:Deter_Salman}.

The approximations in this derivation are to ignore the impact of
$h\in[1,2)$, the noise, as well as all bits $[x]_{n+1}, [x]_{n+2},
\ldots$ with exponent less than zero.  These bits with exponent less
than zero are approximated as being completely corrupted by noise,
whereas the bits with higher exponent are approximated as being received
noise free.  Therefore, we can approximate the Gaussian channel with a
deterministic channel consisting of $n$ parallel error-free links
from the transmitter to the receiver, each carrying one bit per channel
use. 

Having reviewed the signal-strength model for the point-to-point case,
we now turn to the Gaussian multiple-access channel
\begin{equation}
    \label{eq:mac}
    y[t] \triangleq 2^{n}h_1 x_1[t]+ 2^{n} h_2x_2[t] +z[t],
\end{equation}
where $z[t]\sim\mc{N}(0,1)$ is additive white Gaussian
noise.\footnote{For ease of exposition, we consider here the symmetric
case where both links have the same approximate strength $2^n$.} As
before, we impose a unit average transmit power constraint on $x_1[t]$
and $x_2[t]$. Moreover, $n$ is a nonnegative integer, and $h_1, h_2 \in
[1, 2)$.  The signal-strength deterministic model corresponding to the
Gaussian channel~\eqref{eq:mac} is
\begin{equation}
    \label{eq:mac_det1}
    [y]_{j-n} 
    \approx [x_1]_{j} \oplus [x_2]_{j}, \text{ for } 1 \leq j \leq n,
\end{equation}
where $\oplus$ denotes addition over $\Z_2$, i.e., modulo two. 

We note that in this model the contributions of $h_1$ and $h_2$ are
entirely ignored, real addition is replaced with bit-wise modulo-two
addition, and noise is eliminated. As mentioned earlier, this simple
model has been used to characterize the capacity region of several
challenging problems in network information theory to within a constant
gap. However, it falls short for some other settings. For example, for
certain relay networks with specific channel parameters, this model
incorrectly predicts capacity zero. Similarly, for interference channels
with more than two users and for X-channels, this model fails to predict
the correct behavior for the Gaussian case.

\subsection{Lower-Triangular Deterministic Model}
\label{sec:deterministic_triangular}

The signal-strength deterministic model recalled in the last section
ignores the contribution of $h\in[1,2)$ in the Gaussian point-to-point
channel~\eqref{eq:point-to-point}. Indeed, $h$ is
approximated by $1$. In this section, we introduce a new deterministic
channel model, termed \emph{lower-triangular deterministic model}, in
which the effect of $h$ is preserved. As we will see later, the new
deterministic model admits a wider range of solutions---a fact that will
be critical for the approximation of Gaussian networks with multiple
interfering signals.

Consider again the Gaussian point-to-point
channel~\eqref{eq:point-to-point}. Write the channel parameter $h\in[1,2)$
in terms of its binary expansion
\begin{equation*}
    h =\sum_{j=0}^{\infty }[h]_j 2^{-j}=[h]_{0}. [h]_{1} [h]_{2} [h]_{3} \ldots.
\end{equation*}
Observe that $[h]_{0}=1$, due to the  assumption that $h \in [1, 2)$. Then,
from~\eqref{eq:point-to-point} and~\eqref{eq:BE},
we have
\begin{align*}
    y 
    & = \sum_{j=-\infty}^{\infty}\!\![y]_j 2^{-j} \nonumber\\
    & = 2^n\biggl(\sum_{j=0}^{\infty}\, [h]_j 2^{-j}\biggr)
    \biggl(\sum_{i=1}^{\infty}\, [x]_i 2^{-i}\biggr)
    + \sum_{j=1}^{\infty}\, [z]_j 2^{-j} \nonumber\\
    & =
    \sum_{j=1}^{n}\biggl(\sum_{i=1}^{j} [h]_{j-i} [x]_i\biggr)  2^{n-j} + 
    \sum_{j=1}^{\infty}\!\biggl(\sum_{i=1}^{j+n} \, [h]_{j+n-i} [x]_i 
    + [z]_j\biggr)2^{-j}\nonumber\\
    & \approx
    \sum_{j=1}^{n}\Big(\sum_{i=1}^{j} 
    \, [h]_{j-i} [x]_i \Big)2^{n-j},
\end{align*}
so that
\begin{equation*}
    [y]_{j-n} \approx \sum_{i=1}^{j}\, [h]_{j-i} [x]_i, 
    \text{ for } 1 \leq j \leq n.
\end{equation*}
The approximation here is to ignore the
noise as well as all bits in the convolution of $1.
[h]_{1} [h]_{2} \ldots$ and $0 . [x]_{1} [x]_2 \ldots$ with
exponent less than zero. These bits with exponent less than zero are
approximated as being completely corrupted by noise, whereas the bits
with higher exponent are approximated as being received noise free.

This suggests to approximate the Gaussian point-to-point
channel~\eqref{eq:point-to-point} by a deterministic channel between the
binary input vector
\begin{equation*}
    \bar{\bm{x}} \triangleq
    \begin{pmatrix}
        \bar{x}_{1} & \bar{x}_{2} & \ldots & \bar{x}_{n}
    \end{pmatrix}^\T
\end{equation*}
and the binary output vector
\begin{equation*}
    \bar{\bm{y}} \triangleq 
    \begin{pmatrix}
        \bar{y}_{1} & \bar{y}_{2} & \ldots & \bar{y}_{n}
    \end{pmatrix}^\T
\end{equation*}
connected through the channel operation
\begin{equation}
    \label{eq:p2p_det2}
    \bar{\bm{y}} \triangleq
    \bar{\bm{H}}\bar{\bm{x}},
\end{equation}
with
\begin{equation*}
    \bar{\bm{H}} \triangleq
    \begin{pmatrix}
        1 & 0 & \cdots & 0 & 0 \\
        [h]_{1\phantom{-1}} & 1 & \cdots & 0 & 0 \\
        \vdots & \vdots & \ddots & \vdots & \vdots \\
        [h]_{n-2} & [h]_{n-3} &  \cdots & 1 & 0 \\
        [h]_{n-1} & [h]_{n-2} &  \cdots & [h]_{1} & 1_{\phantom{1}} 
    \end{pmatrix},
\end{equation*}
as depicted in Fig.~\ref{fig:Deter_LT}.  Here, we have normalized the received
vector $\bar{\bm{y}}$ to contain the bits from $1$ to $n$. This is a
deterministic channel with finite input and output alphabets. Note that
all operations in \eqref{eq:p2p_det2} are over $\Z_2$, i.e., modulo two.
Similarly, the Gaussian multiple-access channel~\eqref{eq:mac} can be
approximated by the deterministic channel model
\begin{equation}
    \label{eq:mac_det2}
    \bar{\bm{y}} \triangleq
    \bar{\bm{H}}_1\bar{\bm{x}}_1
    \oplus
    \bar{\bm{H}}_2\bar{\bm{x}}_2.
\end{equation}

\begin{example}
    \label{eg:detp2p}
    For a concrete example, consider the Gaussian point-to-point
    channel~\eqref{eq:point-to-point} with channel gain $21$, so that
    $n=4$ and  $h=1.3125$. The bits in the binary expansion of $h$
    are $[h]_0 = 1$, $[h]_1 = 0$, $[h]_2 = 1$, $[h]_3 = 0$, $[h]_4 = 1$,
    $[h]_5 = [h]_6 = \dots = 0$, and the corresponding lower-triangular
    deterministic model is depicted in Fig.~\ref{fig:detp2p}. For
    channel input $\bar{\bm{x}}$, the channel output is
    \begin{equation*}
        \bar{\bm{y}} = 
        \begin{pmatrix}
            \bar{x}_1 & \bar{x}_2 & \bar{x}_1\oplus\bar{x}_3 & \bar{x}_2\oplus\bar{x}_4
        \end{pmatrix}.
    \end{equation*}
    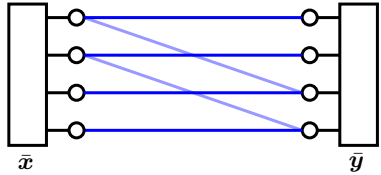
\begin{figure*}[htbp]
        \centering
        % Generated with LaTeXDraw 2.0.8
% Tue Feb 12 11:15:08 EST 2013
% \usepackage[usenames,dvipsnames]{pstricks}
% \usepackage{epsfig}
% \usepackage{pst-grad} % For gradients
% \usepackage{pst-plot} % For axes
\scalebox{1} % Change this value to rescale the drawing.
{
\begin{pspicture}(0,-1.1845312)(6.4071875,1.1445312)
\definecolor{color1888}{rgb}{0.6,0.6,1.0}
\psline[linewidth=0.04cm,linecolor=color1888](1.743125,0.44453126)(4.643125,-0.55546874)
\psline[linewidth=0.04cm,linecolor=color1888](1.743125,0.94453126)(4.643125,-0.05546875)
\psframe[linewidth=0.04,dimen=outer](1.263125,1.1445312)(0.723125,-0.77546877)
\usefont{T1}{ptm}{m}{n}
\rput(0.96671873,-0.98046875){\footnotesize $\bar{\bm{x}}$}
\usefont{T1}{ptm}{m}{n}
\rput(5.366719,-0.98046875){\footnotesize $\bar{\bm{y}}$}
\psline[linewidth=0.04cm](1.243125,0.94453126)(1.543125,0.94453126)
\pscircle[linewidth=0.04,dimen=outer](1.643125,0.94453126){0.12}
\psline[linewidth=0.04cm](1.243125,0.44453126)(1.543125,0.44453126)
\pscircle[linewidth=0.04,dimen=outer](1.643125,0.44453126){0.12}
\psline[linewidth=0.04cm](1.243125,-0.05546875)(1.543125,-0.05546875)
\pscircle[linewidth=0.04,dimen=outer](1.643125,-0.05546875){0.12}
\psframe[linewidth=0.04,dimen=outer](5.663125,1.1445312)(5.123125,-0.77546877)
\psline[linewidth=0.04cm](4.843125,0.94453126)(5.143125,0.94453126)
\pscircle[linewidth=0.04,dimen=outer](4.743125,0.94453126){0.12}
\psline[linewidth=0.04cm](4.843125,0.44453126)(5.143125,0.44453126)
\pscircle[linewidth=0.04,dimen=outer](4.743125,0.44453126){0.12}
\psline[linewidth=0.04cm](4.843125,-0.05546875)(5.143125,-0.05546875)
\pscircle[linewidth=0.04,dimen=outer](4.743125,-0.05546875){0.12}
\psline[linewidth=0.04cm](4.843125,-0.55546874)(5.143125,-0.55546874)
\pscircle[linewidth=0.04,dimen=outer](4.743125,-0.55546874){0.12}
\psline[linewidth=0.04cm,linecolor=blue](1.743125,0.44453126)(4.643125,0.44453126)
\psline[linewidth=0.04cm,linecolor=blue](1.743125,0.94453126)(4.643125,0.94453126)
\psline[linewidth=0.04cm,linecolor=blue](1.743125,-0.05546875)(4.643125,-0.05546875)
\psline[linewidth=0.04cm,linecolor=blue](1.743125,-0.55546874)(4.643125,-0.55546874)
\psline[linewidth=0.04cm](1.243125,-0.55546874)(1.543125,-0.55546874)
\pscircle[linewidth=0.04,dimen=outer](1.643125,-0.55546874){0.12}
\end{pspicture} 
}%
        \caption{Lower-triangular deterministic model for a
        point-to-point channel with $n=4$ and $h=1.3125$.}
        \label{fig:detp2p}
    \end{figure*}
\end{example}

\subsection{Comparison of Deterministic Models}
\label{sec:deterministic_comparison}

We now compare the signal-strength deterministic model reviewed in
Section~\ref{sec:deterministic_salman} and the lower-triangular
deterministic model introduced in
Section~\ref{sec:deterministic_triangular}. As an example, we consider
the Gaussian multiple-access channel \eqref{eq:mac} with signal
strength $n=4$. The corresponding deterministic models are given
by \eqref{eq:mac_det1} and \eqref{eq:mac_det2}.  Assume that transmitter
one wants to send three bits $a_1$, $a_2$, and $a_3$ to the receiver. At
the same time, transmitter two wants to send one bit $b_1$. 

\begin{figure*}[!t]
    \centering
    \subfigure[Signal-strength deterministic model]
    {%
    % Generated with LaTeXDraw 2.0.8
% Mon Nov 07 09:34:06 EST 2011
% \usepackage[usenames,dvipsnames]{pstricks}
% \usepackage{epsfig}
% \usepackage{pst-grad} % For gradients
% \usepackage{pst-plot} % For axes
\scalebox{1} % Change this value to rescale the drawing.
{
\begin{pspicture}(0,-2.4945312)(6.5071874,2.4545312)
\psframe[linewidth=0.04,dimen=outer](1.363125,2.4545312)(0.823125,0.51453125)
\usefont{T1}{ptm}{m}{n}
\rput(1.1267188,0.30953124){\footnotesize $\bar{\bm{x}}_1$}
\psline[linewidth=0.04cm](1.343125,2.2345312)(1.643125,2.2345312)
\pscircle[linewidth=0.04,dimen=outer](1.743125,2.2345312){0.12}
\psline[linewidth=0.04cm](1.343125,1.7345313)(1.643125,1.7345313)
\pscircle[linewidth=0.04,dimen=outer](1.743125,1.7345313){0.12}
\psline[linewidth=0.04cm](1.343125,1.2345313)(1.643125,1.2345313)
\pscircle[linewidth=0.04,dimen=outer](1.743125,1.2345313){0.12}
\psline[linewidth=0.04cm](1.343125,0.7345312)(1.643125,0.7345312)
\pscircle[linewidth=0.04,dimen=outer](1.743125,0.7345312){0.12}
\usefont{T1}{ptm}{m}{n}
\rput(5.4667187,-0.99046874){\footnotesize $\bar{\bm{y}}$}
\psframe[linewidth=0.04,dimen=outer](5.763125,1.1545312)(5.223125,-0.78546876)
\psline[linewidth=0.04cm](4.943125,0.9345313)(5.243125,0.9345313)
\pscircle[linewidth=0.04,dimen=outer](4.843125,0.9345313){0.12}
\psline[linewidth=0.04cm](4.943125,0.43453124)(5.243125,0.43453124)
\pscircle[linewidth=0.04,dimen=outer](4.843125,0.43453124){0.12}
\psline[linewidth=0.04cm](4.943125,-0.06546875)(5.243125,-0.06546875)
\pscircle[linewidth=0.04,dimen=outer](4.843125,-0.06546875){0.12}
\psline[linewidth=0.04cm](4.943125,-0.5654687)(5.243125,-0.5654687)
\pscircle[linewidth=0.04,dimen=outer](4.843125,-0.5654687){0.12}
\psline[linewidth=0.04cm,linecolor=blue](1.843125,0.7345312)(4.743125,-0.5654687)
\psframe[linewidth=0.04,dimen=outer](1.363125,-0.14546876)(0.823125,-2.0854688)
\usefont{T1}{ptm}{m}{n}
\rput(1.1267188,-2.2904687){\footnotesize $\bar{\bm{x}}_2$}
\psline[linewidth=0.04cm](1.343125,-0.36546874)(1.643125,-0.36546874)
\pscircle[linewidth=0.04,dimen=outer](1.743125,-0.36546874){0.12}
\psline[linewidth=0.04cm](1.343125,-0.86546874)(1.643125,-0.86546874)
\pscircle[linewidth=0.04,dimen=outer](1.743125,-0.86546874){0.12}
\psline[linewidth=0.04cm](1.343125,-1.3654687)(1.643125,-1.3654687)
\pscircle[linewidth=0.04,dimen=outer](1.743125,-1.3654687){0.12}
\psline[linewidth=0.04cm](1.343125,-1.8654687)(1.643125,-1.8654687)
\pscircle[linewidth=0.04,dimen=outer](1.743125,-1.8654687){0.12}
\psline[linewidth=0.04cm,linecolor=blue](1.843125,1.2345313)(4.743125,-0.06546875)
\psline[linewidth=0.04cm,linecolor=blue](1.843125,1.7345313)(4.743125,0.43453124)
\psline[linewidth=0.04cm,linecolor=blue](1.843125,2.2345312)(4.743125,0.9345313)
\psline[linewidth=0.04cm,linecolor=red](1.843125,-0.36546874)(4.743125,0.9345313)
\psline[linewidth=0.04cm,linecolor=red](1.843125,-0.86546874)(4.743125,0.43453124)
\psline[linewidth=0.04cm,linecolor=red](1.843125,-1.3654687)(4.743125,-0.06546875)
\psline[linewidth=0.04cm,linecolor=red](1.843125,-1.8654687)(4.743125,-0.5654687)
\usefont{T1}{ptm}{m}{n}
\rput(1.0967188,2.2295313){\footnotesize $a_1$}
\usefont{T1}{ptm}{m}{n}
\rput(1.0967188,1.7295313){\footnotesize $a_2$}
\usefont{T1}{ptm}{m}{n}
\rput(1.0967188,1.2295313){\footnotesize $a_3$}
\usefont{T1}{ptm}{m}{n}
\rput(1.1067188,-1.8704687){\footnotesize $b_1$}
\end{pspicture} 
}%
    \label{fig:Deter_Salman_MAC_Ach1}%
    }
    \hfill
    \subfigure[Lower-triangular deterministic model]
    {%
    % Generated with LaTeXDraw 2.0.8
% Mon Nov 07 09:33:32 EST 2011
% \usepackage[usenames,dvipsnames]{pstricks}
% \usepackage{epsfig}
% \usepackage{pst-grad} % For gradients
% \usepackage{pst-plot} % For axes
\scalebox{1} % Change this value to rescale the drawing.
{
\begin{pspicture}(0,-2.4845312)(6.5071874,2.4645312)
\definecolor{color2754}{rgb}{0.8,0.8,1.0}
\definecolor{color2745}{rgb}{0.6,0.6,1.0}
\definecolor{color2730}{rgb}{0.4,0.4,1.0}
\definecolor{color2779}{rgb}{1.0,0.8,0.8}
\definecolor{color2771}{rgb}{1.0,0.6,0.6}
\definecolor{color2756}{rgb}{1.0,0.4,0.4}
\psline[linewidth=0.04cm,linecolor=color2754,linestyle=dashed,dash=0.16cm 0.16cm](1.843125,2.2445312)(4.743125,-0.55546874)
\psline[linewidth=0.04cm,linecolor=color2745,linestyle=dashed,dash=0.16cm 0.16cm](1.843125,2.2445312)(4.743125,-0.05546875)
\psline[linewidth=0.04cm,linecolor=color2745,linestyle=dashed,dash=0.16cm 0.16cm](1.843125,1.7445313)(4.743125,-0.55546874)
\psline[linewidth=0.04cm,linecolor=color2730,linestyle=dashed,dash=0.16cm 0.16cm](1.843125,2.2445312)(4.743125,0.44453126)
\psline[linewidth=0.04cm,linecolor=color2730,linestyle=dashed,dash=0.16cm 0.16cm](1.843125,1.7445313)(4.743125,-0.05546875)
\psline[linewidth=0.04cm,linecolor=color2730,linestyle=dashed,dash=0.16cm 0.16cm](1.843125,1.2445313)(4.743125,-0.55546874)
\psline[linewidth=0.04cm,linecolor=blue](1.843125,0.7445313)(4.743125,-0.55546874)
\psline[linewidth=0.04cm,linecolor=blue](1.843125,1.2445313)(4.743125,-0.05546875)
\psline[linewidth=0.04cm,linecolor=blue](1.843125,1.7445313)(4.743125,0.44453126)
\psline[linewidth=0.04cm,linecolor=blue](1.843125,2.2445312)(4.743125,0.94453126)
\psline[linewidth=0.04cm,linecolor=color2779,linestyle=dashed,dash=0.16cm 0.16cm](1.843125,-0.35546875)(4.743125,-0.55546874)
\psline[linewidth=0.04cm,linecolor=color2771,linestyle=dashed,dash=0.16cm 0.16cm](1.843125,-0.35546875)(4.743125,-0.05546875)
\psline[linewidth=0.04cm,linecolor=color2771,linestyle=dashed,dash=0.16cm 0.16cm](1.843125,-0.85546875)(4.743125,-0.55546874)
\psline[linewidth=0.04cm,linecolor=color2756,linestyle=dashed,dash=0.16cm 0.16cm](1.843125,-0.35546875)(4.743125,0.44453126)
\psline[linewidth=0.04cm,linecolor=color2756,linestyle=dashed,dash=0.16cm 0.16cm](1.843125,-0.85546875)(4.743125,-0.05546875)
\psline[linewidth=0.04cm,linecolor=color2756,linestyle=dashed,dash=0.16cm 0.16cm](1.843125,-1.3554688)(4.743125,-0.55546874)
\psframe[linewidth=0.04,dimen=middle](1.343125,2.4445312)(0.843125,0.5445312)
\usefont{T1}{ptm}{m}{n}
\rput(1.1267188,0.31953126){\footnotesize $\bar{\bm{x}}_1$}
\psline[linewidth=0.04cm](1.343125,2.2445312)(1.643125,2.2445312)
\pscircle[linewidth=0.04,dimen=outer](1.743125,2.2445312){0.12}
\psline[linewidth=0.04cm](1.343125,1.7445313)(1.643125,1.7445313)
\pscircle[linewidth=0.04,dimen=outer](1.743125,1.7445313){0.12}
\psline[linewidth=0.04cm](1.343125,1.2445313)(1.643125,1.2445313)
\pscircle[linewidth=0.04,dimen=outer](1.743125,1.2445313){0.12}
\psline[linewidth=0.04cm](1.343125,0.7445313)(1.643125,0.7445313)
\pscircle[linewidth=0.04,dimen=outer](1.743125,0.7445313){0.12}
\usefont{T1}{ptm}{m}{n}
\rput(5.4667187,-0.98046875){\footnotesize $\bar{\bm{y}}$}
\psframe[linewidth=0.04,dimen=middle](5.743125,1.1445312)(5.243125,-0.7554687)
\psline[linewidth=0.04cm](4.943125,0.94453126)(5.243125,0.94453126)
\pscircle[linewidth=0.04,dimen=outer](4.843125,0.94453126){0.12}
\psline[linewidth=0.04cm](4.943125,0.44453126)(5.243125,0.44453126)
\pscircle[linewidth=0.04,dimen=outer](4.843125,0.44453126){0.12}
\psline[linewidth=0.04cm](4.943125,-0.05546875)(5.243125,-0.05546875)
\pscircle[linewidth=0.04,dimen=outer](4.843125,-0.05546875){0.12}
\psline[linewidth=0.04cm](4.943125,-0.55546874)(5.243125,-0.55546874)
\pscircle[linewidth=0.04,dimen=outer](4.843125,-0.55546874){0.12}
\psframe[linewidth=0.04,dimen=middle](1.343125,-0.15546875)(0.843125,-2.0554688)
\usefont{T1}{ptm}{m}{n}
\rput(1.1267188,-2.2804687){\footnotesize $\bar{\bm{x}}_2$}
\psline[linewidth=0.04cm](1.343125,-0.35546875)(1.643125,-0.35546875)
\pscircle[linewidth=0.04,dimen=outer](1.743125,-0.35546875){0.12}
\psline[linewidth=0.04cm](1.343125,-0.85546875)(1.643125,-0.85546875)
\pscircle[linewidth=0.04,dimen=outer](1.743125,-0.85546875){0.12}
\psline[linewidth=0.04cm](1.343125,-1.3554688)(1.643125,-1.3554688)
\pscircle[linewidth=0.04,dimen=outer](1.743125,-1.3554688){0.12}
\psline[linewidth=0.04cm](1.343125,-1.8554688)(1.643125,-1.8554688)
\pscircle[linewidth=0.04,dimen=outer](1.743125,-1.8554688){0.12}
\psline[linewidth=0.04cm,linecolor=red](1.843125,-0.35546875)(4.743125,0.94453126)
\psline[linewidth=0.04cm,linecolor=red](1.843125,-0.85546875)(4.743125,0.44453126)
\psline[linewidth=0.04cm,linecolor=red](1.843125,-1.3554688)(4.743125,-0.05546875)
\psline[linewidth=0.04cm,linecolor=red](1.843125,-1.8554688)(4.743125,-0.55546874)
\usefont{T1}{ptm}{m}{n}
\rput(1.0967188,2.2395313){\footnotesize $a_1$}
\usefont{T1}{ptm}{m}{n}
\rput(1.0967188,1.7395313){\footnotesize $a_2$}
\usefont{T1}{ptm}{m}{n}
\rput(1.0967188,1.2395313){\footnotesize $a_3$}
\usefont{T1}{ptm}{m}{n}
\rput(1.1067188,-1.8604687){\footnotesize $b_1$}
\end{pspicture} 
}%
    \label{fig:Deter_LT_MAC_Ach1}%
    }
    \caption{Permissible signaling schemes for both deterministic models.}
    \label{fig:Deterministic_Similartites}
\end{figure*}
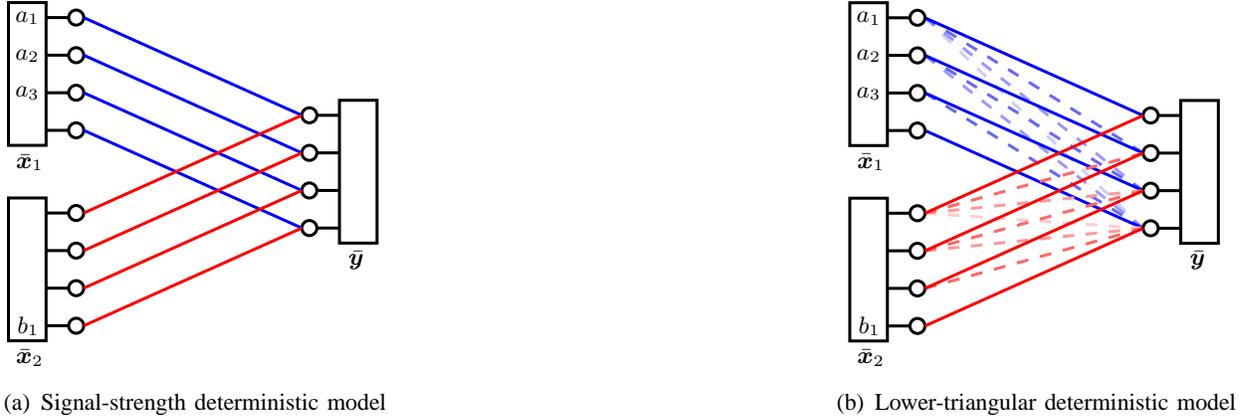

Some signaling schemes work for both deterministic
models~\eqref{eq:mac_det1} and~\eqref{eq:mac_det2}. For example, in both
models transmitter one can use the first three layers to send $a_1$,
$a_2$, and $a_3$, while transmitter two can use the last layer to send
$b_1$, as shown in Fig.~\ref{fig:Deterministic_Similartites}.  For the
signal-strength model, the decoding scheme is trivial. For the
lower-triangular model, the receiver starts by decoding the highest
layer containing only $a_1$. Having recovered $a_1$, the receiver
cancels out its contribution in all lower layers. The decoding process
continues in the same manner with $a_2$ at the second-highest layer,
until all bits are decoded.

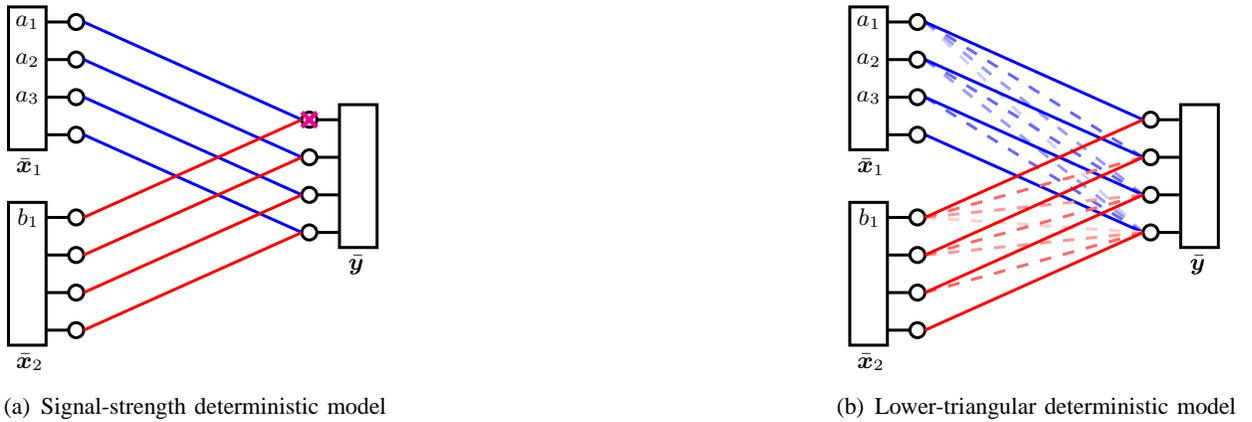
\begin{figure*}[!t]
    \centering
    \subfigure[Signal-strength deterministic model]
    {%
    % Generated with LaTeXDraw 2.0.8
% Mon Nov 07 09:35:36 EST 2011
% \usepackage[usenames,dvipsnames]{pstricks}
% \usepackage{epsfig}
% \usepackage{pst-grad} % For gradients
% \usepackage{pst-plot} % For axes
\scalebox{1} % Change this value to rescale the drawing.
{
\begin{pspicture}(0,-2.4945312)(6.5071874,2.4545312)
\psframe[linewidth=0.04,dimen=outer](1.363125,2.4545312)(0.823125,0.51453125)
\usefont{T1}{ptm}{m}{n}
\rput(1.1267188,0.30953124){\footnotesize $\bar{\bm{x}}_1$}
\psline[linewidth=0.04cm](1.343125,2.2345312)(1.643125,2.2345312)
\pscircle[linewidth=0.04,dimen=outer](1.743125,2.2345312){0.12}
\psline[linewidth=0.04cm](1.343125,1.7345313)(1.643125,1.7345313)
\pscircle[linewidth=0.04,dimen=outer](1.743125,1.7345313){0.12}
\psline[linewidth=0.04cm](1.343125,1.2345313)(1.643125,1.2345313)
\pscircle[linewidth=0.04,dimen=outer](1.743125,1.2345313){0.12}
\psline[linewidth=0.04cm](1.343125,0.7345312)(1.643125,0.7345312)
\pscircle[linewidth=0.04,dimen=outer](1.743125,0.7345312){0.12}
\usefont{T1}{ptm}{m}{n}
\rput(5.4667187,-0.99046874){\footnotesize $\bar{\bm{y}}$}
\psframe[linewidth=0.04,dimen=outer](5.763125,1.1545312)(5.223125,-0.78546876)
\psline[linewidth=0.04cm](4.943125,0.9345313)(5.243125,0.9345313)
\pscircle[linewidth=0.04,dimen=outer](4.843125,0.9345313){0.12}
\psline[linewidth=0.04cm](4.943125,0.43453124)(5.243125,0.43453124)
\pscircle[linewidth=0.04,dimen=outer](4.843125,0.43453124){0.12}
\psline[linewidth=0.04cm](4.943125,-0.06546875)(5.243125,-0.06546875)
\pscircle[linewidth=0.04,dimen=outer](4.843125,-0.06546875){0.12}
\psline[linewidth=0.04cm](4.943125,-0.5654687)(5.243125,-0.5654687)
\pscircle[linewidth=0.04,dimen=outer](4.843125,-0.5654687){0.12}
\psline[linewidth=0.04cm,linecolor=blue](1.843125,0.7345312)(4.743125,-0.5654687)
\psframe[linewidth=0.04,dimen=outer](1.363125,-0.14546876)(0.823125,-2.0854688)
\usefont{T1}{ptm}{m}{n}
\rput(1.1267188,-2.2904687){\footnotesize $\bar{\bm{x}}_2$}
\psline[linewidth=0.04cm](1.343125,-0.36546874)(1.643125,-0.36546874)
\pscircle[linewidth=0.04,dimen=outer](1.743125,-0.36546874){0.12}
\psline[linewidth=0.04cm](1.343125,-0.86546874)(1.643125,-0.86546874)
\pscircle[linewidth=0.04,dimen=outer](1.743125,-0.86546874){0.12}
\psline[linewidth=0.04cm](1.343125,-1.3654687)(1.643125,-1.3654687)
\pscircle[linewidth=0.04,dimen=outer](1.743125,-1.3654687){0.12}
\psline[linewidth=0.04cm](1.343125,-1.8654687)(1.643125,-1.8654687)
\pscircle[linewidth=0.04,dimen=outer](1.743125,-1.8654687){0.12}
\psline[linewidth=0.04cm,linecolor=blue](1.843125,1.2345313)(4.743125,-0.06546875)
\psline[linewidth=0.04cm,linecolor=blue](1.843125,1.7345313)(4.743125,0.43453124)
\psline[linewidth=0.04cm,linecolor=blue](1.843125,2.2345312)(4.743125,0.9345313)
\psline[linewidth=0.04cm,linecolor=red](1.843125,-0.36546874)(4.743125,0.9345313)
\psline[linewidth=0.04cm,linecolor=red](1.843125,-0.86546874)(4.743125,0.43453124)
\psline[linewidth=0.04cm,linecolor=red](1.843125,-1.3654687)(4.743125,-0.06546875)
\psline[linewidth=0.04cm,linecolor=red](1.843125,-1.8654687)(4.743125,-0.5654687)
\usefont{T1}{ptm}{m}{n}
\rput(1.0967188,2.2295313){\footnotesize $a_1$}
\usefont{T1}{ptm}{m}{n}
\rput(1.0967188,1.7295313){\footnotesize $a_2$}
\usefont{T1}{ptm}{m}{n}
\rput(1.0967188,1.2295313){\footnotesize $a_3$}
\usefont{T1}{ptm}{m}{n}
\rput(1.1067188,-0.37046874){\footnotesize $b_1$}
\psline[linewidth=0.06cm,linecolor=magenta](4.743125,1.0345312)(4.943125,0.83453125)
\psline[linewidth=0.06cm,linecolor=magenta](4.743125,0.83453125)(4.943125,1.0345312)
\end{pspicture} 
}%
    \label{fig:Deter_Salman_MAC_Ach2}%
    }
    \hfill
    \subfigure[Lower-triangular deterministic model]
    {%
    % Generated with LaTeXDraw 2.0.8
% Mon Nov 07 09:36:02 EST 2011
% \usepackage[usenames,dvipsnames]{pstricks}
% \usepackage{epsfig}
% \usepackage{pst-grad} % For gradients
% \usepackage{pst-plot} % For axes
\scalebox{1} % Change this value to rescale the drawing.
{
\begin{pspicture}(0,-2.4945312)(6.5071874,2.4545312)
\definecolor{color2930}{rgb}{0.8,0.8,1.0}
\definecolor{color2931}{rgb}{0.6,0.6,1.0}
\definecolor{color2933}{rgb}{0.4,0.4,1.0}
\definecolor{color2940}{rgb}{1.0,0.8,0.8}
\definecolor{color2941}{rgb}{1.0,0.6,0.6}
\definecolor{color2943}{rgb}{1.0,0.4,0.4}
\psline[linewidth=0.04cm,linecolor=color2930,linestyle=dashed,dash=0.16cm 0.16cm](1.843125,2.2345312)(4.743125,-0.5654687)
\psline[linewidth=0.04cm,linecolor=color2931,linestyle=dashed,dash=0.16cm 0.16cm](1.843125,2.2345312)(4.743125,-0.06546875)
\psline[linewidth=0.04cm,linecolor=color2931,linestyle=dashed,dash=0.16cm 0.16cm](1.843125,1.7345313)(4.743125,-0.5654687)
\psline[linewidth=0.04cm,linecolor=color2933,linestyle=dashed,dash=0.16cm 0.16cm](1.843125,2.2345312)(4.743125,0.43453124)
\psline[linewidth=0.04cm,linecolor=color2933,linestyle=dashed,dash=0.16cm 0.16cm](1.843125,1.7345313)(4.743125,-0.06546875)
\psline[linewidth=0.04cm,linecolor=color2933,linestyle=dashed,dash=0.16cm 0.16cm](1.843125,1.2345313)(4.743125,-0.5654687)
\psline[linewidth=0.04cm,linecolor=blue](1.843125,0.7345312)(4.743125,-0.5654687)
\psline[linewidth=0.04cm,linecolor=blue](1.843125,1.2345313)(4.743125,-0.06546875)
\psline[linewidth=0.04cm,linecolor=blue](1.843125,1.7345313)(4.743125,0.43453124)
\psline[linewidth=0.04cm,linecolor=blue](1.843125,2.2345312)(4.743125,0.9345313)
\psline[linewidth=0.04cm,linecolor=color2940,linestyle=dashed,dash=0.16cm 0.16cm](1.843125,-0.36546874)(4.743125,-0.5654687)
\psline[linewidth=0.04cm,linecolor=color2941,linestyle=dashed,dash=0.16cm 0.16cm](1.843125,-0.36546874)(4.743125,-0.06546875)
\psline[linewidth=0.04cm,linecolor=color2941,linestyle=dashed,dash=0.16cm 0.16cm](1.843125,-0.86546874)(4.743125,-0.5654687)
\psline[linewidth=0.04cm,linecolor=color2943,linestyle=dashed,dash=0.16cm 0.16cm](1.843125,-0.36546874)(4.743125,0.43453124)
\psline[linewidth=0.04cm,linecolor=color2943,linestyle=dashed,dash=0.16cm 0.16cm](1.843125,-0.86546874)(4.743125,-0.06546875)
\psline[linewidth=0.04cm,linecolor=color2943,linestyle=dashed,dash=0.16cm 0.16cm](1.843125,-1.3654687)(4.743125,-0.5654687)
\psframe[linewidth=0.04,dimen=outer](1.363125,2.4545312)(0.823125,0.51453125)
\usefont{T1}{ptm}{m}{n}
\rput(1.1267188,0.30953124){\footnotesize $\bar{\bm{x}}_1$}
\psline[linewidth=0.04cm](1.343125,2.2345312)(1.643125,2.2345312)
\pscircle[linewidth=0.04,dimen=outer](1.743125,2.2345312){0.12}
\psline[linewidth=0.04cm](1.343125,1.7345313)(1.643125,1.7345313)
\pscircle[linewidth=0.04,dimen=outer](1.743125,1.7345313){0.12}
\psline[linewidth=0.04cm](1.343125,1.2345313)(1.643125,1.2345313)
\pscircle[linewidth=0.04,dimen=outer](1.743125,1.2345313){0.12}
\psline[linewidth=0.04cm](1.343125,0.7345312)(1.643125,0.7345312)
\pscircle[linewidth=0.04,dimen=outer](1.743125,0.7345312){0.12}
\usefont{T1}{ptm}{m}{n}
\rput(5.4667187,-0.99046874){\footnotesize $\bar{\bm{y}}$}
\psframe[linewidth=0.04,dimen=outer](5.763125,1.1545312)(5.223125,-0.78546876)
\psline[linewidth=0.04cm](4.943125,0.9345313)(5.243125,0.9345313)
\pscircle[linewidth=0.04,dimen=outer](4.843125,0.9345313){0.12}
\psline[linewidth=0.04cm](4.943125,0.43453124)(5.243125,0.43453124)
\pscircle[linewidth=0.04,dimen=outer](4.843125,0.43453124){0.12}
\psline[linewidth=0.04cm](4.943125,-0.06546875)(5.243125,-0.06546875)
\pscircle[linewidth=0.04,dimen=outer](4.843125,-0.06546875){0.12}
\psline[linewidth=0.04cm](4.943125,-0.5654687)(5.243125,-0.5654687)
\pscircle[linewidth=0.04,dimen=outer](4.843125,-0.5654687){0.12}
\psframe[linewidth=0.04,dimen=outer](1.363125,-0.14546876)(0.823125,-2.0854688)
\usefont{T1}{ptm}{m}{n}
\rput(1.1267188,-2.2904687){\footnotesize $\bar{\bm{x}}_2$}
\psline[linewidth=0.04cm](1.343125,-0.36546874)(1.643125,-0.36546874)
\pscircle[linewidth=0.04,dimen=outer](1.743125,-0.36546874){0.12}
\psline[linewidth=0.04cm](1.343125,-0.86546874)(1.643125,-0.86546874)
\pscircle[linewidth=0.04,dimen=outer](1.743125,-0.86546874){0.12}
\psline[linewidth=0.04cm](1.343125,-1.3654687)(1.643125,-1.3654687)
\pscircle[linewidth=0.04,dimen=outer](1.743125,-1.3654687){0.12}
\psline[linewidth=0.04cm](1.343125,-1.8654687)(1.643125,-1.8654687)
\pscircle[linewidth=0.04,dimen=outer](1.743125,-1.8654687){0.12}
\psline[linewidth=0.04cm,linecolor=red](1.843125,-0.36546874)(4.743125,0.9345313)
\psline[linewidth=0.04cm,linecolor=red](1.843125,-0.86546874)(4.743125,0.43453124)
\psline[linewidth=0.04cm,linecolor=red](1.843125,-1.3654687)(4.743125,-0.06546875)
\psline[linewidth=0.04cm,linecolor=red](1.843125,-1.8654687)(4.743125,-0.5654687)
\usefont{T1}{ptm}{m}{n}
\rput(1.0967188,2.2295313){\footnotesize $a_1$}
\usefont{T1}{ptm}{m}{n}
\rput(1.0967188,1.7295313){\footnotesize $a_2$}
\usefont{T1}{ptm}{m}{n}
\rput(1.0967188,1.2295313){\footnotesize $a_3$}
\usefont{T1}{ptm}{m}{n}
\rput(1.1067188,-0.37046874){\footnotesize $b_1$}
\end{pspicture} 
}%
    \label{fig:Deter_LT_MAC_Ach2}%
    }
    \caption{Illustration of a signaling scheme that succeeds for the
    lower-triangular model (assuming the subspace
    condition~\eqref{eq:Ex_matrix_condition} holds), but fails for the
    signal-strength model.}
    \label{fig:Deterministic_Difference}
\end{figure*}

There are, however, some signaling schemes that are only decodable in the
lower-triangular model, but not in the signal-strength model. An example
of such a signaling scheme is depicted in
Fig.~\ref{fig:Deterministic_Difference}. In this scheme, transmitter one
uses again the first three layers to send $a_1$, $a_2$, and $a_3$.
Unlike before, transmitter two now also uses the first layer to send $b_1$.
From Fig.~\ref{fig:Deter_Salman_MAC_Ach2}, we can see that, in the
signal-strength model, receiver one observes $a_1\oplus b_1$ and cannot
recover $a_1$ and $b_1$ from the received signal. However, this scheme
can be utilized successfully in the lower-triangular model as long as
the subspaces spanned by the message bits at the receivers are linearly
independent. In this case, the subspace spanned by the first three
columns of $\bar{\bm{H}}_1$ and the subspace spanned by the first
columns of $\bar{\bm{H}}_2$ need to be linearly independent.
This is the case if and only if
\begin{equation}
    \label{eq:Ex_matrix_condition}
    \det
    \begin{pmatrix}
        1         & 0         & 0         & 1   \\
        [h_1]_{1} & 1         & 0         & [h_2]_{1} \\
        [h_1]_{2} & [h_1]_{1} & 1         & [h_2]_{2} \\
        [h_1]_{3} & [h_1]_{2} & [h_1]_{1} & [h_2]_{3} 
    \end{pmatrix}
    \neq 0.
\end{equation}

The event \eqref{eq:Ex_matrix_condition} depends not only on $n$, but
also on the bits in the binary expansion of $h_1$ and $h_2$.  Thus, this
scheme is successful for all channel gains $(h_1,
h_2)\in(1,2]^2\setminus B$, where $B$ is the event that
\eqref{eq:Ex_matrix_condition} does not hold. The set $B$ can be
understood as an outage event: if the channel gains are in $B$, the
achievable scheme fails to deliver the desired target rate of $4$ bits
per channel use.

Noting that the scheme depicted in Fig.~\ref{fig:Deter_LT_MAC_Ach1}
always works while the scheme depicted in
Fig.~\ref{fig:Deter_LT_MAC_Ach2} only works under some conditions, one
might question the relevance of the second class of solutions. The
answer is that this second class of solutions make use of the
``diversity'' provided by the lower-order bits of the channel gains. It
is precisely this diversity that is required for efficient communication
over the X-channel to be investigated in Section~\ref{sec:main}.

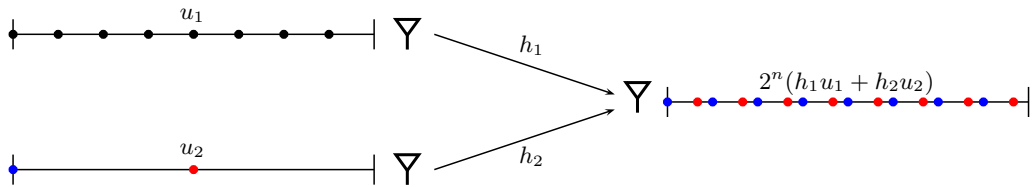
\begin{figure}[htbp]
    \centering
    %\hspace{-0.2cm}
    % Generated with LaTeXDraw 2.0.8
% Tue Nov 08 09:34:17 EST 2011
% \usepackage[usenames,dvipsnames]{pstricks}
% \usepackage{epsfig}
% \usepackage{pst-grad} % For gradients
% \usepackage{pst-plot} % For axes
\scalebox{1} % Change this value to rescale the drawing.
{
\begin{pspicture}(0,-1.2061554)(13.57,1.2461554)
\psline[linewidth=0.02cm](0.06,-0.78353214)(0.06,-1.1835321)
\psline[linewidth=0.02cm](0.06,-0.98353213)(4.86,-0.98353213)
\pspolygon[linewidth=0.04](8.208567,0.17420271)(8.514077,0.17420271)(8.361322,-0.035985287)
\psline[linewidth=0.04cm](8.361322,-0.015967382)(8.361322,-0.22615537)
\pspolygon[linewidth=0.04](5.148567,1.0142027)(5.454077,1.0142027)(5.301322,0.80401474)
\psline[linewidth=0.04cm](5.301322,0.8240326)(5.301322,0.61384463)
\psline[linewidth=0.02cm,arrowsize=0.05291667cm 2.0,arrowlength=1.4,arrowinset=0.4]{->}(5.66,0.8164679)(8.06,0.016467877)
\psline[linewidth=0.02cm,arrowsize=0.05291667cm 2.0,arrowlength=1.4,arrowinset=0.4]{->}(5.66,-0.98353213)(8.06,-0.18353212)
\usefont{T1}{ptm}{m}{n}
\rput(6.9535937,0.6714679){\footnotesize $h_1$}
\psline[linewidth=0.02cm](0.06,0.8164679)(4.86,0.8164679)
\usefont{T1}{ptm}{m}{n}
\rput(2.443125,1.0714679){\footnotesize $u_1$}
\usefont{T1}{ptm}{m}{n}
\rput(11.143125,0.17146787){\footnotesize $2^n( h_1 u_1 + h_2 u_2 )$}
\pscircle[linewidth=0.06,dimen=outer,fillstyle=solid,fillcolor=black](0.06,0.8164679){0.06}
\pscircle[linewidth=0.06,dimen=outer,fillstyle=solid,fillcolor=black](0.66,0.8164679){0.06}
\pscircle[linewidth=0.06,dimen=outer,fillstyle=solid,fillcolor=black](4.26,0.8164679){0.06}
\pscircle[linewidth=0.06,dimen=outer,fillstyle=solid,fillcolor=black](1.26,0.8164679){0.06}
\pscircle[linewidth=0.06,dimen=outer,fillstyle=solid,fillcolor=black](1.86,0.8164679){0.06}
\pscircle[linewidth=0.06,dimen=outer,fillstyle=solid,fillcolor=black](3.66,0.8164679){0.06}
\pscircle[linewidth=0.06,dimen=outer,fillstyle=solid,fillcolor=black](2.46,0.8164679){0.06}
\pscircle[linewidth=0.06,dimen=outer,fillstyle=solid,fillcolor=black](3.06,0.8164679){0.06}
\psline[linewidth=0.02cm](0.06,1.0164679)(0.06,0.6164679)
\psline[linewidth=0.02cm](4.86,1.0164679)(4.86,0.6164679)
\pscircle[linewidth=0.06,linecolor=blue,dimen=outer,fillstyle=solid,fillcolor=black](0.06,-0.98353213){0.06}
\pscircle[linewidth=0.06,linecolor=red,dimen=outer,fillstyle=solid,fillcolor=black](2.46,-0.98353213){0.06}
\psline[linewidth=0.02cm](4.86,-0.78353214)(4.86,-1.1835321)
\pspolygon[linewidth=0.04](5.148567,-0.7857973)(5.454077,-0.7857973)(5.301322,-0.99598527)
\psline[linewidth=0.04cm](5.301322,-0.9759674)(5.301322,-1.1861553)
\usefont{T1}{ptm}{m}{n}
\rput(6.9535937,-0.8085321){\footnotesize $h_2$}
\usefont{T1}{ptm}{m}{n}
\rput(2.443125,-0.72853214){\footnotesize $u_2$}
\psline[linewidth=0.02cm](8.76,-0.083532125)(13.56,-0.083532125)
\psline[linewidth=0.02cm](8.76,0.11646788)(8.76,-0.2835321)
\psline[linewidth=0.02cm](13.56,0.11646788)(13.56,-0.2835321)
\pscircle[linewidth=0.06,linecolor=red,dimen=outer,fillstyle=solid,fillcolor=black](9.16,-0.083532125){0.06}
\pscircle[linewidth=0.06,linecolor=red,dimen=outer,fillstyle=solid,fillcolor=black](9.76,-0.083532125){0.06}
\pscircle[linewidth=0.06,linecolor=red,dimen=outer,fillstyle=solid,fillcolor=black](13.36,-0.083532125){0.06}
\pscircle[linewidth=0.06,linecolor=red,dimen=outer,fillstyle=solid,fillcolor=black](10.36,-0.083532125){0.06}
\pscircle[linewidth=0.06,linecolor=red,dimen=outer,fillstyle=solid,fillcolor=black](10.96,-0.083532125){0.06}
\pscircle[linewidth=0.06,linecolor=red,dimen=outer,fillstyle=solid,fillcolor=black](12.76,-0.083532125){0.06}
\pscircle[linewidth=0.06,linecolor=red,dimen=outer,fillstyle=solid,fillcolor=black](11.56,-0.083532125){0.06}
\pscircle[linewidth=0.06,linecolor=red,dimen=outer,fillstyle=solid,fillcolor=black](12.16,-0.083532125){0.06}
\pscircle[linewidth=0.06,linecolor=blue,dimen=outer,fillstyle=solid,fillcolor=black](8.76,-0.083532125){0.06}
\pscircle[linewidth=0.06,linecolor=blue,dimen=outer,fillstyle=solid,fillcolor=black](9.36,-0.083532125){0.06}
\pscircle[linewidth=0.06,linecolor=blue,dimen=outer,fillstyle=solid,fillcolor=black](12.96,-0.083532125){0.06}
\pscircle[linewidth=0.06,linecolor=blue,dimen=outer,fillstyle=solid,fillcolor=black](9.96,-0.083532125){0.06}
\pscircle[linewidth=0.06,linecolor=blue,dimen=outer,fillstyle=solid,fillcolor=black](10.56,-0.083532125){0.06}
\pscircle[linewidth=0.06,linecolor=blue,dimen=outer,fillstyle=solid,fillcolor=black](12.36,-0.083532125){0.06}
\pscircle[linewidth=0.06,linecolor=blue,dimen=outer,fillstyle=solid,fillcolor=black](11.16,-0.083532125){0.06}
\pscircle[linewidth=0.06,linecolor=blue,dimen=outer,fillstyle=solid,fillcolor=black](11.76,-0.083532125){0.06}
\end{pspicture} 
}
    \caption{Modulation scheme for the Gaussian model suggested by the
    signaling scheme for the lower-triangular deterministic model
    depicted in Fig.~\ref{fig:Deter_LT_MAC_Ach2}. At the decoder, blue
    dots correspond to input tuples $(u_1,0)$ with
    $u_1\in\{0,1/8,\ldots,7/8\}$, and red dots correspond to input
    tuples $(u_1,1/2)$ with $u_1\in\{0,1/8,\ldots,7/8\}$. Here,
    $n=h_1=1$ and $h_2=1/6$.} 
    \label{fig:Mac_Scheme}
\end{figure}

As pointed out earlier, the second step in using the deterministic
approach is to translate the solution for the deterministic model to a
solution for the original Gaussian model. We now show how this can be
done for the signaling scheme shown in Fig.~\ref{fig:Deter_LT_MAC_Ach2}.
The proposed scheme for the Gaussian multiple-access channel is depicted
in Fig.~\ref{fig:Mac_Scheme}. In this scheme, the input constellation
at transmitter one is the set $\{0,1/8,\ldots, 7/8\}$, and the
input constellation at transmitter two is the set $\{0,1/2\}$.
Since the additive Gaussian receiver noise has unit variance, we expect
the receiver to be able to recover the coded input signals roughly when
\begin{equation}
    \label{eq:Ex_Gaussian_condition}
    2^n| h_1 u_1+h_1 u_2-h_1 u'_1-h_2 u'_2|>2
\end{equation} 
for all $u_1, u'_1\in \{0,1/8,\ldots, 7/8\}$, $u_2,u'_2 \in \{0,1/2\}$ such
that $(u_1,u_2) \neq (u'_1,u'_2)$. In words, we require the minimum
constellation distance as seen at the receiver to be greater than two.

\begin{figure}[htbp]
    \centering
    %\hspace{-0.2cm}
    \scalebox{0.8}{\input{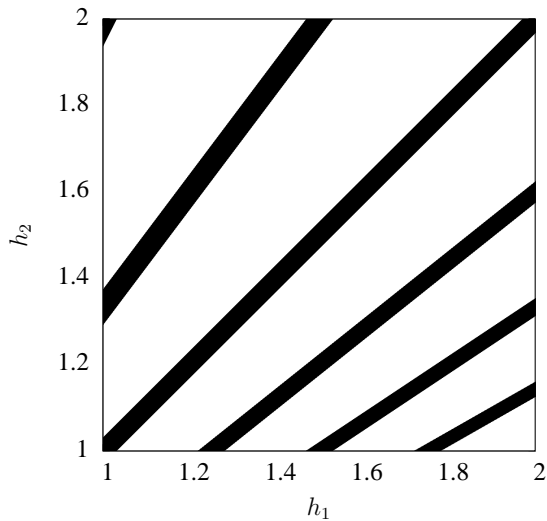}}
    \caption{Outage set $B$ (indicated in black) for the modulation
    scheme in Fig.~\ref{fig:Mac_Scheme} with $n=7$. The set $B$ consists
    of all channel gains $(h_1,h_2)$ such that~\eqref{eq:Ex_Gaussian_condition}
    fails to hold for some channel inputs. The figure makes clear that,
    for finite SNR $n$, the outage set $B$ is not determined
    by the rationality or irrationality of the channel gains $(h_1,h_2)$.} 
    \label{fig:B}
\end{figure}

We note that condition~\eqref{eq:Ex_Gaussian_condition} for the Gaussian
channel corresponds to condition~\eqref{eq:Ex_matrix_condition} for the
deterministic model. As in the deterministic case, this scheme fails to
work whenever the channel gains are in the set $B$ not satisfying
\eqref{eq:Ex_Gaussian_condition}, and one can bound the Lebesgue measure
of this outage event $B$. It is worth emphasizing that condition
\eqref{eq:Ex_Gaussian_condition} has nothing to do with the rationality
or irrationality of the channel coefficients as can be seen from
Fig.~\ref{fig:B}.

\begin{remark}
    In the special case in which each transmitter has the same message
    size, the modulation scheme shown in Fig.~\ref{fig:Mac_Scheme} is
    the same as the modulation scheme used in \emph{real interference
    alignment}~\cite{motahari09,etkin09}. The objective
    in~\cite{motahari09} is to achieve only the degrees-of-freedom of
    the channel, and therefore the scheme there is designed and
    calibrated for the high-${\SNR}$ regime. As a result, the modulation
    scheme in~\cite{motahari09} is not sufficient to prove a
    constant-gap capacity approximation.  Rather, as we will see in
    Section~\ref{sec:main}, asymmetric message sizes and judicious layer
    selection guided by the proposed lower-triangular deterministic
    model together with a more careful and more general analysis of the
    receivers are required to move from a degrees-of-freedom to a
    constant-gap capacity approximation.
\end{remark}

\section{Network Model}
\label{sec:model}

In the remainder of the paper, we focus on the X-channel, which is
formally introduced in this section. We start with notational
conventions in Section~\ref{sec:model_notation}. We introduce the
Gaussian X-channel in Section~\ref{sec:model_gauss}, and the
corresponding lower-triangular deterministic X-channel in
Section~\ref{sec:model_det}.

\subsection{Notation}
\label{sec:model_notation}

Throughout this paper, we use small and capital bold font to denote
vectors and matrices, i.e., $\bm{x}$ and $\bm{H}$. For a real number
$a\in\R$, we use $(a)^+$ to denote $\max\{a,0\}$. For a set $B\subset
\R^d$, $\mu(B)=\mu_d(B)$ denotes $d$-dimensional Lebesgue measure.
Finally, all logarithms are expressed to the base two and 
capacities are expressed in bits per channel use.

\subsection{Gaussian X-Channel}
\label{sec:model_gauss}

The Gaussian X-channel consists of two transmitters and two receivers. 
The channel output $y_m$ at receiver $m\in\{1,2\}$ and time $t\in\N$ is 
\begin{equation}
    \label{eq:xc_arb}
    y_m[t] \triangleq 2^{n_{m1}}h_{m1}x_1[t]+ 2^{n_{m2}}h_{m2}x_2[t] + z_m[t],
\end{equation}
where $x_k[t]$ is the channel input at transmitter $k\in\{1,2\}$, where
$2^{n_{mk}}h_{mk}$ is the channel gain from transmitter $k$ to receiver
$m$, and where $z_m[t]\sim\mc{N}(0,1)$ is additive white Gaussian
receiver noise.  The channel gains consist of two parts, $2^{n_{mk}}$
and $h_{mk}$. We assume that $n_{mk}\in\Zp$ and that $h_{mk}\in(1,2]$
for each $m,k$. Since $2^{n_{mk}}h_{mk}$ varies over $(2^{n_{mk}},
2^{n_{mk}+1}]$ as $h_{mk}$ varies over $(1,2]$, we see that any real
channel gain greater than one can be written in this form. As discussed
in Section~\ref{sec:deterministic_salman}, this implies
that~\eqref{eq:xc_arb} models essentially the general Gaussian
X-channel.\footnote{Indeed, channel gains with magnitude less than one
are not relevant for a constant-gap capacity approximation, and can
hence be ignored. Similarly, negative channel gains have no effect on
the achievable schemes and outer bounds presented later, and can
therefore be ignored as well.} 

Writing the channel gains in the form $2^{n_{mk}}h_{mk}$ decomposes them
into two parts capturing different aspects. The parameter $n_{mk}$
captures the magnitude or coarse structure of the channel gain. Indeed,
the $\SNR$ of the link from transmitter $k$ to receiver $m$ is
approximately $2^{2n_{mk}}$. On the other hand, the parameter $h_{mk}$
captures the fine structure of the channel gain. As we will see soon,
the impact of these two parameters on the behavior of channel capacity
is quite different. We denote by
\begin{equation*}
    \bm{N} \triangleq
    \begin{pmatrix}
        n_{11} & n_{12} \\
        n_{21} & n_{22}
    \end{pmatrix}
\end{equation*}
the collection of $n_{mk}$.

Each transmitter has one message to communicate to each receiver.
So there are a total of four mutually independent messages $w_{mk}$ with
$m,k\in\{1,2\}$. We impose a unit average power constraint on
each of the two encoders. Denote by $R_{mk}$ the rate of message
$w_{mk}$ and by $C(\bm{N})$ the sum capacity of the Gaussian
X-channel. 

An important special case of this setting is the symmetric Gaussian
X-channel, for which $n_{mk}=n$ for all $m,k$ so that
\begin{equation}
    \label{eq:xc}
    y_m[t] \triangleq 2^{n}h_{m1}x_1[t]+ 2^{n}h_{m2}x_2[t] + z_m[t].
\end{equation}
With slight abuse of notation, we denote the sum capacity of the
symmetric Gaussian X-channel by $C(n)$. 

In the following, we will be interested in a particular modulation
scheme for the Gaussian channel, which we describe next. Fix a time slot
$t$; to simplify notation, we will drop the dependence of variables on
$t$ whenever there is no risk of confusion.  Assume each message
$w_{mk}$ is modulated into the signal $u_{mk}$. Transmitter one forms
the channel input
\begin{subequations}
    \label{eq:modulation}
    \begin{equation}
        x_1 \triangleq h_{22} u_{11}+ h_{12} u_{21}.
    \end{equation}
    Similarly, transmitter two forms the channel input
    \begin{equation}
        x_2 \triangleq h_{11} u_{22}+ h_{21} u_{12}.
    \end{equation}
\end{subequations}

The received signals are then given by
\begin{subequations}
    \label{eq:xc_ym}
    \begin{align}
        \label{eq:xc_ym1}
        y_1  & = h_{11}h_{22}2^{n_{11}}u_{11}+ h_{12}h_{21}2^{n_{12}}u_{12} 
        +  h_{11}h_{12}  (2^{n_{11}}  u_{21}+ 2^{n_{12}}  u_{22}) +z_1,\\
        y_2  & = h_{22}h_{11}2^{n_{22}}u_{22}+ h_{21}h_{12}2^{n_{21}}u_{21} 
        +  h_{22}h_{21}(2^{n_{22}}  u_{12}+ 2^{n_{21}}  u_{11}) +z_2.  
    \end{align}
\end{subequations}
Receiver one is interested in the signals $u_{11}$ and $u_{12}$. The
other two signals $u_{21}$ and $u_{22}$ are interference.  We see from
\eqref{eq:xc_ym1} that the interfering signals $u_{21}$ and $u_{22}$ are
received with the same coefficient $h_{11}h_{12}$. The situation is
similar for receiver two.

It will be convenient in the following to refer to the effective channel
gains including the modulation scheme as $g_{mk}$, i.e., 
\begin{subequations}
    \label{eq:gdef}
    \begin{align}
        g_{10} & \triangleq h_{11}h_{12}, &  \qquad\qquad\qquad g_{20} & \triangleq h_{22}h_{21}, \\
        g_{11} & \triangleq h_{11}h_{22}, &  \qquad\qquad\qquad g_{21} & \triangleq h_{21}h_{12}, \\
        g_{12} & \triangleq h_{12}h_{21}, &  \qquad\qquad\qquad g_{22} & \triangleq h_{22}h_{11}.
    \end{align}
\end{subequations}
Here $g_{mk}$ for $m,k\in\{1,2\}$ corresponds to the desired signal
$u_{mk}$, and $g_{m0}$ for $m\in\{1,2\}$ corresponds to the
interference terms. Since $h_{mk}\in(1,2]$, we have
$g_{mk}\in(1,4]$. We can then rewrite~\eqref{eq:xc_ym} as
\begin{subequations}
    \label{eq:xc_ymg}
    \begin{align}
        y_1  & = g_{11}2^{n_{11}} u_{11}+ g_{12}2^{n_{12}}u_{12} 
        +  g_{10} (2^{n_{11}}  u_{21}+ 2^{n_{12}}  u_{22}) +z_1,\\
        y_2  & = g_{22}2^{n_{22}} u_{22}+  g_{21}2^{n_{21}}u_{21} 
        +  g_{20}(2^{n_{22}}  u_{12}+ 2^{n_{21}}  u_{11}) +z_2.  
    \end{align}
\end{subequations}

\subsection{Deterministic X-Channel}
\label{sec:model_det}

As in the discussion in Section~\ref{sec:deterministic_triangular}, it
is insightful to consider the lower-triangular deterministic equivalent
of the modulated Gaussian X-channel~\eqref{eq:xc_ymg}. To simplify the
discussion, we assume for the derivation and analysis of the
deterministic channel model that the channel gains $g_{mk}$ defined
in~\eqref{eq:gdef} are in $(1,2]$ instead of $(1,4]$---the Gaussian
setting will be analyzed for the general case.

Let us first consider the symmetric X-channel~\eqref{eq:xc}, i.e.,
$n_{mk} = n$ for all $m$ and $k$. Let
\begin{equation}
    \label{eq:dbarhmk}
    \bar{\bm{G}}_{mk}\triangleq
    \begin{pmatrix}
        1 & 0 & \cdots & 0 & 0 \\
        [g_{mk}]_{1\phantom{-1}} & 1 & \cdots & 0 & 0 \\
        \vdots & \vdots & \ddots & \vdots & \vdots \\
        [g_{mk}]_{n-2} & [g_{mk}]_{n-3} &  \cdots & 1 & 0 \\
        [g_{mk}]_{n-1} & [g_{mk}]_{n-2} &  \cdots & [g_{mk}]_{1} & 1
    \end{pmatrix}
\end{equation}
be the deterministic channel matrix corresponding to the binary
expansion of the channel gain $g_{mk}$ with $m\in\{1,2\}$ and
$k\in\{0,1,2\}$. Since $g_{mk}\in(1,2]$ by assumption so that
$[g_{mk}]_0=1$, the diagonal entries of $\bar{\bm{G}}_{mk}$ are equal to
one.  

The lower-triangular deterministic equivalent of the
modulated Gaussian X-channel~\eqref{eq:xc_ymg} is then given by
\begin{subequations}
    \label{eq:xc_det2}
    \begin{align}
        \bar{\bm{y}}_1
        & \triangleq \bar{\bm{G}}_{11}\bar{\bm{u}}_{11} \oplus \bar{\bm{G}}_{12}\bar{\bm{u}}_{12}
        \oplus \bar{\bm{G}}_{10}(\bar{\bm{u}}_{21} \oplus \bar{\bm{u}}_{22}), \\
        \bar{\bm{y}}_2
        & \triangleq \bar{\bm{G}}_{22}\bar{\bm{u}}_{22} \oplus \bar{\bm{G}}_{21}\bar{\bm{u}}_{21}
        \oplus \bar{\bm{G}}_{20}(\bar{\bm{u}}_{12} \oplus \bar{\bm{u}}_{11}),
    \end{align}
\end{subequations}
where the channel input $\bar{\bm{u}}_{mk}$ and the channel
output $\bar{\bm{y}}_m$ are all binary vectors of length $n$, and where
all operations are over $\Z_2$.\footnote{This definition of the
deterministic model corresponds to a power constraint of $16$ in the
Gaussian model. This is mainly for convenience of notation. Since the
additional factor $16$ in power only increases capacity by a constant
number of bits per channel use, this does not significantly affect the
quality of approximation.}

Let us then consider the general X-channel~\eqref{eq:xc_arb}. To
simplify the presentation, we focus in the following on the case where
the direct links are stronger than the cross links\footnote{This
assumption is made for ease of exposition. Since the labeling of the
receivers is arbitrary, all results carry immediately over to the case
$\min\{ n_{12}, n_{21} \} \geq \max\{ n_{11}, n_{22} \}$. The models and
tools developed in this paper for these two cases can be applied to the
other cases as well.}, i.e.,
\begin{equation*}
    \min\{  n_{11}, n_{22} \} 
    \geq \max\{ n_{12}, n_{21} \}.
\end{equation*}

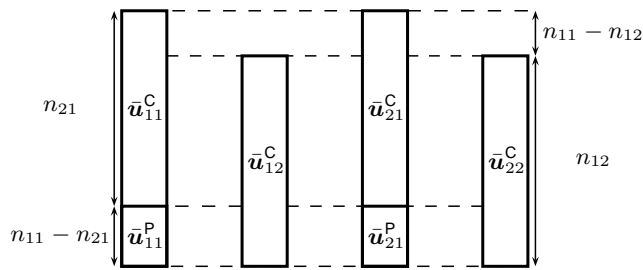
\begin{figure}[htbp]
    \centering
    %\hspace{-0.5cm}
    % Generated with LaTeXDraw 2.0.8
% Tue Sep 20 15:19:44 EDT 2011
% \usepackage[usenames,dvipsnames]{pstricks}
% \usepackage{epsfig}
% \usepackage{pst-grad} % For gradients
% \usepackage{pst-plot} % For axes
\scalebox{1} % Change this value to rescale the drawing.
{
\begin{pspicture}(0,-1.72)(9.3671875,1.72)
\psframe[linewidth=0.04,dimen=outer](2.533125,1.72)(1.893125,-1.72)
\psframe[linewidth=0.04,dimen=outer](4.133125,1.12)(3.493125,-1.72)
\psframe[linewidth=0.04,dimen=outer](5.733125,1.72)(5.093125,-1.72)
\psframe[linewidth=0.04,dimen=outer](7.333125,1.12)(6.693125,-1.72)
\psline[linewidth=0.02cm,linestyle=dashed,dash=0.16cm 0.16cm](1.913125,-0.9)(3.513125,-0.9)
\usefont{T1}{ptm}{m}{n}
\rput(1.1067188,-1.2853125){\footnotesize $n_{11}-n_{21}$}
\psline[linewidth=0.02cm,linestyle=dashed,dash=0.16cm 0.16cm](2.513125,1.1)(5.113125,1.1)
\usefont{T1}{ptm}{m}{n}
\rput(8.186719,-0.2853125){\footnotesize $n_{12}$}
\psline[linewidth=0.02cm,linestyle=dashed,dash=0.16cm 0.16cm](1.913125,1.7)(5.113125,1.7)
\usefont{T1}{ptm}{m}{n}
\rput(1.1067188,0.4146875){\footnotesize $n_{21}$}
\psline[linewidth=0.02cm,linestyle=dashed,dash=0.16cm 0.16cm](1.913125,-1.7)(7.313125,-1.7)
\usefont{T1}{ptm}{m}{n}
\rput(2.2367187,0.4146875){\footnotesize $\bar{\bm{u}}_{11}^\sC$}
\usefont{T1}{ptm}{m}{n}
\rput(3.8367188,-0.2853125){\footnotesize $\bar{\bm{u}}_{12}^\sC$}
\usefont{T1}{ptm}{m}{n}
\rput(5.436719,0.4146875){\footnotesize $\bar{\bm{u}}_{21}^\sC$}
\usefont{T1}{ptm}{m}{n}
\rput(7.036719,-0.2853125){\footnotesize $\bar{\bm{u}}_{22}^\sC$}
\psframe[linewidth=0.04,dimen=outer](2.533125,-0.88)(1.893125,-1.72)
\psframe[linewidth=0.04,dimen=outer](5.733125,-0.88)(5.093125,-1.72)
\psline[linewidth=0.02cm,linestyle=dashed,dash=0.16cm 0.16cm](4.113125,-0.9)(6.713125,-0.9)
\usefont{T1}{ptm}{m}{n}
\rput(2.2367187,-1.2853125){\footnotesize $\bar{\bm{u}}_{11}^\sP$}
\usefont{T1}{ptm}{m}{n}
\rput(5.436719,-1.2853125){\footnotesize $\bar{\bm{u}}_{21}^\sP$}
\psline[linewidth=0.02cm,linestyle=dashed,dash=0.16cm 0.16cm](5.713125,1.1)(6.713125,1.1)
\psline[linewidth=0.02cm,linestyle=dashed,dash=0.16cm 0.16cm](5.713125,1.7)(7.313125,1.7)
\psline[linewidth=0.02cm,arrowsize=0.05291667cm 2.0,arrowlength=1.4,arrowinset=0.4]{<->}(1.813125,1.7)(1.813125,-0.9)
\psline[linewidth=0.02cm,arrowsize=0.05291667cm 2.0,arrowlength=1.4,arrowinset=0.4]{<->}(7.413125,1.1)(7.413125,-1.7)
\psline[linewidth=0.02cm,arrowsize=0.05291667cm 2.0,arrowlength=1.4,arrowinset=0.4]{<->}(1.813125,-0.9)(1.813125,-1.7)
\psline[linewidth=0.02cm,arrowsize=0.05291667cm 2.0,arrowlength=1.4,arrowinset=0.4]{<->}(7.413125,1.7)(7.413125,1.1)
\usefont{T1}{ptm}{m}{n}
\rput(8.186719,1.4146875){\footnotesize $n_{11}-n_{12}$}
\end{pspicture} 
}
    \caption{Deterministic model at receiver one. The figure shows the
    signal $\bar{\bm{y}}_1$ observed at receiver one decomposed into its four components
    (see~\eqref{eq:xc_det_arb1}).  For simplicity, the matrices
    $\bar{\bm{G}}_{mk}$ are omitted. The interference terms
    $\bar{\bm{u}}_{21}$ and $\bar{\bm{u}}_{22}$ are observed at receiver
    one multiplied by the same matrix $\bar{\bm{G}}_{10}$. The desired
    terms $\bar{\bm{u}}_{11}$ and $\bar{\bm{u}}_{12}$ are multiplied by
    different matrices $\bar{\bm{G}}_{11}$ and $\bar{\bm{G}}_{12}$,
    respectively.} 
    \label{fig:Rx1-deterministic0}
\end{figure}

\begin{figure}[htbp]
    \centering
    %\hspace{-0.5cm}
    % Generated with LaTeXDraw 2.0.8
% Mon Sep 12 14:20:10 EDT 2011
% \usepackage[usenames,dvipsnames]{pstricks}
% \usepackage{epsfig}
% \usepackage{pst-grad} % For gradients
% \usepackage{pst-plot} % For axes
\scalebox{1} % Change this value to rescale the drawing.
{
\begin{pspicture}(0,-1.62)(9.307187,1.62)
\psframe[linewidth=0.04,dimen=outer](2.513125,1.02)(1.873125,-1.62)
\usefont{T1}{ptm}{m}{n}
\rput(2.2167187,-0.2853125){\footnotesize $\bar{\bm{u}}_{11}^\sC$}
\psframe[linewidth=0.04,dimen=outer](4.113125,1.62)(3.473125,-1.62)
\psframe[linewidth=0.04,dimen=outer](5.713125,1.02)(5.073125,-1.62)
\psframe[linewidth=0.04,dimen=outer](7.313125,1.62)(6.673125,-1.62)
\usefont{T1}{ptm}{m}{n}
\rput(8.1267185,-1.1853125){\footnotesize $n_{22}-n_{12}$}
\psline[linewidth=0.02cm,linestyle=dashed,dash=0.16cm 0.16cm](2.493125,-0.8)(5.093125,-0.8)
\psline[linewidth=0.02cm,linestyle=dashed,dash=0.16cm 0.16cm](1.893125,1.0)(3.493125,1.0)
\usefont{T1}{ptm}{m}{n}
\rput(1.1067188,-0.2853125){\footnotesize $n_{21}$}
\psline[linewidth=0.02cm,linestyle=dashed,dash=0.16cm 0.16cm](2.493125,-1.6)(3.493125,-1.6)
\psline[linewidth=0.02cm,linestyle=dashed,dash=0.16cm 0.16cm](1.893125,1.6)(3.493125,1.6)
\usefont{T1}{ptm}{m}{n}
\rput(8.1267185,0.4146875){\footnotesize $n_{12}$}
\usefont{T1}{ptm}{m}{n}
\rput(3.8167188,0.4146875){\footnotesize $\bar{\bm{u}}_{12}^\sC$}
\usefont{T1}{ptm}{m}{n}
\rput(5.416719,-0.2853125){\footnotesize $\bar{\bm{u}}_{21}^\sC$}
\usefont{T1}{ptm}{m}{n}
\rput(7.016719,0.4146875){\footnotesize $\bar{\bm{u}}_{22}^\sC$}
\psframe[linewidth=0.04,dimen=outer](4.113125,-0.78)(3.473125,-1.62)
\psframe[linewidth=0.04,dimen=outer](7.313125,-0.78)(6.673125,-1.62)
\psline[linewidth=0.02cm,linestyle=dashed,dash=0.16cm 0.16cm](5.693125,-0.8)(7.293125,-0.8)
\psline[linewidth=0.02cm,linestyle=dashed,dash=0.16cm 0.16cm](4.093125,1.0)(6.693125,1.0)
\psline[linewidth=0.02cm,linestyle=dashed,dash=0.16cm 0.16cm](4.093125,1.6)(7.293125,1.6)
\psline[linewidth=0.02cm,linestyle=dashed,dash=0.16cm 0.16cm](4.093125,-1.6)(5.093125,-1.6)
\psline[linewidth=0.02cm,linestyle=dashed,dash=0.16cm 0.16cm](5.693125,-1.6)(6.693125,-1.6)
\usefont{T1}{ptm}{m}{n}
\rput(3.8167188,-1.1853125){\footnotesize $\bar{\bm{u}}_{12}^\sP$}
\usefont{T1}{ptm}{m}{n}
\rput(7.016719,-1.1853125){\footnotesize $\bar{\bm{u}}_{22}^\sP$}
\psline[linewidth=0.02cm,arrowsize=0.05291667cm 2.0,arrowlength=1.4,arrowinset=0.4]{<->}(1.793125,1.0)(1.793125,-1.6)
\psline[linewidth=0.02cm,arrowsize=0.05291667cm 2.0,arrowlength=1.4,arrowinset=0.4]{<->}(7.393125,1.6)(7.393125,-0.8)
\psline[linewidth=0.02cm,arrowsize=0.05291667cm 2.0,arrowlength=1.4,arrowinset=0.4]{<->}(7.393125,-0.8)(7.393125,-1.6)
\psline[linewidth=0.02cm,arrowsize=0.05291667cm 2.0,arrowlength=1.4,arrowinset=0.4]{<->}(1.793125,1.6)(1.793125,1.0)
\usefont{T1}{ptm}{m}{n}
\rput(1.1067188,1.3146875){\footnotesize $n_{22}-n_{21}$}
\end{pspicture} 
}
    \caption{Deterministic model at receiver two
    (see~\eqref{eq:xc_det_arb2}). The matrices $\bar{\bm{G}}_{mk}$ are
    again omitted. The interference terms $\bar{\bm{u}}_{11}$ and
    $\bar{\bm{u}}_{12}$ are observed at receiver one multiplied by the
    same matrix $\bar{\bm{G}}_{20}$. The desired terms
    $\bar{\bm{u}}_{21}$ and $\bar{\bm{u}}_{22}$ are multiplied by
    different matrices $\bar{\bm{G}}_{21}$ and $\bar{\bm{G}}_{22}$,
    respectively.} 
    \label{fig:Rx2-deterministic0}
\end{figure}
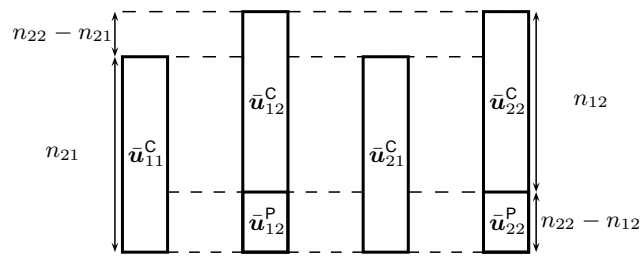

It will be convenient to split the channel input into ``common'' and ``private''
portions, i.e., 
\begin{equation*}
    \bar{\bm{u}}_{mk} \triangleq
    \begin{pmatrix}
        \bar{\bm{u}}_{mk}^{\sC} \\
        \bar{\bm{u}}_{mk}^{\sP}
    \end{pmatrix},
\end{equation*}
where $\bar{\bm{u}}_{m1}^{\sC}\in\Z_2^{n_{21}}$ and
$\bar{\bm{u}}_{m2}^{\sC}\in\Z_2^{n_{12}}$ for $m\in\{1,2\}$. The
lower-triangular deterministic equivalent of the modulated Gaussian
X-channel~\eqref{eq:xc_ymg} is then 
\begin{subequations}
    \label{eq:xc_det_arb}
    \begin{align}
        \label{eq:xc_det_arb1}
        \bar{\bm{y}}_1
        & \triangleq \bar{\bm{G}}_{11} \bar{\bm{u}}_{11} 
        \oplus 
        \bar{\bm{G}}_{12} 
        \begin{pmatrix}
            \bm{0}  \\
            \bar{\bm{u}}_{12}^{\sC}
        \end{pmatrix}
        \oplus 
        \bar{\bm{G}}_{10} \left( \bar{\bm{u}}_{21} \oplus 
        \begin{pmatrix}
            \bm{0} \\
            \bar{\bm{u}}_{22}^{\sC}
        \end{pmatrix}
        \right), \\
        \label{eq:xc_det_arb2}
        \bar{\bm{y}}_2
        & \triangleq \bar{\bm{G}}_{22} \bar{\bm{u}}_{22} 
        \oplus 
        \bar{\bm{G}}_{21} 
        \begin{pmatrix}
            \bm{0}  \\
            \bar{\bm{u}}_{21}^{\sC}
        \end{pmatrix}
        \oplus 
        \bar{\bm{G}}_{20} \left( \bar{\bm{u}}_{12} \oplus 
        \begin{pmatrix}
            \bm{0} \\
            \bar{\bm{u}}_{11}^{\sC}
        \end{pmatrix}
        \right),
    \end{align}
\end{subequations}
where all operations are again over $\Z_2$, see
Figs.~\ref{fig:Rx1-deterministic0} and~\ref{fig:Rx2-deterministic0}.
Here, the lower-triangular binary matrices $\bar{\bm{G}}_{mk}$ are
defined in analogy to \eqref{eq:dbarhmk}. The matrix $\bar{\bm{G}}_{1k}$
is of dimension $n_{11}\times n_{11}$  and $\bar{\bm{G}}_{2k}$ is of
dimension $n_{22}\times n_{22}$ for all $k\in\{0,1,2\}$.  Comparing the
general deterministic model~\eqref{eq:xc_det_arb} to the symmetric
one~\eqref{eq:xc_det2}, we see that the difference in the values of
$n_{mk}$ results in the inputs $\bar{\bm{u}}_{mk}$ observed over the
cross links to be shifted down. As a consequence, the private portions
of the channel inputs are visible at only the intended receiver, whereas
the common portions are visible at both receivers.


As before, there are four independent messages $w_{mk}$. Each
transmitter $k$ consists of two\footnote{Observe that in the definition
of capacity $\bar{C}(n)$ of the modulated deterministic X-channel
\eqref{eq:xc_det_arb} we use two encoders at each transmitter (one for 
each of the two messages).  This differs from the definition of capacity
$C(n)$ of the Gaussian X-channel \eqref{eq:xc}, where we use a single
encoder. Thus, in the deterministic case, we force the messages to be
encoded separately, while we allow joint encoding of the two messages in
the Gaussian case.  This restriction is introduced because the aim of
the deterministic model is to better understand the modulated Gaussian
X-channel~\eqref{eq:xc_ym}, which already handles the joint encoding of
the messages through the modulation process.} encoders mapping one of
the two messages $w_{mk}$ to a sequence of channel inputs
$\bar{\bm{u}}_{mk}$.  Denote by $\bar{R}_{mk}$ the rate of message
$w_{mk}$ and by $\bar{C}(\bm{N})$ the sum capacity of the (modulated)
deterministic X-channel~\eqref{eq:xc_det_arb}. For the special case of
the symmetric deterministic X-channel~\eqref{eq:xc_det2}, the sum
capacity is denoted by $\bar{C}(n)$.

\section{Main Results}
\label{sec:main}

The main result of this paper is  a constant-gap approximation for the
capacity of the Gaussian X-channel. To simplify the presentation of the
relevant concepts and results, we start with the analysis of the
Gaussian X-channel with symmetric $\SNR$s in
Section~\ref{sec:main_symmetric}. We then consider the Gaussian
X-channel with arbitrary $\SNR$s in Section~\ref{sec:main_arbitrary}.

\subsection{X-Channel with Symmetric $\SNR$s}
\label{sec:main_symmetric}

We start with the analysis of the deterministic X-channel---as we will
see in the following, the insights obtained for this model carry over to
the Gaussian X-channel.  The capacity $\bar{C}(n)$ of the symmetric
deterministic X-channel is characterized by the next theorem.
\begin{theorem}
    \label{thm:symmetric_det}
    For every $\delta\in(0,1]$ and $n\in\Zp$, there exists a set
    $B_n\subseteq (1,2]^{2\times 3}$ of Lebesgue measure
    \begin{equation*}
        \mu(B_n) \leq \delta
    \end{equation*}
    such that for all channel gains
    $(g_{mk})\in(1,2]^{2\times 3}\setminus B_n$ the sum
    capacity $\bar{C}(n)$ of the (modulated) symmetric deterministic 
    X-channel \eqref{eq:xc_det2} satisfies
    \begin{equation*}
        \tfrac{4}{3}n-2\log(c_1/\delta)
        \leq \bar{C}(n) 
        \leq \tfrac{4}{3}n
    \end{equation*}
    for some positive universal constant $c_1$. 
\end{theorem}

Theorem~\ref{thm:symmetric_det} is a special case of
Theorem~\ref{thm:arbitrary_det} presented in
Section~\ref{sec:main_arbitrary}. We hence omit its proof.

Theorem~\ref{thm:symmetric_det} approximates the capacity of the
modulated deterministic X-channel \eqref{eq:xc_det2} up to a constant
gap for all channel gains $g_{mk}\in(1,2]$ outside the set $B_n$ of
arbitrarily small measure. The event $(g_{mk})\in B_n$ can be
interpreted as an outage event, as in this case the proposed achievable
scheme fails to deliver the target rate of
$\tfrac{4}{3}n-2\log(c_1/\delta)$. Here $\delta$ parametrizes the
trade-off between the measure of the outage set $B_n$ and the target
rate: decreasing $\delta$ decreases the measure of the outage event
$B_n$, but at the same time also decreases the target rate
$\tfrac{4}{3}n-2\log(c_1/\delta)$. We point out that $\delta$ can be
chosen independently of the number of input bits $n$, hence the
approximation gap is uniform in $n$.

Theorem~\ref{thm:symmetric_det} can be used to derive the more familiar
result on the degrees-of-freedom $\lim_{n\to\infty}\bar{C}(n)/n$ of the
deterministic X-channel. Setting $\delta = n^{-2}$ results in the
measures $\mu(B_n)\leq n^{-2}$ to be summable over $n\in\Zp$. Applying
the Borel-Cantelli lemma yields then the following corollary to
Theorem~\ref{thm:symmetric_det}. 

\begin{corollary}
    \label{thm:symmetric_det_cor}
    For almost all channel gains $(g_{mk})\in(1,2]^{2\times 3}$ the
    (modulated) symmetric deterministic X-channel \eqref{eq:xc_det2}
    has $4/3$ degrees-of-freedom, i.e., 
    \begin{equation*}
        \lim_{n\to\infty}\frac{\bar{C}(n)}{n} = 4/3.
    \end{equation*}
\end{corollary}

We emphasize that, while Corollary~\ref{thm:symmetric_det_cor} is
simpler to state and perhaps more familiar in form,
Theorem~\ref{thm:symmetric_det} is considerably stronger. Indeed,
Theorem~\ref{thm:symmetric_det} provides the stronger \emph{constant
gap} capacity approximation for the sum capacity $\bar{C}(n)$, whereas
Corollary~\ref{thm:symmetric_det_cor} provides the weaker
\emph{degrees-of-freedom} capacity approximation. Moreover,
Theorem~\ref{thm:symmetric_det} provides bounds for \emph{finite} $n$ on
the measure of the outage event $B_n$, whereas
Corollary~\ref{thm:symmetric_det_cor} provides only \emph{asymptotic}
information about its size.

\begin{figure}[htbp]
    \centering
    \scalebox{0.667}{\input{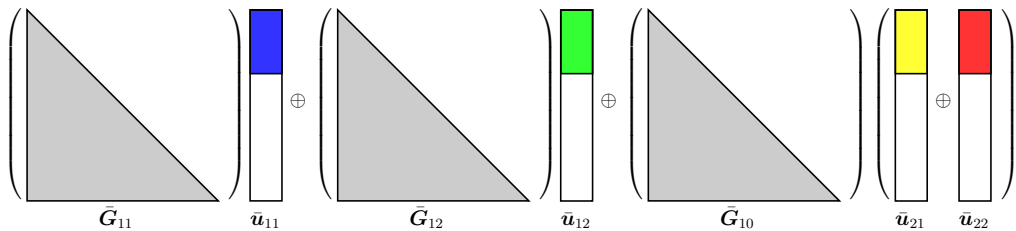}} 

    \caption{Allocation of bits for the deterministic X-channel with
    symmetric $\SNR$s as seen at receiver one. The white
    regions correspond to zero bits; the shaded regions carry
    information. Observe that the interference signals $\bar{\bm{u}}_{21}$
    and $\bar{\bm{u}}_{22}$ are aligned.}

    \label{fig:allocation1}
\end{figure}

We now describe the communication scheme achieving the lower bound in
Theorem~\ref{thm:symmetric_det} (see Fig.~\ref{fig:allocation1}).  Use
the first $\bar{R}$ components of each vector $\bar{\bm{u}}_{mk}$ to
transmit information, and set the last $n-\bar{R}$ components to zero.
The sum rate of this communication scheme is hence $4\bar{R}$.  Receiver one
is interested in $\bar{\bm{u}}_{11}$ and $\bar{\bm{u}}_{12}$. These
vectors are received in the subspace spanned by the first $\bar{R}$
columns of $\bar{\bm{G}}_{11}$ and $\bar{\bm{G}}_{12}$, respectively. On
the other hand, the messages $\bar{\bm{u}}_{21}$ and $\bar{\bm{u}}_{22}$
that receiver one is not interested in, and that can hence be regarded
as interference, are both received in the same subspace spanned by the
first $\bar{R}$ columns of $\bar{\bm{G}}_{10}$. Thus, the two
interference vectors are aligned in a subspace of dimension $\bar{R}$.
The situation at receiver two is similar.

Assume that the three subspaces spanned by the first $\bar{R}$ columns
of $\bar{\bm{G}}_{11}$, $\bar{\bm{G}}_{12}$, and $\bar{\bm{G}}_{10}$ are
linearly independent. Then receiver one can recover the two desired
vectors by projecting the received vector into the corresponding
subspaces in order to zero force the two interfering vectors. We show
that for most channel gains this linear independence of the three
subspaces holds for $\bar{R}\approx n/3$.  The outage event $B_n$ in
Theorem~\ref{thm:symmetric_det} is thus precisely the event that at
either of the two receivers the three subspaces spanned by the first
$\bar{R}$ columns of $\bar{\bm{G}}_{m1}$, $\bar{\bm{G}}_{m2}$, and
$\bar{\bm{G}}_{m0}$ are not linearly independent.

We now turn to the Gaussian X-channel.  The results for the
deterministic X-channel suggest that the modulation
scheme~\eqref{eq:modulation} should achieve a sum rate of 
\begin{equation*}
    3\bar{R} \approx \tfrac{4}{3}n\pm O(1)
\end{equation*}
over the Gaussian channel as $n\to\infty$. Furthermore, it suggests that
a $n$-bit quantization of the channel gains $h_{mk}$ available at both
transmitters and receivers should be sufficient to achieve this
asymptotic rate. This intuition turns out to be correct, as the next
theorem shows.

\begin{theorem}
    \label{thm:symmetric_gaussian}
    For every $\delta\in(0,1]$ and $n\in\Zp$, there exists a set
    $B_n\subseteq(1,2]^{2\times 2}$ of Lebesgue measure at most
    \begin{equation*}
        \mu(B_n) \leq \delta
    \end{equation*}
    such that for all channel gains $(h_{mk})\in(1,2]^{2\times
    2}\setminus B_n$ the sum capacity of the symmetric Gaussian
    X-channel \eqref{eq:xc} satisfies
    \begin{equation*}
        \tfrac{4}{3}n-2\log(c_2/\delta)
        \leq C(n) 
        \leq \tfrac{4}{3}n+4
    \end{equation*}
    for some positive universal constant $c_2$. Moreover, the lower bound
    is achievable with a $n$-bit quantization of the channel gains
    $h_{mk}$ available at both transmitters and receivers.
\end{theorem}

Theorem~\ref{thm:symmetric_gaussian} is a special case of
Theorem~\ref{thm:arbitrary_gaussian} presented in
Section~\ref{sec:main_arbitrary}. We hence omit its proof.

Theorem~\ref{thm:symmetric_gaussian} provides a constant-gap capacity
approximation for the symmetric Gaussian X-channel~\eqref{eq:xc}. The constant in the
approximation is uniform in the channel gains $h_{mk}\in(1,2]$ outside
the set $B_n$ of arbitrarily small measure, and uniform in $n$. The
event $(h_{mk})\in B_n$ can again be interpreted as an outage event, and
$\delta$ parametrizes the trade-off between the measure of the outage
set $B_n$ and the target rate of the achievable scheme.  Since
$\delta$ can be chosen independently of $n$, the approximation gap is
uniform in the $\SNR$, i.e., uniform in $2^{2n}$. 

\begin{remark}
    It is worth pointing out that the outage set $B_n$ can be explicitly
    computed: given channel gains $h_{mk}$, there is an algorithm that
    can determine in bounded time if these channel gains are in the
    outage set $B_n$. More precisely, $B_n$ is the union of
    $2^{\Theta(n)}$ ``strips'' similar to Fig.~\ref{fig:B} in
    Section~\ref{sec:deterministic_comparison}. Membership of $(h_{mk})$
    in the outage set $B_n$ is mostly determined by the $n$
    most-significant bits in the binary expansion of the channel gains
    $h_{mk}$. In particular, for any finite $n$ (and hence finite
    $\SNR$), the question of rationality or irrationality of the channel
    gains $h_{mk}$ is largely irrelevant to determining membership in
    $B_n$.
\end{remark}

The theorem shows furthermore that the proposed achievable scheme for
the Gaussian X-channel is not dependent on the exact knowledge of the
channel gains, and a quantized version, available at all transmitters
and receivers, is sufficient. In fact, the scheme achieving the lower
bound uses mismatched encoders and decoders.  The encoders perform
modulation with respect to the \emph{wrong} channel model
\begin{equation}
    \label{eq:wrong}
    y_m[t] = 2^n\hat{h}_{m1}x_1[t]+2^n\hat{h}_{m2} x_2[t]+z_m[t],
\end{equation}
where $\hat{h}_{mk}$ is a $n$-bit (or, equivalently,
$\tfrac{1}{2}\log(\SNR)$-bit) quantization of the true channel gain
$h_{mk}$. In other words, the channel inputs are
\begin{align*}
    x_1[t] & = \hat{h}_{22}u_{11}[t]+\hat{h}_{12}u_{21}[t], \\
    x_2[t] & = \hat{h}_{11}u_{22}[t]+\hat{h}_{21}u_{12}[t].
\end{align*}
The decoders perform maximum-likelihood decoding also with respect to
the wrong channel model \eqref{eq:wrong}. Thus, both the encoders and
the decoders treat the channel estimates as if they were the true channel
gains. This shows that the proposed achievable scheme is actually quite
robust with respect to channel estimation and quantization errors.

As before, we can use Theorem~\ref{thm:symmetric_gaussian} to derive
more familiar results on the degrees-of-freedom of the Gaussian
X-channel. Consider a sequence of $\SNR$s $2^{2n}$ indexed by $n\in\Zp$,
and set $\delta = n^{-2}$. Then the measures $\mu(B_{n}) \leq n^{-2}$
are summable over $n\in\Zp$. Applying the Borel-Cantelli lemma as before
yields the following corollary to Theorem~\ref{thm:symmetric_gaussian}.
\begin{corollary}
    \label{thm:symmetric_gaussian_cor}
    For almost all channel gains $(h_{mk})\in(1,2]^{2\times 2}$ the
    symmetric Gaussian X-channel \eqref{eq:xc} has $4/3$
    degrees-of-freedom, i.e., 
    \begin{equation*}
        \lim_{n\to\infty}\frac{C(n)}{n} = 4/3.
    \end{equation*}
\end{corollary}

Since the $\SNR$ of the channel is approximately $2^{2n}$ so that $n
\approx \tfrac{1}{2}\log(\SNR)$, the quantity $\lim_{n\to\infty}C(n)/n$
in Corollary~\ref{thm:symmetric_gaussian_cor} is indeed the
degrees-of-freedom limit. Corollary~\ref{thm:symmetric_gaussian_cor}
recovers the result in~\cite{motahari09}. We emphasize again that
Theorem~\ref{thm:symmetric_gaussian} is considerably stronger than
Corollary~\ref{thm:symmetric_gaussian_cor}. Indeed,
Theorem~\ref{thm:symmetric_gaussian} proves the \emph{constant-gap}
capacity approximation 
\begin{equation*}
    \abs{C(n)-\tfrac{4}{3}n} \leq O(1)
\end{equation*}
with pre-constant in the $O(1)$ term uniform in the channel gains
$h_{mk}$ outside $B_n$. This is considerably stronger than the
\emph{degrees-of-freedom} capacity approximation in
Corollary~\ref{thm:symmetric_gaussian_cor}, which shows only that
\begin{equation*}
    \abs{C(n)-\tfrac{4}{3}n} \leq o(n)
\end{equation*}
with pre-constant in the $o(n)$ term depending on $h_{mk}$.  Moreover,
Theorem~\ref{thm:symmetric_gaussian} provides bounds on the measure of
the outage event for \emph{finite} $\SNR$s, not just \emph{asymptotic}
guarantees as in Corollary~\ref{thm:symmetric_gaussian_cor}.

\subsection{X-Channel with Arbitrary $\SNR$s}
\label{sec:main_arbitrary}

In the last section, we considered the Gaussian X-channel with $\SNR$s
across each link of order $2^{2n}$. Thus, all links had approximately
the same strength. We now turn to the Gaussian X-channel with arbitrary
$\SNR$s. As before, we start with the analysis of the deterministic
X-channel. The next theorem provides an approximate characterization of
the sum capacity $\bar{C}(\bm{N})$ of the general deterministic
X-channel with bit levels $\bm{N}$. 

\begin{theorem}
    \label{thm:arbitrary_det}
    For every $\delta\in(0,1]$ and $\bm{N}\in\Zp^{2\times 2}$ with
    $\min\{ n_{11}, n_{22}  \} \geq \max \{ n_{12}, n_{21} \}$
    there exists a set $B\subseteq(1,2]^{2\times 3}$ of Lebesgue
    measure 
    \begin{equation*}
        \mu(B) \leq \delta
    \end{equation*}
    such that for all channel gains
    $(g_{mk})\in(1,2]^{2\times 3}\setminus B$ the sum
    capacity $\bar{C}(\bm{N})$ of the (modulated) general deterministic
    X-channel~\eqref{eq:xc_det_arb} satisfies
    \begin{equation*}
        D(\bm{N})-2\log(c_1/\delta)
        \leq \bar{C}(\bm{N}) 
        \leq D(\bm{N})
    \end{equation*}
    for some positive universal constant $c_1$, and where
    \begin{equation*}
        D(\bm{N}) 
        \triangleq \min \big\{ D_1(\bm{N}), D_2(\bm{N}), 
        D_3(\bm{N}), D_4(\bm{N}) \big\} 
        + (n_{11}-n_{21})+(n_{22}-n_{12})
    \end{equation*}
    and 
    \begin{align*}
        D_1(\bm{N}) & 
        \triangleq (n_{12}+n_{21}-n_{11})^+ +  (n_{12}+n_{21}-n_{22})^+, \\
        D_2(\bm{N}) 
        & \triangleq \tfrac{1}{2}\big( n_{12}+n_{21}+ (n_{12}+n_{21}- n_{22})^+\big), \\
        D_3(\bm{N}) 
        & \triangleq \tfrac{1}{2}\big( n_{12}+n_{21}+ (n_{12}+n_{21}- n_{11})^+\big), \\
        D_4(\bm{N}) 
        & \triangleq \tfrac{2}{3}(n_{12}+n_{21}).
    \end{align*}
\end{theorem}  

The proof of Theorem~\ref{thm:arbitrary_det} is presented in
Section~\ref{sec:proofs_det}. For the special case of symmetric channel
$\SNR$s, $n_{mk}=n$ for all $m,k$, Theorem~\ref{thm:arbitrary_det} reduces to
Theorem~\ref{thm:symmetric_det} in Section~\ref{sec:main_symmetric}.

We now provide a sketch of the communication scheme achieving the lower
bound in Theorem~\ref{thm:arbitrary_det} (see Figs.~\ref{fig:Rx1-deterministic}
and~\ref{fig:Rx2-deterministic}). 
\begin{figure}[htbp]
    \centering
    %\hspace{-0.2cm}
    % Generated with LaTeXDraw 2.0.8
% Mon Oct 31 15:05:52 EDT 2011
% \usepackage[usenames,dvipsnames]{pstricks}
% \usepackage{epsfig}
% \usepackage{pst-grad} % For gradients
% \usepackage{pst-plot} % For axes
\scalebox{1} % Change this value to rescale the drawing.
{
\begin{pspicture}(0,-1.9845313)(8.457188,1.9445312)
\definecolor{color803b}{rgb}{0.2,0.2,1.0}
\definecolor{color804b}{rgb}{0.8,0.8,1.0}
\definecolor{color806b}{rgb}{0.2,1.0,0.2}
\definecolor{color808b}{rgb}{1.0,1.0,0.2}
\definecolor{color810b}{rgb}{1.0,0.2,0.2}
\psframe[linewidth=0.04,dimen=outer](1.663125,1.9445312)(1.023125,-1.4954687)
\psframe[linewidth=0.04,dimen=outer,fillstyle=solid,fillcolor=color803b](1.663125,1.9445312)(1.023125,0.90453124)
\psframe[linewidth=0.04,dimen=outer,fillstyle=solid,fillcolor=color804b](1.663125,-0.65546876)(1.023125,-1.4954687)
\psframe[linewidth=0.04,dimen=outer](3.263125,1.3445313)(2.623125,-1.4954687)
\psframe[linewidth=0.04,dimen=outer,fillstyle=solid,fillcolor=color806b](3.263125,0.7445313)(2.623125,0.10453125)
\psframe[linewidth=0.04,dimen=outer](4.863125,1.9445312)(4.223125,-1.4954687)
\psframe[linewidth=0.04,dimen=outer,fillstyle=solid,fillcolor=color808b](4.863125,1.3445313)(4.223125,0.50453126)
\psframe[linewidth=0.04,dimen=outer](6.463125,1.3445313)(5.823125,-1.4954687)
\psframe[linewidth=0.04,dimen=outer,fillstyle=solid,fillcolor=color810b](6.463125,1.3445313)(5.823125,0.30453125)
\psline[linewidth=0.02cm,arrowsize=0.05291667cm 2.0,arrowlength=1.4,arrowinset=0.4]{<->}(5.743125,1.3245312)(5.743125,0.32453126)
\usefont{T1}{ptm}{m}{n}
\rput(5.4667187,0.95953125){\footnotesize $\bar{R}_{22}^\sC$}
\psline[linewidth=0.02cm,arrowsize=0.05291667cm 2.0,arrowlength=1.4,arrowinset=0.4]{<->}(1.743125,1.9245312)(1.743125,0.9245312)
\usefont{T1}{ptm}{m}{n}
\rput(2.0267189,1.5592188){\footnotesize $\bar{R}_{11}^\sC$}
\psline[linewidth=0.02cm,arrowsize=0.05291667cm 2.0,arrowlength=1.4,arrowinset=0.4]{<->}(1.743125,-0.67546874)(1.743125,-1.4754688)
\usefont{T1}{ptm}{m}{n}
\rput(2.0467188,-1.0607812){\footnotesize $\bar{R}_{11}^\sP$}
\psline[linewidth=0.02cm,arrowsize=0.05291667cm 2.0,arrowlength=1.4,arrowinset=0.4]{<->}(3.343125,0.72453123)(3.343125,0.12453125)
\usefont{T1}{ptm}{m}{n}
\rput(3.6367188,0.43921876){\footnotesize $\bar{R}_{12}$}
\psline[linewidth=0.02cm,arrowsize=0.05291667cm 2.0,arrowlength=1.4,arrowinset=0.4]{<->}(4.143125,1.3245312)(4.143125,0.52453125)
\usefont{T1}{ptm}{m}{n}
\rput(3.8767188,0.95953125){\footnotesize $\bar{R}_{21}$}
\psline[linewidth=0.02cm,linestyle=dashed,dash=0.16cm 0.16cm](1.043125,-0.67546874)(6.443125,-0.67546874)
\usefont{T1}{ptm}{m}{n}
\rput(7.2767186,1.6392188){\footnotesize $n_{11}-n_{12}$}
\psline[linewidth=0.02cm,linestyle=dashed,dash=0.16cm 0.16cm](1.043125,1.3245312)(6.443125,1.3245312)
\usefont{T1}{ptm}{m}{n}
\rput(0.68671876,0.63921875){\footnotesize $n_{21}$}
\psline[linewidth=0.02cm,linestyle=dashed,dash=0.16cm 0.16cm](1.043125,1.9245312)(6.443125,1.9245312)
\psline[linewidth=0.02cm,linestyle=dashed,dash=0.16cm 0.16cm](1.043125,0.72453123)(6.443125,0.72453123)
\usefont{T1}{ptm}{m}{n}
\rput(7.266719,1.0392188){\footnotesize $n_{22}-n_{21}$}
\psline[linewidth=0.02cm,linestyle=dashed,dash=0.16cm 0.16cm](1.043125,-1.4754688)(6.443125,-1.4754688)
\usefont{T1}{ptm}{m}{n}
\rput(1.3567188,-1.7804687){\footnotesize $\bar{\bm{u}}_{11}$}
\usefont{T1}{ptm}{m}{n}
\rput(2.9567187,-1.7804687){\footnotesize $\bar{\bm{u}}_{12}$}
\usefont{T1}{ptm}{m}{n}
\rput(4.556719,-1.7804687){\footnotesize $\bar{\bm{u}}_{21}$}
\usefont{T1}{ptm}{m}{n}
\rput(6.1567187,-1.7804687){\footnotesize $\bar{\bm{u}}_{22}$}
\psline[linewidth=0.02cm,arrowsize=0.05291667cm 2.0,arrowlength=1.4,arrowinset=0.4]{<->}(0.943125,1.9245312)(0.943125,-0.67546874)
\psline[linewidth=0.02cm,arrowsize=0.05291667cm 2.0,arrowlength=1.4,arrowinset=0.4]{<->}(6.543125,1.3245312)(6.543125,0.72453123)
\psline[linewidth=0.02cm,arrowsize=0.05291667cm 2.0,arrowlength=1.4,arrowinset=0.4]{<->}(6.543125,1.9245312)(6.543125,1.3245312)
\end{pspicture} 
}
    \caption{Allocation of bits as seen at receiver one. Here,
    $\bar{\bm{u}}_{11}$ and $\bar{\bm{u}}_{12}$ are the desired bits
    and are received multiplied by the matrices $\bar{\bm{G}}_{11}$ and
    $\bar{\bm{G}}_{12}$ (not shown in the figure), respectively. The
    vectors $\bar{\bm{u}}_{21}$ and $\bar{\bm{u}}_{22}$ are
    interference and are both received multiplied by the same matrix
    $\bar{\bm{G}}_{10}$.} 
    \label{fig:Rx1-deterministic}
\end{figure}
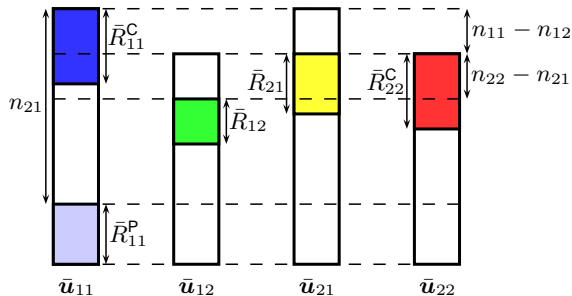
\begin{figure}[htbp]
    \centering
    %\hspace{-0.2cm}
    % Generated with LaTeXDraw 2.0.8
% Mon Sep 12 14:19:23 EDT 2011
% \usepackage[usenames,dvipsnames]{pstricks}
% \usepackage{epsfig}
% \usepackage{pst-grad} % For gradients
% \usepackage{pst-plot} % For axes
\scalebox{1} % Change this value to rescale the drawing.
{
\begin{pspicture}(0,-1.8845313)(8.447187,1.8445313)
\definecolor{color175b}{rgb}{0.2,0.2,1.0}
\definecolor{color176b}{rgb}{1.0,0.8,0.8}
\definecolor{color178b}{rgb}{0.2,1.0,0.2}
\definecolor{color183b}{rgb}{1.0,1.0,0.2}
\definecolor{color185b}{rgb}{1.0,0.2,0.2}
\psframe[linewidth=0.04,dimen=outer](2.523125,1.2445313)(1.883125,-1.3954687)
\psframe[linewidth=0.04,dimen=outer,fillstyle=solid,fillcolor=color175b](2.523125,1.2445313)(1.883125,0.20453125)
\psframe[linewidth=0.04,dimen=outer,fillstyle=solid,fillcolor=color176b](7.323125,-0.55546874)(6.683125,-1.3954687)
\psframe[linewidth=0.04,dimen=outer](4.123125,1.8445313)(3.483125,-1.3954687)
\psframe[linewidth=0.04,dimen=outer,fillstyle=solid,fillcolor=color178b](4.123125,1.2445313)(3.483125,0.70453125)
\psline[linewidth=0.02cm,arrowsize=0.05291667cm 2.0,arrowlength=1.4,arrowinset=0.4]{<->}(4.203125,1.2245313)(4.203125,0.72453123)
\usefont{T1}{ptm}{m}{n}
\rput(4.516719,0.9892188){\footnotesize $\bar{R}_{12}$}
\psframe[linewidth=0.04,dimen=outer](5.723125,1.2445313)(5.083125,-1.3954687)
\psframe[linewidth=0.04,dimen=outer,fillstyle=solid,fillcolor=color183b](5.723125,0.64453125)(5.083125,-0.19546875)
\psframe[linewidth=0.04,dimen=outer](7.323125,1.8445313)(6.683125,-1.3954687)
\psframe[linewidth=0.04,dimen=outer,fillstyle=solid,fillcolor=color185b](7.323125,1.8445313)(6.683125,0.8045313)
\psline[linewidth=0.02cm,arrowsize=0.05291667cm 2.0,arrowlength=1.4,arrowinset=0.4]{<->}(6.603125,1.8245312)(6.603125,0.82453126)
\usefont{T1}{ptm}{m}{n}
\rput(6.306719,1.4395312){\footnotesize $\bar{R}_{22}^\sC$}
\usefont{T1}{ptm}{m}{n}
\rput(1.1067188,0.93921876){\footnotesize $n_{11}-n_{12}$}
\psline[linewidth=0.02cm,arrowsize=0.05291667cm 2.0,arrowlength=1.4,arrowinset=0.4]{<->}(2.603125,1.2245313)(2.603125,0.22453125)
\usefont{T1}{ptm}{m}{n}
\rput(2.9067187,0.82921875){\footnotesize $\bar{R}_{11}^\sC$}
\psline[linewidth=0.02cm,arrowsize=0.05291667cm 2.0,arrowlength=1.4,arrowinset=0.4]{<->}(5.003125,0.62453127)(5.003125,-0.17546874)
\usefont{T1}{ptm}{m}{n}
\rput(4.7367187,0.23921876){\footnotesize $\bar{R}_{21}$}
\usefont{T1}{ptm}{m}{n}
\rput(1.1167188,1.5392188){\footnotesize $n_{22}-n_{21}$}
\psline[linewidth=0.02cm,linestyle=dashed,dash=0.16cm 0.16cm](1.903125,-0.5754688)(7.303125,-0.5754688)
\psline[linewidth=0.02cm,linestyle=dashed,dash=0.16cm 0.16cm](1.903125,1.2245313)(7.303125,1.2245313)
\usefont{T1}{ptm}{m}{n}
\rput(7.726719,0.63921875){\footnotesize $n_{12}$}
\psline[linewidth=0.02cm,linestyle=dashed,dash=0.16cm 0.16cm](1.903125,-1.3754687)(7.303125,-1.3754687)
\psline[linewidth=0.02cm,linestyle=dashed,dash=0.16cm 0.16cm](1.903125,0.62453127)(7.303125,0.62453127)
\psline[linewidth=0.02cm,linestyle=dashed,dash=0.16cm 0.16cm](1.903125,1.8245312)(7.303125,1.8245312)
\usefont{T1}{ptm}{m}{n}
\rput(2.2167187,-1.6804688){\footnotesize $\bar{\bm{u}}_{11}$}
\usefont{T1}{ptm}{m}{n}
\rput(3.8167188,-1.6804688){\footnotesize $\bar{\bm{u}}_{12}$}
\usefont{T1}{ptm}{m}{n}
\rput(5.416719,-1.6804688){\footnotesize $\bar{\bm{u}}_{21}$}
\usefont{T1}{ptm}{m}{n}
\rput(7.016719,-1.6804688){\footnotesize $\bar{\bm{u}}_{22}$}
\psline[linewidth=0.02cm,arrowsize=0.05291667cm 2.0,arrowlength=1.4,arrowinset=0.4]{<->}(6.603125,-0.5754688)(6.603125,-1.3754687)
\usefont{T1}{ptm}{m}{n}
\rput(6.306719,-0.9607813){\footnotesize $\bar{R}_{22}^\sP$}
\psline[linewidth=0.02cm,arrowsize=0.05291667cm 2.0,arrowlength=1.4,arrowinset=0.4]{<->}(1.803125,1.8245312)(1.803125,1.2245313)
\psline[linewidth=0.02cm,arrowsize=0.05291667cm 2.0,arrowlength=1.4,arrowinset=0.4]{<->}(1.803125,1.2245313)(1.803125,0.62453127)
\psline[linewidth=0.02cm,arrowsize=0.05291667cm 2.0,arrowlength=1.4,arrowinset=0.4]{<->}(7.403125,1.8245312)(7.403125,-0.5754688)
\end{pspicture} 
}
    \caption{Allocation of bits as seen at receiver two. Here,
    $\bar{\bm{u}}_{21}$ and $\bar{\bm{u}}_{22}$ are the desired bits
    and are received multiplied by the matrices $\bar{\bm{G}}_{21}$ and
    $\bar{\bm{G}}_{22}$ (not shown in the figure), respectively. The
    vectors $\bar{\bm{u}}_{11}$ and $\bar{\bm{u}}_{12}$ are
    interference and are both received multiplied by the same matrix
    $\bar{\bm{G}}_{20}$.
    } 
    \label{fig:Rx2-deterministic}
\end{figure}
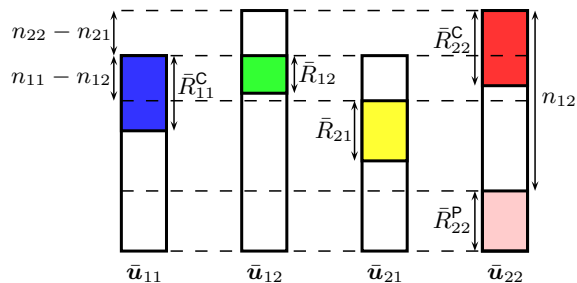
Observe from Figs.~\ref{fig:Rx1-deterministic0}
and~\ref{fig:Rx2-deterministic0} in Section~\ref{sec:model_det} that the
$n_{11}-n_{21}$ least-significant bits $\bar{\bm{u}}_{11}^{\sP}$ of
$\bar{\bm{u}}_{11}$ are not visible at the second receiver.  Therefore,
we can use these bits to privately carry $n_{11}-n_{21}$ bits from the
first transmitter to the first receiver without affecting the second
receiver. The rate of this private message is denoted by
$\bar{R}_{11}^{\sP}$.  The remaining rate is denoted by
$\bar{R}_{11}^{\sC}$, i.e.,
\begin{align*}
    \bar{R}_{11} \triangleq \bar{R}_{11}^{\sC}+\bar{R}_{11}^{\sP},
\end{align*}
where 
\begin{align*}
    \bar{R}_{11}^{\sP} \triangleq n_{11}-n_{21}.
\end{align*}

Similarly, the $n_{22}-n_{12}$ least-significant bits
$\bar{\bm{u}}_{22}^{\sP}$ of $\bar{\bm{u}}_{22}$ are not visible at the
first receiver.  Therefore, we can use this part to privately carry
$n_{22}-n_{12}$ bits from the second transmitter to the second receiver
without affecting the first receiver.  The rate of this private message
is denoted by $\bar{R}_{22}^{\sP}$. The remaining rate is denoted by
$\bar{R}_{22}^{\sC}$, i.e.,
\begin{align*}
    \bar{R}_{22} \triangleq \bar{R}_{22}^{\sC}+\bar{R}_{22}^{\sP},
\end{align*}
where
\begin{align*}
    \bar{R}_{22}^{\sP} \triangleq n_{22}-n_{12}.
\end{align*}

It remains to choose the values of $\bar{R}_{11}^{\sC}$, $\bar{R}_{22}^{\sC}$,
$\bar{R}_{12}$, and $\bar{R}_{21}$. Our proposed design rules are as follows.
\begin{itemize}
    \item We dedicate the $\bar{R}_{11}^{\sC}$ most-significant bits of
        $\bar{\bm{u}}_{11}$ to carry information from transmitter one to
        receiver one. 
    \item Similarly, we dedicate the $\bar{R}_{22}^{\sC}$ most-significant
        bits of $\bar{\bm{u}}_{22}$ to carry information from
        transmitter two to receiver two. 
    \item We always set the $n_{22}-n_{21}$ most-significant bits of
        $\bar{\bm{u}}_{12}$ to zero. The next $\bar{R}_{12}$ bits
        of $\bar{\bm{u}}_{12}$
        carry information from transmitter two to receiver one. As shown
        in Fig.~\ref{fig:Rx2-deterministic}, this guarantees 
        the (partial) alignment of $\bar{\bm{u}}_{12}$ with
        $\bar{\bm{u}}_{11}$ at the second receiver. 
    \item We always set the $n_{11}-n_{12}$ most-significant bits of
        $\bar{\bm{u}}_{21}$ to zero. The next $\bar{R}_{21}$ 
        bits of $\bar{\bm{u}}_{21}$ carry information from transmitter one to
        receiver two. As shown in Fig.~\ref{fig:Rx1-deterministic}, this
        guarantees the (partial) alignment of $\bar{\bm{u}}_{21}$ with
        $\bar{\bm{u}}_{22}$ at the first receiver. 
\end{itemize} 

Optimizing the values of the rates $\bar{R}_{mk}$ subject to the condition
that both receivers can decode the desired messages yields the lower
bound in Theorem~\ref{thm:arbitrary_det}. The details of this analysis can
be found in Section~\ref{sec:proofs_det_lower}.

Generalizing these ideas from the deterministic to the Gaussian model,
we obtain the following constant-gap capacity approximation for the Gaussian
X-channel with general asymmetric channel gains.
\begin{theorem}
    \label{thm:arbitrary_gaussian}
    For every $\delta \in(0,1]$ and $\bm{N}\in\Zp^{2\times 2}$ with
    $\min\{ n_{11}, n_{22}  \} \geq \max \{ n_{12}, n_{21} \}$
    there exists a set $B\subseteq(1,2]^{2\times 2}$ of Lebesgue
    measure 
    \begin{equation*}
        \mu(B) \leq \delta
    \end{equation*}
    such that for all channel gains
    $(h_{mk})\in(1,2]^{2\times 2}\setminus B$ the sum
    capacity $C(\bm{N})$ of the general Gaussian X-channel~\eqref{eq:xc_arb}
    satisfies
    \begin{equation*}
        D(\bm{N})-2\log(c_2/\delta)
        \leq C(\bm{N}) 
        \leq D(\bm{N})+4
    \end{equation*}
    for some positive universal constant $c_2$, and where $D(\bm{N})$ is
    as defined in Theorem~\ref{thm:arbitrary_det}.  Moreover, the lower
    bound on $C(\bm{N})$ is achievable with a $\max\{n_{mk}\}$-bit
    quantization of the channel gains $h_{mk}$ available at both
    transmitters and receivers.
\end{theorem}  

The proof of Theorem~\ref{thm:arbitrary_gaussian} is presented in
Section~\ref{sec:proofs_gaussian}. For the special case of symmetric
channel $\SNR$s, $n_{mk}=n$ for all $m,k$,
Theorem~\ref{thm:arbitrary_gaussian} reduces to
Theorem~\ref{thm:symmetric_gaussian} in
Section~\ref{sec:main_symmetric}. Comparing
Theorems~\ref{thm:arbitrary_gaussian} and~\ref{thm:arbitrary_det}, we
see that, up to a constant gap, the Gaussian X-channel and its
lower-triangular deterministic approximation have the same capacity.
Thus, the lower-triangular deterministic model captures the relevant
features of the Gaussian X-channel.

The lower bound in Theorem~\ref{thm:arbitrary_gaussian} is achieved by
encoders and decoders that have access to only a $\max\{n_{mk}\}$-bit
quantization $\hat{h}_{mk}$ of the channel gains $h_{mk}$. As before,
the encoders and decoders are mismatched, in the sense that they are
operating under the assumption that $\hat{h}_{mk}$ is the correct
channel gain. This shows again that the proposed communication scheme is
quite robust with respect to channel estimation and quantization errors.

\section{Proof of Theorem~\ref{thm:arbitrary_det} (Deterministic X-Channel)}
\label{sec:proofs_det}

This section contains the proof of the capacity approximation for the
deterministic X-channel in Theorem~\ref{thm:arbitrary_det}. Achievability of
the lower bound in the theorem is proved in Section
\ref{sec:proofs_det_lower}; the upper bound is proved in
Section~\ref{sec:proofs_det_upper}.

\subsection{Achievability for the Deterministic X-Channel}
\label{sec:proofs_det_lower}

This section contains the proof of the lower bound in
Theorem~\ref{thm:arbitrary_det}.  Without loss of generality, we assume that
$n_{22} \geq n_{11}$.  We use the achievable scheme outlined in
Section~\ref{sec:main_arbitrary} (see
Figs.~\ref{fig:Rx1-deterministic} and~\ref{fig:Rx2-deterministic}
there). We want to maximize the sum rate
\begin{equation*}
    \bar{R}_{11}^{\sC}+\bar{R}_{11}^{\sP}+\bar{R}_{22}^{\sC}+\bar{R}_{22}^{\sP}+\bar{R}_{12}+\bar{R}_{21},
\end{equation*}
where
\begin{equation*}
    \bar{R}_{kk}^{\sC}+\bar{R}_{kk}^{\sP} = \bar{R}_{kk}
\end{equation*}
is the total rate from transmitter $k$ to receiver $k$. The constraint
is that each receiver can solve for its own desired messages plus the
visible parts of the aligned interference bits. 

If the subspaces spanned by the columns of $\bar{\bm{G}}_{mk}$
corresponding to information-bearing bits of $\bar{\bm{u}}_{mk}$ are
linearly independent, then there exists a unique channel input to the
deterministic X-channel that results in the observed channel output.
The decoder declares that this unique channel input was sent. The next
lemma provides a sufficient condition for this linear independence to
hold and hence for decoding to be successful.

\begin{lemma}
    \label{thm:decoding_det}
    Let $\delta\in(0,1]$ and $\bm{N}\in\Zp^{2\times 2}$ such that 
    $\min\{n_{11}, n_{22}\}\geq\max\{n_{12},n_{21}\}$.
    Assume $\bar{R}_{11}^\sP$,$\bar{R}_{11}^\sC$, $\bar{R}_{12}$, $\bar{R}_{21}$, $\bar{R}_{22}^\sP$,
    $\bar{R}_{22}^\sC\in\Zp$ satisfy
    \begin{subequations}
        \label{eq:decoding1}
        \begin{align}
            \label{eq:decoding1a}
            \bar{R}_{11}^{\sC} +\max\{\bar{R}_{21}, \bar{R}_{22}^{\sC}\} +\bar{R}_{12} +\bar{R}_{11}^{\sP}
            & \leq n_{11}-\log(32/\delta), \\
            \label{eq:decoding1b}
            \max\{\bar{R}_{21}, \bar{R}_{22}^{\sC}\} +\bar{R}_{12} +\bar{R}_{11}^{\sP}
            & \leq n_{12}-\log(32/\delta), \\
            \label{eq:decoding1c}
            \bar{R}_{12}+\bar{R}_{11}^{\sP}
            & \leq n_{12}+n_{21}-n_{22},
        \end{align}
    \end{subequations}
    and
    \begin{subequations}
        \label{eq:decoding2}
        \begin{align}
            \label{eq:decoding2a}
            \bar{R}_{22}^{\sC} +\max\{\bar{R}_{12}, \bar{R}_{11}^{\sC}\} +\bar{R}_{21} +\bar{R}_{22}^{\sP}
            & \leq n_{22}-\log(32/\delta), \\
            \label{eq:decoding2b}
            \max\{\bar{R}_{12}, \bar{R}_{11}^{\sC}\} +\bar{R}_{21} +\bar{R}_{22}^{\sP}
            & \leq n_{21}-\log(32/\delta), \\
            \label{eq:decoding2c}
            \bar{R}_{21}+\bar{R}_{22}^{\sP}
            & \leq n_{12}+n_{21}-n_{11}.
        \end{align}
    \end{subequations}
    Then the bit allocation in Section~\ref{sec:main_arbitrary} for
    the (modulated) deterministic X-channel~\eqref{eq:xc_det_arb}
    allows successful decoding at both receivers for all channel gains
    $(g_{mk})\in(1,2]^{2\times 3}$ except for a set 
    $B\subset(1,2]^{2\times 3}$ of Lebesgue measure
    \begin{equation*}
        \mu(B) \leq \delta.
    \end{equation*}
    If $\max\{\bar{R}_{21},\bar{R}_{22}^\sC\}=0$, then
    \eqref{eq:decoding1b} can be removed (i.e., does not need to be
    verified); and if $\bar{R}_{12}=0$, \eqref{eq:decoding1c} can
    be removed. Similarly, if $\max\{\bar{R}_{12},\bar{R}_{11}^\sC\}=0$,
    \eqref{eq:decoding2b} can be removed; and if 
    $\bar{R}_{21}=0$, \eqref{eq:decoding2c} can be removed.

\end{lemma}

The proof of Lemma~\ref{thm:decoding_det} is reported in
Section~\ref{sec:foundations_det}.

We now choose rates satisfying these decoding conditions.
For ease of notation, we will ignore the $\log(32/\delta)$ terms
throughout---the reduction in sum rate due to this additional
requirement is at most $2\log(32/\delta)$. The optimal allocation of
bits at the transmitters depends on the value $n_{12}+n_{21}$. We treat
the cases 
\begin{align*}
    \text{I:} \quad n_{12}+n_{21} & \in \big[0, n_{11}\big] \\
    \text{II:} \quad n_{12}+n_{21} & \in \big(n_{11}, n_{22}\big] \\
    \text{III:} \quad n_{12}+n_{21} & \in \big(n_{22}, n_{11}+\tfrac{1}{2}n_{22}\big] \\
    \text{IV:} \quad n_{12}+n_{21} & \in \big(n_{11}+\tfrac{1}{2}n_{22}, 
        \tfrac{3}{2}n_{22}\big] \\
    \text{V:} \quad n_{12}+n_{21} & \in \big(\tfrac{3}{2}n_{22}, n_{11}+n_{22}\big]
\end{align*}
separately. Since $n_{12}+n_{21} \leq n_{11}+n_{22}$ by the assumption
$\max\{n_{12},n_{21}\}\leq \min\{n_{11},n_{22}\}$,
this covers all possible values of $\bm{N}$.

Case I ($0 \leq n_{12}+n_{21} \leq n_{11}$):
We set
\begin{align*}
    \bar{R}_{11}^{\sP} & \triangleq n_{11}-n_{21}, \\
    \bar{R}_{22}^{\sP} & \triangleq n_{22}-n_{12}, \\
    \bar{R}_{22}^{\sC} & \triangleq \bar{R}_{11}^{\sC} \triangleq \bar{R}_{12} \triangleq \bar{R}_{21} \triangleq 0.
\end{align*}
In words, we solely communicate using the private channel inputs
$\bar{\bm{u}}_{11}^{\sP}$ and $\bar{\bm{u}}_{22}^{\sP}$. Recall that, by
our assumptions throughout this section, $\max\{n_{12}, n_{21}\} \leq
n_{11} \leq n_{22}$. Hence, $\bar{R}_{11}^{\sP} \geq 0$ and
$\bar{R}_{22}^{\sP} \geq 0$, so that this rate allocation is valid. The
calculation in Appendix~\ref{sec:appendix_decoding} verifies that this
rate allocation satisfies the decoding conditions \eqref{eq:decoding1}
and \eqref{eq:decoding2} in Lemma~\ref{thm:decoding_det}.  Hence both
receivers can recover the desired messages. The sum rate can be verified
to be
\begin{align}
    \label{eq:case1}
    (n_{11} -n_{21})+(n_{22}-n_{12}) 
    & = D_1(\bm{N})+ (n_{11}-n_{21})+(n_{22}-n_{12}) \nonumber\\
    & \geq D(\bm{N}).
\end{align}

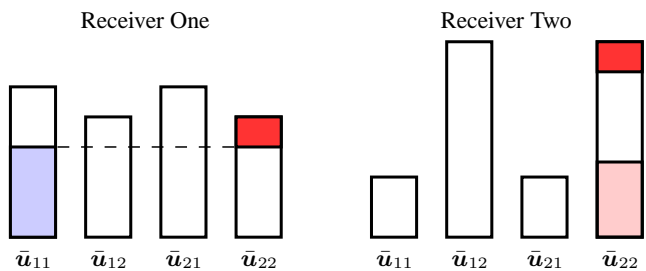
\begin{figure}[htbp]
    \centering
    %\hspace{-1cm}
    % Generated with LaTeXDraw 2.0.8
% Mon Sep 12 10:27:41 EDT 2011
% \usepackage[usenames,dvipsnames]{pstricks}
% \usepackage{epsfig}
% \usepackage{pst-grad} % For gradients
% \usepackage{pst-plot} % For axes
\scalebox{1} % Change this value to rescale the drawing.
{
\begin{pspicture}(0,-1.785625)(10.587188,1.785625)
\definecolor{color449b}{rgb}{1.0,0.8,0.8}
\definecolor{color466b}{rgb}{0.8,0.8,1.0}
\definecolor{color488b}{rgb}{1.0,0.2,0.2}
\psframe[linewidth=0.04,dimen=outer](6.463125,-0.4565625)(5.823125,-1.2965626)
\psframe[linewidth=0.04,dimen=outer,fillstyle=solid,fillcolor=color449b](9.463125,-0.2565625)(8.823125,-1.2965626)
\psframe[linewidth=0.04,dimen=outer](9.463125,1.3434376)(8.823125,-1.2965626)
\psframe[linewidth=0.04,dimen=outer](7.463125,1.3434376)(6.823125,-1.2965626)
\psframe[linewidth=0.04,dimen=outer](8.463125,-0.4565625)(7.823125,-1.2965626)
\usefont{T1}{ptm}{m}{n}
\rput(6.1567187,-1.5815625){\footnotesize $\bar{\bm{u}}_{11}$}
\usefont{T1}{ptm}{m}{n}
\rput(7.1567187,-1.5815625){\footnotesize $\bar{\bm{u}}_{12}$}
\usefont{T1}{ptm}{m}{n}
\rput(8.156719,-1.5815625){\footnotesize $\bar{\bm{u}}_{21}$}
\usefont{T1}{ptm}{m}{n}
\rput(9.156719,-1.5815625){\footnotesize $\bar{\bm{u}}_{22}$}
\psframe[linewidth=0.04,dimen=outer](1.663125,0.7434375)(1.023125,-1.2965626)
\psframe[linewidth=0.04,dimen=outer,fillstyle=solid,fillcolor=color466b](1.663125,-0.0565625)(1.023125,-1.2965626)
\psframe[linewidth=0.04,dimen=outer](4.663125,0.3434375)(4.023125,-1.2965626)
\psframe[linewidth=0.04,dimen=outer](2.663125,0.3434375)(2.023125,-1.2965626)
\psframe[linewidth=0.04,dimen=outer](3.663125,0.7434375)(3.023125,-1.2965626)
\usefont{T1}{ptm}{m}{n}
\rput(1.3567188,-1.5815625){\footnotesize $\bar{\bm{u}}_{11}$}
\usefont{T1}{ptm}{m}{n}
\rput(2.3567188,-1.5815625){\footnotesize $\bar{\bm{u}}_{12}$}
\usefont{T1}{ptm}{m}{n}
\rput(3.3567188,-1.5815625){\footnotesize $\bar{\bm{u}}_{21}$}
\usefont{T1}{ptm}{m}{n}
\rput(4.3567185,-1.5815625){\footnotesize $\bar{\bm{u}}_{22}$}
\usefont{T1}{ptm}{m}{n}
\rput(2.8375,1.6184375){\footnotesize Receiver One}
\usefont{T1}{ptm}{m}{n}
\rput(7.6375,1.6184375){\footnotesize Receiver Two}
\psframe[linewidth=0.04,dimen=outer,fillstyle=solid,fillcolor=color488b](4.663125,0.3434375)(4.023125,-0.0965625)
\psline[linewidth=0.02cm,linestyle=dashed,dash=0.16cm 0.16cm](1.043125,-0.0765625)(4.643125,-0.0765625)
\psframe[linewidth=0.04,dimen=outer,fillstyle=solid,fillcolor=color488b](9.463125,1.3434376)(8.823125,0.9034375)
\end{pspicture} 
}
    \caption{Allocation of bits in case II. Here $n_{11}=10, n_{22}=13,
    n_{12}=8, n_{21}=4$. The transmitters send private messages at rates
    $\bar{R}_{11}^{\sP}=6$ and $\bar{R}_{22}^{\sP}=5$. Transmitter two
    sends a common message to receiver two at rate
    $\bar{R}_{22}^{\sC}=2$.} 
    \label{fig:Det_Case_II}
\end{figure}

Case II ($n_{11} < n_{12}+n_{21} \leq n_{22}$): 
We set
\begin{align*}
    \bar{R}_{11}^{\sP} & \triangleq n_{11}-n_{21}, \\
    \bar{R}_{22}^{\sP} & \triangleq n_{22}-n_{12}, \\
    \bar{R}_{22}^{\sC} & \triangleq n_{12}-\bar{R}_{11}^{\sP}, \\
    \bar{R}_{11}^{\sC} & \triangleq \bar{R}_{12} \triangleq \bar{R}_{21} \triangleq 0,
\end{align*}
as shown in Fig.~\ref{fig:Det_Case_II}. Since $n_{12}+n_{21} > n_{11}$,
we have $\bar{R}_{22}^{\sC} \geq 0$, and hence this rate allocation is valid.

The calculation in Appendix~\ref{sec:appendix_decoding} verifies that
this rate allocation satisfies the decoding conditions
\eqref{eq:decoding1} and \eqref{eq:decoding2} in
Lemma~\ref{thm:decoding_det}. Hence both receivers can decode
successfully. The sum rate can be verified to be 
\begin{align}
    \label{eq:case2}
    (n_{12} +n_{21}-  n_{11})+ (n_{11}-n_{21})+(n_{22}-n_{12}) 
    & = D_1(\bm{N})+ (n_{11}-n_{21})+(n_{22}-n_{12}) \nonumber\\
    & \geq D(\bm{N}).
\end{align}

\begin{figure}[htbp]
    \centering
    %\hspace{-1cm}
    % Generated with LaTeXDraw 2.0.8
% Mon Sep 12 10:28:09 EDT 2011
% \usepackage[usenames,dvipsnames]{pstricks}
% \usepackage{epsfig}
% \usepackage{pst-grad} % For gradients
% \usepackage{pst-plot} % For axes
\scalebox{1} % Change this value to rescale the drawing.
{
\begin{pspicture}(0,-1.785625)(10.587188,1.785625)
\definecolor{color100b}{rgb}{1.0,0.8,0.8}
\definecolor{color102b}{rgb}{1.0,0.2,0.2}
\definecolor{color116b}{rgb}{0.8,0.8,1.0}
\definecolor{color128b}{rgb}{1.0,1.0,0.2}
\definecolor{color129b}{rgb}{0.2,0.2,1.0}
\definecolor{color130b}{rgb}{0.2,1.0,0.2}
\psframe[linewidth=0.04,dimen=outer](6.463125,0.5434375)(5.823125,-1.2965626)
\psframe[linewidth=0.04,dimen=outer,fillstyle=solid,fillcolor=color100b](9.463125,-0.2565625)(8.823125,-1.2965626)
\psframe[linewidth=0.04,dimen=outer](9.463125,1.3434376)(8.823125,-1.2965626)
\psframe[linewidth=0.04,dimen=outer,fillstyle=solid,fillcolor=color102b](9.463125,1.3434376)(8.823125,0.5034375)
\psframe[linewidth=0.04,dimen=outer](7.463125,1.3434376)(6.823125,-1.2965626)
\psframe[linewidth=0.04,dimen=outer](8.463125,0.5434375)(7.823125,-1.2965626)
\usefont{T1}{ptm}{m}{n}
\rput(6.1567187,-1.5815625){\footnotesize $\bar{\bm{u}}_{11}$}
\usefont{T1}{ptm}{m}{n}
\rput(7.1567187,-1.5815625){\footnotesize $\bar{\bm{u}}_{12}$}
\usefont{T1}{ptm}{m}{n}
\rput(8.156719,-1.5815625){\footnotesize $\bar{\bm{u}}_{21}$}
\usefont{T1}{ptm}{m}{n}
\rput(9.156719,-1.5815625){\footnotesize $\bar{\bm{u}}_{22}$}
\psline[linewidth=0.02cm,linestyle=dashed,dash=0.16cm 0.16cm](5.843125,-0.0765625)(9.443125,-0.0765625)
\psline[linewidth=0.02cm,linestyle=dashed,dash=0.16cm 0.16cm](5.843125,0.5234375)(9.443125,0.5234375)
\psframe[linewidth=0.04,dimen=outer](1.663125,0.9434375)(1.023125,-1.2965626)
\psframe[linewidth=0.04,dimen=outer,fillstyle=solid,fillcolor=color116b](1.663125,-0.8565625)(1.023125,-1.2965626)
\psframe[linewidth=0.04,dimen=outer](4.663125,0.3434375)(4.023125,-1.2965626)
\psframe[linewidth=0.04,dimen=outer](2.663125,0.3434375)(2.023125,-1.2965626)
\psframe[linewidth=0.04,dimen=outer](3.663125,0.9434375)(3.023125,-1.2965626)
\usefont{T1}{ptm}{m}{n}
\rput(1.3567188,-1.5815625){\footnotesize $\bar{\bm{u}}_{11}$}
\usefont{T1}{ptm}{m}{n}
\rput(2.3567188,-1.5815625){\footnotesize $\bar{\bm{u}}_{12}$}
\usefont{T1}{ptm}{m}{n}
\rput(3.3567188,-1.5815625){\footnotesize $\bar{\bm{u}}_{21}$}
\usefont{T1}{ptm}{m}{n}
\rput(4.3567185,-1.5815625){\footnotesize $\bar{\bm{u}}_{22}$}
\psframe[linewidth=0.04,dimen=outer,fillstyle=solid,fillcolor=color128b](8.463125,-0.0565625)(7.823125,-0.2965625)
\psframe[linewidth=0.04,dimen=outer,fillstyle=solid,fillcolor=color129b](1.663125,0.9434375)(1.023125,0.3034375)
\psframe[linewidth=0.04,dimen=outer,fillstyle=solid,fillcolor=color130b](2.663125,-0.4565625)(2.023125,-0.8965625)
\psline[linewidth=0.02cm,linestyle=dashed,dash=0.16cm 0.16cm](1.043125,0.3234375)(4.643125,0.3234375)
\usefont{T1}{ptm}{m}{n}
\rput(2.8375,1.6184375){\footnotesize Receiver One}
\usefont{T1}{ptm}{m}{n}
\rput(7.6375,1.6184375){\footnotesize Receiver Two}
\psframe[linewidth=0.04,dimen=outer,fillstyle=solid,fillcolor=color129b](6.463125,0.5434375)(5.823125,-0.0965625)
\psline[linewidth=0.02cm,linestyle=dashed,dash=0.16cm 0.16cm](5.843125,-0.2765625)(9.443125,-0.2765625)
\psframe[linewidth=0.04,dimen=outer,fillstyle=solid,fillcolor=color102b](4.663125,0.3434375)(4.023125,-0.4965625)
\psline[linewidth=0.02cm,linestyle=dashed,dash=0.16cm 0.16cm](1.043125,-0.4765625)(4.643125,-0.4765625)
\psframe[linewidth=0.04,dimen=outer,fillstyle=solid,fillcolor=color130b](7.463125,0.5434375)(6.823125,0.1034375)
\psframe[linewidth=0.04,dimen=outer,fillstyle=solid,fillcolor=color128b](3.663125,0.3434375)(3.023125,0.1034375)
\psline[linewidth=0.02cm,linestyle=dashed,dash=0.16cm 0.16cm](1.043125,-0.8765625)(4.643125,-0.8765625)
\end{pspicture} 
}
    \caption{Allocation of bits in case III. Here $n_{11}=11, n_{22}=13,
    n_{12}=8, n_{21}=9$. The transmitters send private messages at
    rates $\bar{R}_{11}^{\sP}=2$ and $\bar{R}_{22}^{\sP}=5$. Transmitter one sends a
    common message to receiver one at rate $\bar{R}_{11}^{\sC}=3$.
    Transmitter two sends a common message to receiver two at rate $\bar{R}_{22}^{\sC}=4$.
    The rates over the cross links are $\bar{R}_{12}=2$ and $\bar{R}_{21}=1$.
    Observe that the interference terms are partially aligned at each
    receiver.} 
    \label{fig:Det_Case_III}
\end{figure}
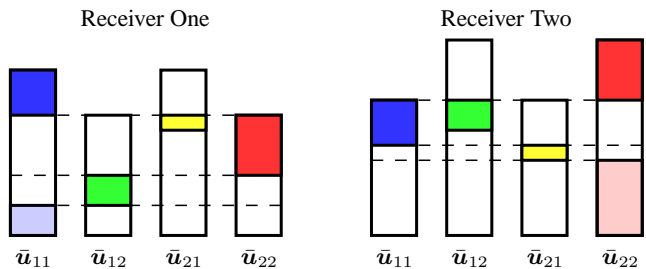

Case III ($n_{22} < n_{12}+n_{21} \leq n_{11}+\tfrac{1}{2}n_{22}$):
We set
\begin{align*}
    \bar{R}_{11}^{\sP} & \triangleq n_{11}-n_{21}, \\
    \bar{R}_{22}^{\sP} & \triangleq n_{22}-n_{12}, \\
    \bar{R}_{12} & \triangleq (n_{12}+2n_{21}-n_{11}-n_{22})^+, \\
    \bar{R}_{21} & \triangleq (n_{21}+2n_{12}-n_{11}-n_{22})^+, \\
    \bar{R}_{11}^{\sC} & \triangleq n_{21}-\bar{R}_{22}^{\sP}-\bar{R}_{21}, \\
    \bar{R}_{22}^{\sC} & \triangleq n_{12}-\bar{R}_{11}^{\sP}-\bar{R}_{12},
\end{align*}
as depicted in Fig.~\ref{fig:Det_Case_III}. Using $n_{12}+n_{21} >
n_{22}$ and $n_{22} \geq n_{11} \geq \max\{n_{12}, n_{21}\}$, it can be
verified that $\bar{R}_{11}^{\sC} \geq 0$ and $\bar{R}_{22}^{\sC} \geq 0$, and hence
this rate allocation is valid. 

The calculation in Appendix~\ref{sec:appendix_decoding} verifies the
decoding conditions \eqref{eq:decoding1} and \eqref{eq:decoding2} in
Lemma~\ref{thm:decoding_det}. The sum rate can be verified to be
\begin{align}
    \label{eq:case3}
    n_{12}+n_{21}
    & = D_1(\bm{N})+ (n_{11}-n_{21})+(n_{22}-n_{12}) \nonumber\\
    & \geq D(\bm{N}).
\end{align}

\begin{figure}[htbp]
    \centering
    %\hspace{-1cm}
    % Generated with LaTeXDraw 2.0.8
% Mon Sep 12 10:28:28 EDT 2011
% \usepackage[usenames,dvipsnames]{pstricks}
% \usepackage{epsfig}
% \usepackage{pst-grad} % For gradients
% \usepackage{pst-plot} % For axes
\scalebox{1} % Change this value to rescale the drawing.
{
\begin{pspicture}(0,-1.785625)(10.587188,1.785625)
\definecolor{color225b}{rgb}{1.0,0.8,0.8}
\definecolor{color227b}{rgb}{1.0,0.2,0.2}
\definecolor{color229b}{rgb}{0.2,1.0,0.2}
\definecolor{color241b}{rgb}{0.8,0.8,1.0}
\definecolor{color253b}{rgb}{1.0,1.0,0.2}
\definecolor{color254b}{rgb}{0.2,0.2,1.0}
\psframe[linewidth=0.04,dimen=outer](6.463125,0.3434375)(5.823125,-1.2965626)
\psframe[linewidth=0.04,dimen=outer,fillstyle=solid,fillcolor=color225b](9.463125,-0.2565625)(8.823125,-1.2965626)
\psframe[linewidth=0.04,dimen=outer](9.463125,1.3434376)(8.823125,-1.2965626)
\psframe[linewidth=0.04,dimen=outer,fillstyle=solid,fillcolor=color227b](9.463125,1.3434376)(8.823125,0.3034375)
\psframe[linewidth=0.04,dimen=outer](7.463125,1.3434376)(6.823125,-1.2965626)
\psframe[linewidth=0.04,dimen=outer,fillstyle=solid,fillcolor=color229b](7.463125,0.3434375)(6.823125,0.0034375)
\psframe[linewidth=0.04,dimen=outer](8.463125,0.3434375)(7.823125,-1.2965626)
\usefont{T1}{ptm}{m}{n}
\rput(6.1567187,-1.5815625){\footnotesize $\bar{\bm{u}}_{11}$}
\usefont{T1}{ptm}{m}{n}
\rput(7.1567187,-1.5815625){\footnotesize $\bar{\bm{u}}_{12}$}
\usefont{T1}{ptm}{m}{n}
\rput(8.156719,-1.5815625){\footnotesize $\bar{\bm{u}}_{21}$}
\usefont{T1}{ptm}{m}{n}
\rput(9.156719,-1.5815625){\footnotesize $\bar{\bm{u}}_{22}$}
\psline[linewidth=0.02cm,linestyle=dashed,dash=0.16cm 0.16cm](5.843125,0.3234375)(9.443125,0.3234375)
\psframe[linewidth=0.04,dimen=outer](1.663125,0.5434375)(1.023125,-1.2965626)
\psframe[linewidth=0.04,dimen=outer,fillstyle=solid,fillcolor=color241b](1.663125,-1.0565625)(1.023125,-1.2965626)
\psframe[linewidth=0.04,dimen=outer](4.663125,0.3434375)(4.023125,-1.2965626)
\psframe[linewidth=0.04,dimen=outer](2.663125,0.3434375)(2.023125,-1.2965626)
\psframe[linewidth=0.04,dimen=outer](3.663125,0.5434375)(3.023125,-1.2965626)
\usefont{T1}{ptm}{m}{n}
\rput(1.3567188,-1.5815625){\footnotesize $\bar{\bm{u}}_{11}$}
\usefont{T1}{ptm}{m}{n}
\rput(2.3567188,-1.5815625){\footnotesize $\bar{\bm{u}}_{12}$}
\usefont{T1}{ptm}{m}{n}
\rput(3.3567188,-1.5815625){\footnotesize $\bar{\bm{u}}_{21}$}
\usefont{T1}{ptm}{m}{n}
\rput(4.3567185,-1.5815625){\footnotesize $\bar{\bm{u}}_{22}$}
\psframe[linewidth=0.04,dimen=outer,fillstyle=solid,fillcolor=color253b](8.463125,0.1434375)(7.823125,-0.1965625)
\psframe[linewidth=0.04,dimen=outer,fillstyle=solid,fillcolor=color254b](1.663125,0.5434375)(1.023125,0.2234375)
\psframe[linewidth=0.04,dimen=outer,fillstyle=solid,fillcolor=color229b](2.663125,-0.0565625)(2.023125,-0.3765625)
\psline[linewidth=0.02cm,linestyle=dashed,dash=0.16cm 0.16cm](1.043125,0.3234375)(4.643125,0.3234375)
\usefont{T1}{ptm}{m}{n}
\rput(2.8375,1.6184375){\footnotesize Receiver One}
\usefont{T1}{ptm}{m}{n}
\rput(7.6375,1.6184375){\footnotesize Receiver Two}
\psframe[linewidth=0.04,dimen=outer,fillstyle=solid,fillcolor=color254b](6.463125,0.3434375)(5.823125,0.0034375)
\psline[linewidth=0.02cm,linestyle=dashed,dash=0.16cm 0.16cm](5.843125,-0.2765625)(9.443125,-0.2765625)
\psframe[linewidth=0.04,dimen=outer,fillstyle=solid,fillcolor=color253b](3.663125,0.3434375)(3.023125,0.0034375)
\psframe[linewidth=0.04,dimen=outer,fillstyle=solid,fillcolor=color227b](4.663125,0.3434375)(4.023125,-0.6765625)
\psline[linewidth=0.02cm,linestyle=dashed,dash=0.16cm 0.16cm](1.043125,-0.0765625)(4.643125,-0.0765625)
\psline[linewidth=0.02cm,linestyle=dashed,dash=0.16cm 0.16cm](1.043125,-1.0765625)(4.643125,-1.0765625)
\psline[linewidth=0.02cm,linestyle=dashed,dash=0.16cm 0.16cm](5.843125,0.1234375)(9.443125,0.1234375)
\end{pspicture} 
}
    \caption{Allocation of bits in case IV. Here $n_{11}=18, n_{22}=26,
    n_{12}=n_{21}=16$. The transmitters send private messages at rates
    $\bar{R}_{11}^{\sP} = 2$ and $\bar{R}_{22}^{\sP} = 10$. Transmitter one sends a common
    message to receiver one at rate $\bar{R}_{11}^{\sC} = 3$, and transmitter two
    sends a common message to receiver two at rate $\bar{R}_{22}^{\sC} = 10$. The
    rates over the cross links are $\bar{R}_{12} = 3$ and $\bar{R}_{21} = 3$. In case
    IV, the interference terms are completely aligned at receiver two,
    but only partially aligned at receiver one.}
    \label{fig:Det_Case_IV}
\end{figure}
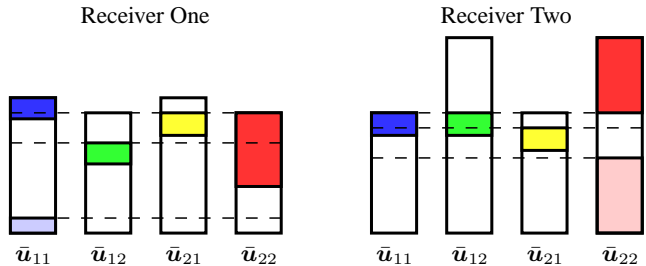

Case IV ($n_{11}+\tfrac{1}{2}n_{22} < n_{12}+n_{21} \leq \tfrac{3}{2}n_{22}$):
We set
\begin{align*}
    \bar{R}_{11}^{\sP} & \triangleq n_{11}-n_{21}, \\
    \bar{R}_{22}^{\sP} & \triangleq n_{22}-n_{12}, \\
    \bar{R}_{21} & \triangleq \big\lfloor n_{12}-\tfrac{1}{2}n_{22} \big\rfloor, \\
    \bar{R}_{12} & \triangleq \bar{R}_{11}^{\sC} \triangleq \big\lfloor n_{21}-\tfrac{1}{2}n_{22} \big\rfloor, \\
    \bar{R}_{22}^{\sC} & \triangleq n_{22}-n_{21},
\end{align*}
as shown in Fig.~\ref{fig:Det_Case_IV}.  Using that $n_{11}
+\tfrac{1}{2}n_{22} \leq n_{12}+n_{21}$, it can be verified that
$\bar{R}_{21}$, $\bar{R}_{12}$, and $\bar{R}_{11}^{\sC}$ are
nonnegative, so that this rate allocation is valid.

The calculation in Appendix~\ref{sec:appendix_decoding} verifies the
decoding conditions \eqref{eq:decoding1} and \eqref{eq:decoding2} in
Lemma~\ref{thm:decoding_det}. The sum rate can be verified to be at
least
\begin{align}
    \label{eq:case4}
    (n_{12} + n_{21}- & \tfrac{1}{2}n_{22}) + (n_{11}-n_{21})+(n_{22}-n_{12})-3 \nonumber\\
    & = D_2(\bm{N}) + (n_{11}-n_{21}) + (n_{22}-n_{12})-3 \nonumber\\
    & \geq D(\bm{N})-3,
\end{align}
where the loss of three bits is due to the floor operation in the
definition of $\bar{R}_{21}, \bar{R}_{12}, \bar{R}_{11}^\sC$.

\begin{figure}[htbp]
    \centering
    %\hspace{-1cm}
    % Generated with LaTeXDraw 2.0.8
% Mon Sep 12 10:46:40 EDT 2011
% \usepackage[usenames,dvipsnames]{pstricks}
% \usepackage{epsfig}
% \usepackage{pst-grad} % For gradients
% \usepackage{pst-plot} % For axes
\scalebox{1} % Change this value to rescale the drawing.
{
\begin{pspicture}(0,-1.785625)(10.587188,1.785625)
\definecolor{color729b}{rgb}{1.0,0.8,0.8}
\definecolor{color731b}{rgb}{1.0,0.2,0.2}
\definecolor{color733b}{rgb}{0.2,1.0,0.2}
\definecolor{color745b}{rgb}{0.8,0.8,1.0}
\definecolor{color757b}{rgb}{1.0,1.0,0.2}
\definecolor{color758b}{rgb}{0.2,0.2,1.0}
\psframe[linewidth=0.04,dimen=outer](6.463125,0.3434375)(5.823125,-1.2965626)
\psframe[linewidth=0.04,dimen=outer,fillstyle=solid,fillcolor=color729b](9.463125,-1.0565625)(8.823125,-1.2965626)
\psframe[linewidth=0.04,dimen=outer](9.463125,1.3434376)(8.823125,-1.2965626)
\psframe[linewidth=0.04,dimen=outer,fillstyle=solid,fillcolor=color731b](9.463125,1.3434376)(8.823125,0.3034375)
\psframe[linewidth=0.04,dimen=outer](7.463125,1.3434376)(6.823125,-1.2965626)
\psframe[linewidth=0.04,dimen=outer,fillstyle=solid,fillcolor=color733b](7.463125,0.5434375)(6.823125,0.1034375)
\psframe[linewidth=0.04,dimen=outer](8.463125,0.5234375)(7.823125,-1.2965626)
\usefont{T1}{ptm}{m}{n}
\rput(6.1567187,-1.5815625){\footnotesize $\bar{\bm{u}}_{11}$}
\usefont{T1}{ptm}{m}{n}
\rput(7.1567187,-1.5815625){\footnotesize $\bar{\bm{u}}_{12}$}
\usefont{T1}{ptm}{m}{n}
\rput(8.156719,-1.5815625){\footnotesize $\bar{\bm{u}}_{21}$}
\usefont{T1}{ptm}{m}{n}
\rput(9.156719,-1.5815625){\footnotesize $\bar{\bm{u}}_{22}$}
\psline[linewidth=0.02cm,linestyle=dashed,dash=0.16cm 0.16cm](5.843125,0.5234375)(9.443125,0.5234375)
\psframe[linewidth=0.04,dimen=outer](1.663125,1.1234375)(1.023125,-1.2965626)
\psframe[linewidth=0.04,dimen=outer,fillstyle=solid,fillcolor=color745b](1.663125,-0.6565625)(1.023125,-1.2965626)
\psframe[linewidth=0.04,dimen=outer](4.663125,1.1234375)(4.023125,-1.2965626)
\psframe[linewidth=0.04,dimen=middle](2.643125,1.1234375)(2.043125,-1.2765625)
\psframe[linewidth=0.04,dimen=outer](3.663125,1.1234375)(3.023125,-1.2965626)
\usefont{T1}{ptm}{m}{n}
\rput(1.3567188,-1.5815625){\footnotesize $\bar{\bm{u}}_{11}$}
\usefont{T1}{ptm}{m}{n}
\rput(2.3567188,-1.5815625){\footnotesize $\bar{\bm{u}}_{12}$}
\usefont{T1}{ptm}{m}{n}
\rput(3.3567188,-1.5815625){\footnotesize $\bar{\bm{u}}_{21}$}
\usefont{T1}{ptm}{m}{n}
\rput(4.3567185,-1.5815625){\footnotesize $\bar{\bm{u}}_{22}$}
\psframe[linewidth=0.04,dimen=outer,fillstyle=solid,fillcolor=color757b](8.463125,0.5434375)(7.823125,-0.4965625)
\psframe[linewidth=0.04,dimen=outer,fillstyle=solid,fillcolor=color758b](1.663125,1.1434375)(1.023125,0.7034375)
\psframe[linewidth=0.04,dimen=outer,fillstyle=solid,fillcolor=color733b](2.663125,0.3434375)(2.023125,-0.0965625)
\usefont{T1}{ptm}{m}{n}
\rput(2.8375,1.6184375){\footnotesize Receiver One}
\usefont{T1}{ptm}{m}{n}
\rput(7.6375,1.6184375){\footnotesize Receiver Two}
\psframe[linewidth=0.04,dimen=outer,fillstyle=solid,fillcolor=color758b](6.463125,0.5434375)(5.823125,0.1034375)
\psline[linewidth=0.02cm,linestyle=dashed,dash=0.16cm 0.16cm](5.843125,-1.0765625)(9.443125,-1.0765625)
\psframe[linewidth=0.04,dimen=outer,fillstyle=solid,fillcolor=color757b](3.663125,1.1434375)(3.023125,0.1034375)
\psframe[linewidth=0.04,dimen=outer,fillstyle=solid,fillcolor=color731b](4.663125,1.1434375)(4.023125,0.1034375)
\psline[linewidth=0.02cm,linestyle=dashed,dash=0.16cm 0.16cm](1.043125,1.1234375)(4.643125,1.1234375)
\psline[linewidth=0.02cm,linestyle=dashed,dash=0.16cm 0.16cm](1.043125,-0.6765625)(4.643125,-0.6765625)
\psline[linewidth=0.02cm,linestyle=dashed,dash=0.16cm 0.16cm](1.043125,0.3234375)(4.643125,0.3234375)
\end{pspicture} 
}
    \caption{Allocation of bits in case V. Here $n_{11}=12, n_{22}=13,
    n_{12}=12, n_{21}=9$. The private messages to receiver one and two
    have rates $\bar{R}_{11}^{\sP} = 3$ and $\bar{R}_{22}^{\sP}=1$. The remaining messages to 
    receiver one have rate $\bar{R}_{12}=\bar{R}_{11}^{\sC}=2$, and are both entirely aligned at
    receiver two. The remaining messages to receiver two have rate
    $\bar{R}_{21}=\bar{R}_{22}^{\sC}=5$, and are both entirely aligned at receiver one.} 
    \label{fig:Det_Case_V}
\end{figure}
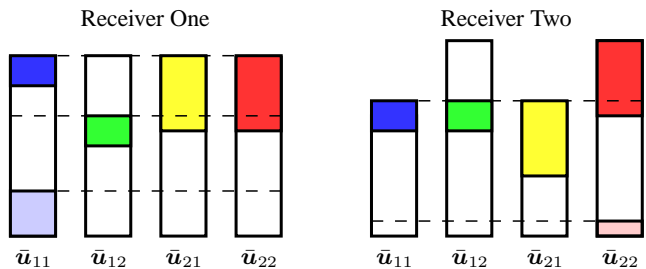

Case V ($\tfrac{3}{2}n_{22} < n_{12}+n_{21} \leq n_{11}+n_{22}$):  
We set
\begin{align*}
    \bar{R}_{11}^{\sP} & \triangleq n_{11}-n_{21}, \\
    \bar{R}_{22}^{\sP} & \triangleq n_{22}-n_{12}, \\
    \bar{R}_{12} & \triangleq \bar{R}_{11}^{\sC} 
    \triangleq \big\lfloor \tfrac{2}{3}n_{21}-\tfrac{1}{3}n_{12} \big\rfloor, \\
    \bar{R}_{21} & \triangleq \bar{R}_{22}^{\sC} 
    \triangleq \big\lfloor \tfrac{2}{3}n_{12}-\tfrac{1}{3}n_{21} \big\rfloor, 
\end{align*}
as shown in Fig.~\ref{fig:Det_Case_V}.  From $\tfrac{3}{2}n_{22} <
n_{12}+n_{21}$, it follows that $\bar{R}_{12}$, $\bar{R}_{11}^{\sC}$,
$\bar{R}_{21}$, and $\bar{R}_{22}^{\sC}$ are nonnegative, so that this
rate allocation is valid.

The calculation in Appendix~\ref{sec:appendix_decoding} verifies the
decoding conditions \eqref{eq:decoding1} and \eqref{eq:decoding2} in
Lemma~\ref{thm:decoding_det}.  The sum rate is at least
\begin{align}
    \label{eq:case5}
    \tfrac{2}{3}(n_{12}+ n_{21})+(n_{11}- & n_{21})+(n_{22}-n_{12})-4 \nonumber\\
    & = D_4(\bm{N})+(n_{11}-n_{21})+(n_{22}-n_{12})-4 \nonumber\\
    & \geq D(\bm{N})-4,
\end{align}
where the loss of four bits is due to floor operation in the definition
of $\bar{R}_{12}, \bar{R}_{11}^\sC, \bar{R}_{21}, \bar{R}_{22}^\sC$.

Combining \eqref{eq:case1}--\eqref{eq:case5}, and accounting for the loss
of $2\log(32/\delta)$ in Lemma~\ref{thm:decoding_det} shows that, 
assuming $n_{22}\geq n_{11}$, 
\begin{align*}
    \bar{C}(\bm{N}) 
    & \geq D(\bm{N})-4-2\log(32/\delta) \\
    & = D(\bm{N})-2\log(c_1/\delta)
\end{align*}
with
\begin{equation*}
    c_1 \triangleq 128.
\end{equation*}
If $n_{11}\geq n_{22}$, we can simply relabel the two transmitters and
receivers, and the same argument holds. This relabeling of receivers
introduces the function $D_3(\bm{N})$ instead of $D_2(\bm{N})$ in the
lower bound. Together, this concludes the proof of the lower bound in
Theorem~\ref{thm:arbitrary_det}.\hfill\IEEEQED

\subsection{Upper Bound for the Deterministic X-Channel}
\label{sec:proofs_det_upper}

The section contains the proof of the upper bound in
Theorem~\ref{thm:arbitrary_det}.  We start with a lemma upper bounding
various linear combinations of achievable rates for the deterministic
X-channel. 

\begin{lemma}
    \label{thm:upper}
    Any achievable rate tuple $(\bar{R}_{11},
    \bar{R}_{12},\bar{R}_{21},\bar{R}_{22})$ for the (modulated) deterministic
    X-channel~\eqref{eq:xc_det_arb} satisfies the following inequalities
    \begin{subequations}
        \label{eq:upper-eliminate}
        \begin{align}
            \label{eq:upper-eliminate-R21}
            \bar{R}_{11}+\bar{R}_{12}+\bar{R}_{22} 
            & \leq \max \{ n_{11}, n_{12} \}+ (n_{22}-n_{12})^+,\\
            \label{eq:upper-eliminate-R12}
            \bar{R}_{11}+\bar{R}_{21}+\bar{R}_{22} 
            & \leq \max \{ n_{21}, n_{22} \}+ (n_{11}-n_{21})^+,\\
            \label{eq:upper-eliminate-R22}
            \bar{R}_{11}+\bar{R}_{12}+\bar{R}_{21} 
            & \leq \max \{ n_{11}, n_{12} \}+ (n_{21}-n_{11})^+,\\
            \label{eq:upper-eliminate-R11}
            \bar{R}_{12}+\bar{R}_{21}+\bar{R}_{22} 
            & \leq \max \{ n_{21}, n_{22} \}+ (n_{12}-n_{22})^+,\\
            \label{eq:upper-bound-X2}
            \bar{R}_{11}+\bar{R}_{12}+\bar{R}_{21}+\bar{R}_{22} 
            & \leq \max\{ n_{12}, n_{11}-n_{21} \}
            +  \max\{ n_{21}, n_{22}-n_{12} \},\\
            \label{eq:sum-rate-up-last}
            \bar{R}_{11}+\bar{R}_{12}+\bar{R}_{21}+\bar{R}_{22} 
            & \leq \max\{ n_{11}, n_{12}-n_{22} \}
            +  \max\{ n_{22}, n_{21}-n_{11} \}, \\
            \label{eq:upper-bound-IC1}
            2\bar{R}_{11}+\bar{R}_{12}+\bar{R}_{21}+\bar{R}_{22}
            & \leq \max\{n_{11},n_{12}\}+\max\{n_{21}, n_{22}-n_{12}\} 
            + (n_{11}-n_{21})^+, \\
            \label{eq:upper-bound-IC2}
            \bar{R}_{11}+2\bar{R}_{12}+\bar{R}_{21}+\bar{R}_{22}
            & \leq \max\{n_{11},n_{12}\}+\max\{n_{22}, n_{21}-n_{11}\} 
            + (n_{12}-n_{22})^+, \\
            \label{eq:upper-bound-IC3}
            \bar{R}_{11}+\bar{R}_{12}+2\bar{R}_{21}+\bar{R}_{22} 
            & \leq \max\{n_{22},n_{21}\}+\max\{n_{11}, n_{12}-n_{22}\} 
            + (n_{21}-n_{11})^+, \\
            \label{eq:upper-bound-IC4}
            \bar{R}_{11}+\bar{R}_{12}+\bar{R}_{21}+2\bar{R}_{22}
            & \leq \max\{n_{22},n_{21}\}+\max\{n_{12}, n_{11}-n_{21}\} 
            + (n_{22}-n_{12})^+,
        \end{align}
    \end{subequations}
\end{lemma}

The proof of Lemma~\ref{thm:upper} is reported in
Appendix~\ref{sec:appendix_upper}. Inequalities
\eqref{eq:upper-eliminate-R21}--\eqref{eq:sum-rate-up-last} are based on
an argument from \cite[Theorem~4.4]{huang08}. Inequalities
\eqref{eq:upper-bound-IC1}--\eqref{eq:upper-bound-IC4} are novel. 

The upper bounds in Lemma~\ref{thm:upper} can be understood intuitively
as multiple-access bounds for a channel where the receivers are forced to
decode certain parts of the interference (see
Figs.~\ref{fig:Rx1-deterministic0} and \ref{fig:Rx2-deterministic0} in
Section~\ref{sec:main_arbitrary}). For example,
inequality~\eqref{eq:upper-eliminate-R21} corresponds to the
multiple-access bound
\begin{equation}
    \label{eq:multiple1}
    \bar{R}_{11}+\bar{R}_{12}+\bar{R}_{22}^\sC \leq \max\{n_{11},n_{12}\}
\end{equation}
at receiver one, combined with the inequality
\begin{equation*}
    \bar{R}_{22}^\sP \leq (n_{22}-n_{12})^+.
\end{equation*}
Similarly, inequality~\eqref{eq:upper-bound-X2}
corresponds to the multiple-access bound
\begin{equation*}
    \bar{R}_{11}^\sP+\bar{R}_{12}+\bar{R}_{22}^\sC \leq \max\{n_{12}, n_{11}-n_{21}\}
\end{equation*}
at receiver one, combined with the multiple-access bound
\begin{equation}
    \label{eq:multiple2}
    \bar{R}_{22}^\sP+\bar{R}_{21}+\bar{R}_{11}^\sC \leq \max\{n_{21}, n_{22}-n_{12}\}
\end{equation}
at receiver two. Finally, inequality~\eqref{eq:upper-bound-IC1}
corresponds to the multiple-access bounds \eqref{eq:multiple1} and
\begin{equation*}
    \bar{R}_{11}^\sP \leq (n_{11}-n_{21})^+
\end{equation*}
at receiver one, combined with the multiple-access bound
\eqref{eq:multiple2} at receiver two. The proof of Lemma~\ref{thm:upper}
makes this intuitive reasoning precise.  A detailed discussion of this
type of cut-set interpretation can be found in \cite{bresler08}.

We proceed with the proof of the upper bound in
Theorem~\ref{thm:arbitrary_det} for the deterministic X-channel. Under the
assumption 
\begin{equation}
    \label{eq:cross}
    \min\{ n_{11}, n_{22}  \} \geq
    \max \{ n_{12}, n_{21} \},
\end{equation} 
the first four inequalities
\eqref{eq:upper-eliminate-R21}--\eqref{eq:upper-eliminate-R11} in
Lemma~\ref{thm:upper} yield the following upper bound on sum capacity
\begin{align}
    \label{eq:xcgen_det_upper1}
    \bar{C}(\bm{N})
    & \leq \tfrac{2}{3} (n_{12}+n_{21}) + (n_{11}-n_{21})+(n_{22}-n_{12}) \nonumber\\
    & = D_4(\bm{N})+ (n_{11}-n_{21})+(n_{22}-n_{12}).
\end{align}
Again using~\eqref{eq:cross}, inequality
\eqref{eq:upper-bound-X2} in Lemma~\ref{thm:upper} shows that
\begin{align}
    \label{eq:xcgen_det_upper2}
    \bar{C}(\bm{N}) 
    & \leq \max\{ n_{12}, n_{11}-n_{21} \} +  \max\{ n_{21}, n_{22}-n_{12} \} \nonumber\\
    & = (n_{12}+n_{21}-n_{11})^+ +  (n_{12}+n_{21}-n_{22})^+ 
    +(n_{11}-n_{21})+(n_{22}-n_{12}) \nonumber\\
    & = D_1(\bm{N})+(n_{11}-n_{21})+(n_{22}-n_{12}).
\end{align}
Inequalities \eqref{eq:upper-eliminate-R11} and \eqref{eq:upper-bound-IC1}
in Lemma~\ref{thm:upper} combined with \eqref{eq:cross} yield
\begin{align}
    \label{eq:xcgen_det_upper3}
    \bar{C}(\bm{N})
    & \leq \tfrac{1}{2}\big( n_{11} + n_{22}+\max\{n_{21}, n_{22}-n_{12}\} 
    + (n_{11}-n_{21})\big) \nonumber\\
    & = \tfrac{1}{2}\big( n_{12}+n_{21}+ (n_{12}+n_{21}- n_{22})^+\big)
    + (n_{11}-n_{21}) + (n_{22}-n_{12}) \nonumber\\
    & = D_2(\bm{N})+ (n_{11}-n_{21}) + (n_{22}-n_{12}).
\end{align}
Similarly, from \eqref{eq:upper-eliminate-R22} and
\eqref{eq:upper-bound-IC4} in Lemma~\ref{thm:upper}, 
\begin{align}
    \label{eq:xcgen_det_upper4}
    \bar{C}(\bm{N})
    & \leq \tfrac{1}{2}\big( n_{12}+n_{21}+ (n_{12}+n_{21}- n_{11})^+\big)
    + (n_{11}-n_{21}) + (n_{22}-n_{12}) \nonumber\\
    & = D_3(\bm{N})+ (n_{11}-n_{21}) + (n_{22}-n_{12}).
\end{align}
The sum capacity is hence at most the minimum of the upper bounds
\eqref{eq:xcgen_det_upper1}--\eqref{eq:xcgen_det_upper4}, i.e., 
\begin{align*}
    \bar{C}(\bm{N})
    & \leq \min\big\{
    D_1(\bm{N}),
    D_2(\bm{N}),
    D_3(\bm{N}),
    D_4(\bm{N})
    \big\}
    +(n_{11}-n_{21}) + (n_{22}-n_{12}) \\
    & = D(\bm{N}),
\end{align*}
concluding the proof.\hfill\IEEEQED

\section{Proof of Theorem~\ref{thm:arbitrary_gaussian} (Gaussian X-Channel)}
\label{sec:proofs_gaussian}

This section contains the proof of the capacity approximation for the
Gaussian X-channel in Theorem~\ref{thm:arbitrary_gaussian}.
Achievability of the lower bound in the theorem is proved in
Section~\ref{sec:proofs_gaussian_lower}; the upper bound is proved in
Section~\ref{sec:proofs_gaussian_upper}.

\subsection{Achievability for the Gaussian X-Channel}
\label{sec:proofs_gaussian_lower}

Here, we prove the lower bound in Theorem~\ref{thm:arbitrary_gaussian}
by translating the achievable scheme for the deterministic model to the
Gaussian model. For ease of exposition, we assume in most of the
analysis that all channels gains $h_{mk}$ are exactly known at the two
transmitters and receivers. The changes in the arguments necessary for
the mismatched case, in which the transmitters and receivers have access
only to a quantized version $\hat{h}_{mk}$ of the channel gain $h_{mk}$,
are reported in Appendix~\ref{sec:appendix_mismatch}.

Recall that each transmitter $k$ has access to two messages, $w_{1k}$ and
$w_{2k}$. The transmitter forms the modulated symbol $u_{mk}$ from the
message $w_{mk}$. From these modulated signals, the channel inputs
\begin{align*}
    x_1 & \triangleq h_{22}u_{11}+h_{12}u_{21}, \\
    x_2 & \triangleq h_{21}u_{12}+h_{11}u_{22}
\end{align*}
are constructed. 

We now describe the modulation process from $w_{mk}$ to $u_{mk}$ in
detail. Each $u_{mk}$ is of the form 
\begin{equation*}
    u_{mk} \triangleq \sum_{i=3}^{n_{kk}}[u_{mk}]_i2^{-i}
\end{equation*}
with $[u_{mk}]_i\in\{0,1\}$. Since $\abs{h_{mk}}\leq 2$ and
$\abs{u_{mk}}\leq 1/4$, the resulting channel input $x_k$ satisfies
the unit average power constraint at the transmitters. 

In analogy to the achievable scheme for the deterministic channel, we
only use certain portions of the bits $[u_{mk}]_i$ in the binary
expansion of $u_{mk}$; the remaining bits are set to zero. The
allocation of information bits depends on the channel strength $\bm{N}$
and is chosen as in the deterministic case described in
Sections~\ref{sec:main_arbitrary} and~\ref{sec:proofs_det_lower}, and as
illustrated in Figs.~\ref{fig:Rx1-deterministic}
and~\ref{fig:Rx2-deterministic}. In particular, the messages $u_{kk}$
are again decomposed into common and private portions, i.e.,
\begin{equation*}
    u_{kk} = u_{kk}^\sP+u_{kk}^\sC.
\end{equation*}
We denote by $\bar{R}_{mk}$ the modulation rate of $u_{mk}$ in bits per
symbol in analogy to the deterministic case.

To satisfy the power constraint (as discussed above), we impose that the
two most significant bits of each common message are zero.  For reasons
that will become clear in the next paragraph, we also impose that the
two most significant bits for each private message are zero. This
reduces the modulation rate by at most $12$ bits per channel use
compared to the deterministic case. 

The channel output at receiver one is
\begin{align*}
    y_1 
    & = 2^{n_{11}}h_{11}x_1+2^{n_{12}}h_{12}x_2+z_1 \\
    & = 
    \big( g_{11}2^{n_{11}}u_{11}+
    g_{12} 2^{n_{12}}u_{12}\big)
    +g_{10}\big( 2^{n_{11}}u_{21} +2^{n_{12}}u_{22}^\sC \big)
    + \big(g_{10}2^{n_{12}}u_{22}^\sP 
    +z_1 \big),
\end{align*}
with $g_{mk}$ denoting the product of two channel gains as defined
in~\eqref{eq:gdef} in Section~\ref{sec:model_gauss}. The situation is
similar at receiver two.  The channel output is grouped into three
parts. The first part contains the two desired signals $u_{11}$ and
$u_{12}$. The second part contains the interference signals
$u_{21}$ and $u_{22}^\sC$. Note that these interference terms are
received with the same coefficient $g_{10}$ and are hence aligned.
The third part contains noise $z_1$ and the private portion $u_{22}^\sP$
of the message $u_{22}$.  By construction, 
\begin{equation*}
    2^{n_{12}}u_{22}^\sP \in [0,1/4)
\end{equation*}
so that
\begin{equation*}
    \abs{g_{10}2^{n_{12}}u_{22}^\sP} 
    \leq 4\cdot\tfrac{1}{4}
    \leq 1.
\end{equation*}
We will treat this part of the interference as noise.

Set
\begin{align*}
    s_{11} & \triangleq 2^{n_{11}}u_{11}, \\
    s_{12} & \triangleq 2^{n_{12}}u_{12}, \\
    s_{10} & \triangleq 2^{n_{11}}u_{21}+2^{n_{12}}u_{22}^\sC.
\end{align*}
The goal of the demodulator at receiver one is to find estimates
$\hat{s}_{1k}$ of $s_{1k}$, from which estimates for the desired channel
inputs $u_{11}$ and $u_{12}$ can be derived. The demodulator searches
for $\hat{s}_{11},\hat{s}_{12},\hat{s}_{10}$
that minimize
\begin{equation*}
    \abs{y_1-g_{11}\hat{s}_{11}-g_{12}\hat{s}_{12}-g_{10}\hat{s}_{10}}.
\end{equation*}
We point out that the demodulator decodes only the \emph{sum}
$\hat{s}_{10}$ of the two interfering symbols, but not the individual
interfering symbols themselves. The demodulator at receiver two works in
analogy.

We now lower bound the minimum distance 
\begin{equation}
    \label{eq:xc_d}
    d 
    \triangleq \min_{\substack{(s_{11}, s_{12}, s_{10}) \\ \neq(s_{11}', s_{12}', s_{10}')}}
    \big\lvert g_{11}(s_{11}-s_{11}')
    +g_{12}(s_{12}-s_{12}')
    +g_{10}(s_{10}-s_{10}')\big\rvert.
\end{equation}
between the noiseless received signal generated by the correct $(s_{11},
s_{12}, s_{10})$ and by any other triple $(s_{11}', s_{12}', s_{10}')$.
The next lemma provides a sufficient condition for this minimum distance
to be large at both receivers.

\begin{lemma}
    \label{thm:decoding_gauss}
    Let $\delta\in(0,1]$ and $\bm{N}\in\Zp^{2\times 2}$ such that 
    $\min\{n_{11}, n_{22}\}\geq\max\{n_{12},n_{21}\}$.
    Assume $\bar{R}_{11}^\sP$,$\bar{R}_{11}^\sC$, $\bar{R}_{12}$, 
    $\bar{R}_{21}$, $\bar{R}_{22}^\sP$, $\bar{R}_{22}^\sC\in\Zp$ satisfy
    \begin{subequations}
        \begin{align}
            \label{eq:decoding1agauss}
            \bar{R}_{11}^{\sC} +\max\{\bar{R}_{21}, \bar{R}_{22}^{\sC}\} 
            +\bar{R}_{12} +\bar{R}_{11}^{\sP}
            & \leq n_{11}-6-\log(13104/\delta), \\
            \label{eq:decoding1bgauss}
            \max\{\bar{R}_{21}, \bar{R}_{22}^{\sC}\} +\bar{R}_{12} +\bar{R}_{11}^{\sP}
            & \leq n_{12}-6-\log(13104/\delta), \\
            \label{eq:decoding1cgauss}
            \bar{R}_{12}+\bar{R}_{11}^{\sP}
            & \leq n_{12}+n_{21}-n_{22}-6,
        \end{align}
    \end{subequations}
    and
    \begin{subequations}
        \begin{align}
            \label{eq:decoding2agauss}
            \bar{R}_{22}^{\sC} +\max\{\bar{R}_{12}, \bar{R}_{11}^{\sC}\} 
            +\bar{R}_{21} +\bar{R}_{22}^{\sP}
            & \leq n_{22}-6-\log(13104/\delta), \\
            \label{eq:decoding2bgauss}
            \max\{\bar{R}_{12}, \bar{R}_{11}^{\sC}\} +\bar{R}_{21} +\bar{R}_{22}^{\sP}
            & \leq n_{21}-6-\log(13104/\delta), \\
            \label{eq:decoding2cgauss}
            \bar{R}_{21}+\bar{R}_{22}^{\sP}
            & \leq n_{12}+n_{21}-n_{11}-6.
        \end{align}
    \end{subequations}
    Then the bit allocation in Section~\ref{sec:main_arbitrary} applied
    to the Gaussian X-channel \eqref{eq:xc_arb} results in a minimum
    constellation distance $d \geq 32$ at each receiver
    for all channel gains $(h_{mk})\in(1,2]^{2\times 2}$ except for a
    set $B\subset(1,2]^{2\times 2}$ of Lebesgue measure
    \begin{equation*}
        \mu(B) \leq \delta.
    \end{equation*}
    If $\max\{\bar{R}_{21},\bar{R}_{22}^\sC\}=0$, then
    \eqref{eq:decoding1bgauss} can be removed (i.e., does not need to be
    verified); and if $\bar{R}_{12}=0$, \eqref{eq:decoding1cgauss} can
    be removed. Similarly, if $\max\{\bar{R}_{12},\bar{R}_{11}^\sC\}=0$,
    \eqref{eq:decoding2bgauss} can be removed; and if 
    $\bar{R}_{21}=0$, \eqref{eq:decoding2cgauss} can be removed.
\end{lemma}

The proof of Lemma~\ref{thm:decoding_gauss} is reported in
Section~\ref{sec:foundations_gaussian}. Observe that, up to the
constants, Lemma~\ref{thm:decoding_gauss} is exactly of the same form as
Lemma~\ref{thm:decoding_det} in Section~\ref{sec:proofs_det_lower} for
the lower-triangular deterministic X-Channel, highlighting again the
close connection between the two models. In the following discussion, we
will assume that the channel gains are outside the outage set, i.e.,
$(h_{mk})\notin B$. 

Recall that we have chosen the same allocation of information bits in
the binary expansion of $u_{mk}$ as in the deterministic case analyzed
in Section~\ref{sec:proofs_det_lower}.  Since the most-significant bit
of each $u_{mk}$ is zero, the binary expansion of $s_{mk}$ is also of
the form analyzed there.  Moreover, since the conditions in
Lemma~\ref{thm:decoding_gauss} used here are the same as the conditions
in Lemma~\ref{thm:decoding_det} used in the deterministic case, we
conclude that Lemma~\ref{thm:decoding_gauss} can be applied if we
further reduce the rates to accommodate the constant
$6+\log(13104/\delta)$ in Lemma~\ref{thm:decoding_gauss}. This can be
achieved for example by reducing the modulation rate by a further
$3+\tfrac{1}{2}\log(13104/\delta)\leq 10+\tfrac{1}{2}\log(1/\delta)$
per symbol. Accounting for the loss of $12$ bits per channel use due to the
power constraint, the sum rate of the modulation scheme is then
\begin{align}
    \label{eq:rmod}
    \sum_{m,k}\bar{R}_{mk}
    & = D(\bm{N})-12-4\cdot10-4\cdot\tfrac{1}{2}\log(1/\delta)-4 \nonumber\\
    & = D(\bm{N})-2\log(1/\delta)-56,
\end{align}
with $D(\bm{N})$ as defined in Theorem~\ref{thm:arbitrary_det} for the
deterministic X-channel, and where the additional loss of $4$ bits
results from rounding in the bit allocation for the deterministic
scheme as discussed in Section~\ref{sec:proofs_det_lower}.

Lemma~\ref{thm:decoding_gauss} is sufficient to show that the
probability of demodulation error is \emph{small}. To achieve a
\emph{vanishing} probability of error, we use an outer code over the
modulated channel. The distribution of $u_{mk}$ is chosen to be uniform
over the set allowed by the modulator constraints and independent of all
other modulator inputs. Let $R_{mk}$ denote the rate of this outer code
from transmitter $k$ to receiver $m$.  We now lower bound the rate
$R_{11}$ as a function of the modulation rate $\bar{R}_{11}$. 

We have
\begin{equation*}
    I\bigl(s_{11}, s_{12}, s_{10}; \hat{s}_{11}, \hat{s}_{12}, \hat{s}_{10}\bigr)
    = H\bigl(s_{11}, s_{12}, s_{10}\bigr)
    -H\bigl(s_{11}, s_{12}, s_{10}\bigm\vert \hat{s}_{11}, \hat{s}_{12}, \hat{s}_{10}\bigr).
\end{equation*}
We will argue below that
\begin{equation}
    \label{eq:const}
    H\bigl(s_{11}, s_{12}, s_{10}\bigm\vert \hat{s}_{11}, \hat{s}_{12}, \hat{s}_{10}\bigr)
    \leq 1.5
\end{equation}
so that
\begin{align}
    \label{eq:I1}
    I\bigl(s_{11}, s_{12}, s_{10}; \hat{s}_{11}, \hat{s}_{12}, \hat{s}_{10}\bigr)
    & \geq H\bigl(s_{11}, s_{12}, s_{10}\bigr)-1.5 \notag\\
    & = H(s_{11})+H(s_{12})+H(s_{10})-1.5.
\end{align}
On the other hand,
\begin{equation*}
    I\bigl(s_{11}, s_{12}, s_{10}; \hat{s}_{11}, \hat{s}_{12}, \hat{s}_{10}\bigr)
    \leq I\bigl(s_{11}; \hat{s}_{11}, \hat{s}_{12}, \hat{s}_{10}\bigr)
    + H(s_{12})+H(s_{10}).
\end{equation*}
Together with~\eqref{eq:I1}, this shows that
\begin{align*}
    I\bigl(s_{11}; \hat{s}_{11}, \hat{s}_{12}, \hat{s}_{10}\bigr)
    & \geq H(s_{11})-1.5 \\
    & = \bar{R}_{11}-1.5.
\end{align*}
Since there is a one-to-one relationship between $u_{11}$ and $s_{11}$,
this implies that the outer code can achieve a rate of
\begin{align*}
    R_{11} 
    & = I\bigl(u_{11}; \hat{s}_{11}, \hat{s}_{12}, \hat{s}_{10}\bigr) \\
    & = I\bigl(s_{11}; \hat{s}_{11}, \hat{s}_{12}, \hat{s}_{10}\bigr) \\
    & \geq \bar{R}_{11}-1.5.
\end{align*}

The same argument can be used for the other rates as well, showing that
\begin{equation*}
    R_{mk} \geq \bar{R}_{mk}-1.5
\end{equation*}
for all $m,k\in\{1,2\}$. Hence the outer codes achieves a sum rate of at
least
\begin{equation*}
    \sum_{m,k}R_{mk} \geq \sum_{m,k} \bar{R}_{mk}-6.
\end{equation*}
Using \eqref{eq:rmod}, this shows that, except for a set $B$ of measure
at most $\delta$, 
\begin{align*}
    C(\bm{N}) 
    & \geq D(\bm{N})-2\log(1/\delta)-62 \\
    & = D(\bm{N})-2\log(c_2/\delta)
\end{align*}
with
\begin{equation*}
    c_2 \triangleq 2^{31},
\end{equation*}
which is what needed to be shown.

\begin{remark}
    The rate $R_{mk}$ of the outer code can be lower bounded in terms of
    the modulation rate $\bar{R}_{mk}$ using Fano's inequality. This is
    the approach taken, for example, in \cite{etkin09,motahari09}.
    However, this approach results in a gap that depends on $\bm{N}$,
    and is hence not strong enough for a constant-gap approximation of
    capacity. Instead, we use a stronger argument (see the proof
    of~\eqref{eq:const} below) that yields a gap independent of
    $\bm{N}$. This argument is a key step in the derivation of
    the lower bound on capacity. 
\end{remark}

It remains to prove~\eqref{eq:const}. It will be convenient to define
\begin{equation*}
    v \triangleq g_{11}s_{11}+g_{12}s_{12}+g_{10}s_{10},
\end{equation*}
and similarly for $\hat{v}$ with respect to $\hat{s}_{1k}$. Observe
that the channel output $y_1$ is then equal to $v$ plus signals treated
as noise. Since we assume that the channel gains are outside the outage
set $B$, Lemma~\ref{thm:decoding_gauss} implies that there is a one-to-one
relationship between $v$ and $(s_{11},s_{12},s_{10})$, and between
$\hat{v}$ and $(\hat{s}_{11},\hat{s}_{12},\hat{s}_{10})$. Hence,
\begin{equation}
    \label{eq:1to1}
    H\bigl(s_{11}, s_{12}, s_{10}\bigm\vert \hat{s}_{11}, \hat{s}_{12}, \hat{s}_{10}\bigr)
    = H(v\mid \hat{v}).
\end{equation}

\begin{figure}[htbp]
    \centering
    % Generated with LaTeXDraw 2.0.8
% Mon Mar 11 15:50:46 EDT 2013
% \usepackage[usenames,dvipsnames]{pstricks}
% \usepackage{epsfig}
% \usepackage{pst-grad} % For gradients
% \usepackage{pst-plot} % For axes
\scalebox{1} % Change this value to rescale the drawing.
{
\begin{pspicture}(0,-0.819375)(6.6671877,0.819375)
\psline[linewidth=0.04cm](0.743125,0.2896875)(5.943125,0.2896875)
\pscircle[linewidth=0.04,dimen=outer,fillstyle=solid,fillcolor=black](0.943125,0.2896875){0.1}
\pscircle[linewidth=0.04,dimen=outer,fillstyle=solid,fillcolor=black](2.543125,0.2896875){0.1}
\pscircle[linewidth=0.04,dimen=outer,fillstyle=solid,fillcolor=black](4.543125,0.2896875){0.1}
\pscircle[linewidth=0.04,dimen=outer,fillstyle=solid,fillcolor=black](5.743125,0.2896875){0.1}
\usefont{T1}{ptm}{m}{n}
\rput(2.5167189,0.6446875){\footnotesize $q(\hat{q},0)$}
\usefont{T1}{ptm}{m}{n}
\rput(2.5167189,-0.0153125){\footnotesize $\hat{q}$}
\usefont{T1}{ptm}{m}{n}
\rput(0.93671876,0.6446875){\footnotesize $q(\hat{q},-1)$}
\usefont{T1}{ptm}{m}{n}
\rput(4.516719,0.6446875){\footnotesize $q(\hat{q},1)$}
\usefont{T1}{ptm}{m}{n}
\rput(5.7167187,0.6446875){\footnotesize $q(\hat{q},2)$}
\psline[linewidth=0.04cm,arrowsize=0.05291667cm 2.0,arrowlength=1.4,arrowinset=0.4]{<->}(2.543125,-0.3103125)(4.543125,-0.3103125)
\psline[linewidth=0.04cm,arrowsize=0.05291667cm 2.0,arrowlength=1.4,arrowinset=0.4]{<->}(4.543125,-0.3103125)(5.743125,-0.3103125)
\psline[linewidth=0.04cm,arrowsize=0.05291667cm 2.0,arrowlength=1.4,arrowinset=0.4]{<->}(2.543125,-0.3103125)(0.943125,-0.3103125)
\usefont{T1}{ptm}{m}{n}
\rput(3.52625,-0.6153125){\footnotesize $\geq 32$}
\usefont{T1}{ptm}{m}{n}
\rput(5.1267185,-0.6153125){\footnotesize $\geq 32$}
\usefont{T1}{ptm}{m}{n}
\rput(1.72625,-0.6153125){\footnotesize $\geq 32$}
\end{pspicture} 
}
    \caption{Illustration of the mapping $q(\hat{q},\ell)$. The parameter $\ell$
    ranges over the integers. The parameter $\hat{q}$ ranges over all
    possible values of $\hat{v}$. Observe that, for each fixed value of $\hat{q}$, 
    $q(\hat{q},\cdot)$ ranges over all possible values of $v$ as a
    function of $\ell$. Similarly, for each fixed value of $\ell$, 
    $q(\cdot,\ell)$ ranges over a subset of the possible values of $v$ as a
    function of $\hat{q}$. The distance between any two points is at
    least $32$.}
    \label{fig:decision}
\end{figure}
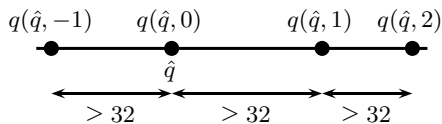

Set
\begin{equation*}
    p_{v\mid\hat{v}}(q\mid\hat{q}) \triangleq \Pp(v = q \mid \hat{v} = \hat{q}).
\end{equation*}
We will show that $H(v\mid \hat{v})$ is small by arguing that
$p_{v\mid\hat{v}}(q\mid\hat{q})$ is close to one for $q=\hat{q}$ and
decays exponentially quickly for $q\neq\hat{q}$. More precisely, define
a mapping $q(\hat{q},\ell)$, with $\hat{q}$ a possible value of
$\hat{v}$ and $\ell$ an integer, as follows. Set $q(\hat{q},0)
\triangleq \hat{q}$. If $\ell$ is a negative integer, set
$q(\hat{q},\ell)$ to be the $\ell$\/th closest possible value of $v$ to
the left of $\hat{q}$. If $\ell$ is a positive integer, set
$q(\hat{q},\ell)$ to be the $\ell$\/th closest possible value of $v$ to
the right of $\hat{q}$. This mapping is illustrated in
Fig.~\ref{fig:decision}. We will show that
$p_{v\mid\hat{v}}(q(\hat{q},\ell)\mid\hat{q})$ decays exponentially in
$\abs{\ell}$.

Rewrite $p_{v\mid\hat{v}}(q(\hat{q},\ell)\mid\hat{q})$ as
\begin{equation*}
    p_{v\mid\hat{v}}(q(\hat{q},\ell)\mid\hat{q})
    = \frac{\Pp(v = q(\hat{q},\ell)) \Pp(\hat{v} = \hat{q} \mid v = q(\hat{q},\ell))}
    {\Pp(\hat{v} = \hat{q})}.
\end{equation*}
Recall that, by Lemma~\ref{thm:decoding_gauss}, the distance between two
possible values of $v$ is at least $d\geq 32$. In order to decode to 
$\hat{v}=\hat{q}$ if the correct value of $v$ is $q(\hat{q},\ell)$, the
noise terms needs to have magnitude at least $16\abs{\ell}$. From this
observation, we can obtain an upper bound on $\Pp(\hat{v} = \hat{q} \mid
v = q(\hat{q},\ell))$.

As mentioned before, this analysis is based on the assumption that both
transmitters and receivers have access to $h_{mk}$. The analysis in
Appendix~\ref{sec:appendix_mismatch} shows that the only difference
under mismatched encoding and decoding, in which the transmitters and
receivers use $\max\{n_{mk}\}$-bit quantized channel gains
$\hat{h}_{mk}$ instead of $h_{mk}$, is a decrease in the minimum
constellation distance $d$. In particular, \eqref{eq:xc_error3_mm} in
Appendix~\ref{sec:appendix_mismatch}, shows that for $\abs{\ell} \geq 1$
\begin{align*}
    \Pp(\hat{v} = \hat{q} \mid v = q(\hat{q},\ell))
    & \leq \Pp(z_1 \geq \abs{\ell} (d-8)/2-3) \\
    & \leq \Pp(z_1 \geq 12\abs{\ell}-3) \nonumber\\
    & \leq \tfrac{1}{2} \exp\bigl(-(12\abs{\ell}-3)^2/2\bigr),
\end{align*}
where in the last inequality we have used the Chernoff bound on the
Q-function. Hence,
\begin{equation}
    \label{eq:xc_error3}
    p_{v\mid\hat{v}}(q(\hat{q},\ell)\mid\hat{q})
    \leq \frac{\Pp(v = q(\hat{q},\ell))} {2\Pp(\hat{v} = \hat{q})}
    \exp\bigl(-(12\abs{\ell}-3)^2/2\bigr),
\end{equation}
showing that $p_{v\mid\hat{v}}(q(\hat{q},\ell)\mid\hat{q})$ decays exponentially in
$\abs{\ell}$. 

We next argue that this exponential decay implies that $H(v\mid\hat{v})$
is small. We have
\begin{equation}
    \label{eq:hcond}
    H(v\mid\hat{v}) = \sum_{\hat{q}}
    \Pp(\hat{v}=\hat{q})H(v\mid\hat{v}=\hat{q}).
\end{equation}
Applying \cite[Theorem~9.7.1]{cover91}, 
\begin{align*}
    H(v\mid \hat{v} = \hat{q})
    & = -\sum_{q} p_{v\mid\hat{v}}(q\mid\hat{q})\log p_{v\mid\hat{v}}(q\mid\hat{q}) \nonumber\\
    & = -\sum_{\ell} p_{v\mid\hat{v}}(q(\hat{q},\ell)\mid\hat{q})
    \log p_{v\mid\hat{v}}(q(\hat{q},\ell)\mid\hat{q}) \nonumber\\
    & \leq \frac{1}{2}\log\bigg((2\pi e)
    \Big(\sum_\ell (2\abs{\ell}+1)^2p_{v\mid\hat{v}}(q(\hat{q},\ell)\mid\hat{q})
    -\frac{11}{12} \Big)
    \bigg) \nonumber\\
    & = \frac{1}{2}\log\bigg((2\pi e)
    \Big(4\sum_\ell (\ell^2+\abs{\ell})p_{v\mid\hat{v}}(q(\hat{q},\ell)\mid\hat{q})
    +\frac{1}{12} \Big)
    \bigg) \nonumber\\
    & \leq \frac{1}{2}\log(2\pi e)
    +\frac{1}{2}\log(e) \biggl( 4\sum_\ell (\ell^2+\abs{\ell})
    p_{v\mid\hat{v}}(q(\hat{q},\ell)\mid\hat{q}) - \frac{11}{12}\biggr) \\
    & = \frac{1}{2}\log(2\pi e^{1/12})
    +2\log(e) \sum_\ell (\ell^2+\abs{\ell})p_{v\mid\hat{v}}(q(\hat{q},\ell)\mid\hat{q}).
\end{align*}
Combined with~\eqref{eq:xc_error3} and \eqref{eq:hcond}, this implies
\begin{align}
    \label{eq:xc_rbound3}
    H(v\mid\hat{v})
    & \leq \frac{1}{2}\log(2\pi e^{1/12})
    +\log(e) \sum_\ell
    (\abs{\ell}^2+\abs{\ell})\exp\bigl(-(12\abs{\ell}-3)^2/2\bigr)
    \sum_{\hat{q}}\Pp(v=q(\hat{q},\ell)).
\end{align}

Now, since for every fixed value of $\ell$, $q(\cdot, \ell)$ takes each
possible value of $v$ at most once as a function of $\hat{q}$ (see
Fig.~\ref{fig:decision}), we have
\begin{equation*}
    \sum_{\hat{q}}\Pp(v=q(\hat{q},\ell)) 
    \leq \sum_{q}\Pp(v=q) 
    = 1.
\end{equation*}
Moreover,
\begin{align*}
    \sum_\ell (\ell^2+\abs{\ell})\exp\bigl(-(12\abs{\ell}-3)^2/2\bigr)
    \leq 2\sum_{\ell=1}^\infty (\ell^2+\ell)\exp\bigl(-(12\ell-3)^2/2\bigr)
    \leq 10^{-16}.
\end{align*}
Substituting this into~\eqref{eq:xc_rbound3} yields
\begin{equation*}
    H(v\mid\hat{v})
    \leq \frac{1}{2}\log(2\pi e^{1/12}) +10^{-16}\log(e)
    \leq 1.5.
\end{equation*}
Together with~\eqref{eq:1to1}, this proves~\eqref{eq:const}.\hfill\IEEEQED

\subsection{Upper Bound for the Gaussian X-Channel}
\label{sec:proofs_gaussian_upper}

This section proves the upper bound in Theorem~\ref{thm:arbitrary_gaussian}.  We
start with a lemma upper bounding various linear combinations of
achievable rates for the Gaussian X-channel.

\begin{lemma}
    \label{thm:upper-G}
    Any achievable rate tuple $(R_{11}, R_{12},R_{21},R_{22})$ for the
    Gaussian X-channel \eqref{eq:xc_arb} satisfies the following
    inequalities
    \begin{subequations}
        \label{eq:upper-eliminate-G}
        \begin{align}
            \label{eq:upper-eliminate-R21-G}
            R_{11}+R_{12}+R_{22} 
            & \leq \tfrac{1}{2}\log(1+2^{2n_{11}}h_{11}^2+2^{2n_{12}}h_{12}^2)
            +\tfrac{1}{2}\log\Big(1+\frac{2^{2n_{22}}h_{22}^2}
            {1+2^{2n_{12}}h_{12}^2}\Big),\\
            \label{eq:upper-eliminate-R12-G}
            R_{11}+R_{21}+R_{22} 
            & \leq \tfrac{1}{2}\log(1+2^{2n_{22}}h_{22}^2+2^{2n_{21}}h_{21}^2)
            +\tfrac{1}{2}\log\Big(1+\frac{2^{2n_{11}}h_{11}^2}
            {1+2^{2n_{21}}h_{21}^2}\Big),\\
            \label{eq:upper-eliminate-R22-G}
            R_{11}+R_{12}+R_{21} 
            & \leq \tfrac{1}{2}\log(1+2^{2n_{11}}h_{11}^2+2^{2n_{12}}h_{12}^2)
            +\tfrac{1}{2}\log\Big(1+\frac{2^{2n_{21}}h_{21}^2}
            {1+2^{2n_{11}}h_{11}^2}\Big),\\
            \label{eq:upper-eliminate-R11-G}
            R_{12}+R_{21}+R_{22} 
            & \leq \tfrac{1}{2}\log(1+2^{2n_{22}}h_{22}^2+2^{2n_{21}}h_{21}^2)
            +\tfrac{1}{2}\log\Big(1+\frac{2^{2n_{12}}h_{12}^2}
            {1+2^{2n_{22}}h_{22}^2}\Big),\\
            \label{eq:upper-bound-X2-G}
            R_{11}+R_{12}+R_{21}+R_{22} 
            & \leq \tfrac{1}{2}\log\Big(1+2^{2n_{12}}h_{12}^2+
            \frac{2^{2n_{11}}h_{11}^2}{1+2^{2n_{21}}h_{21}^2}\Big)
            +\tfrac{1}{2}\log\Big(1+2^{2n_{21}}h_{21}^2
            +\frac{2^{2n_{22}}h_{22}^2}{1+2^{2n_{12}}h_{12}^2}\Big), \\
            \label{eq:sum-rate-up-last-G}
            R_{11}+R_{12}+R_{21}+R_{22} 
            & \leq \tfrac{1}{2}\log\Big(1+2^{2n_{11}}h_{11}^2
            +\frac{2^{2n_{12}}h_{12}^2}{1+2^{2n_{22}}h_{22}^2}\Big)
            +\tfrac{1}{2}\log\Big(1+2^{2n_{22}}h_{22}^2
            +\frac{2^{2n_{21}}h_{21}^2}{1+2^{2n_{11}}h_{11}^2}\Big),\\
            \label{eq:upper-bound-IC1-G}
            2R_{11}+R_{12}+R_{21}+R_{22}&
            \leq \tfrac{1}{2}\log(1+2^{2n_{11}}h_{11}^2+2^{2n_{12}}h_{12}^2) 
            + \tfrac{1}{2}\log\Big(1+2^{2n_{21}}h_{21}^2
            +\frac{2^{2n_{22}}h_{22}^2}{1+2^{2n_{12}}h_{12}^2}\Big) 
            \nonumber\\ 
            & \quad{}+\tfrac{1}{2}\log\Big(1+\frac{2^{2n_{11}}h_{11}^2}
            {1+2^{2n_{21}}h_{21}^2}\Big), \\
            \label{eq:upper-bound-IC2-G}
            R_{11}+2R_{12}+R_{21}+R_{22}&
            \leq \tfrac{1}{2}\log(1+2^{2n_{12}}h_{12}^2+2^{2n_{11}}h_{11}^2) 
            + \tfrac{1}{2}\log\Big(1+2^{2n_{22}}h_{22}^2
            +\frac{2^{2n_{21}}h_{21}^2}{1+2^{2n_{11}}h_{11}^2}\Big) 
            \nonumber\\ 
            & \quad{}+\tfrac{1}{2}\log\Big(1+\frac{2^{2n_{12}}h_{12}^2}
            {1+2^{2n_{22}}h_{22}^2}\Big), \\
            \label{eq:upper-bound-IC3-G}
            R_{11}+R_{12}+2R_{21}+R_{22} 
            & \leq \tfrac{1}{2}\log(1+2^{2n_{21}}h_{21}^2+2^{2n_{22}}h_{22}^2) 
            + \tfrac{1}{2}\log\Big(1+2^{2n_{11}}h_{11}^2
            +\frac{2^{2n_{12}}h_{12}^2}{1+2^{2n_{22}}h_{22}^2}\Big) 
            \nonumber\\ 
            & \quad{}+\tfrac{1}{2}\log\Big(1+\frac{2^{2n_{21}}h_{21}^2}
            {1+2^{2n_{11}}h_{11}^2}\Big), \\
            \label{eq:upper-bound-IC4-G}
            R_{11}+R_{12}+R_{21}+2R_{22} 
            & \leq \tfrac{1}{2}\log(1+2^{2n_{22}}h_{22}^2+2^{2n_{21}}h_{21}^2) 
            + \tfrac{1}{2}\log\Big(1+2^{2n_{12}}h_{12}^2
            +\frac{2^{2n_{11}}h_{11}^2}{1+2^{2n_{21}}h_{21}^2}\Big) \nonumber\\ 
            & \quad{}+\tfrac{1}{2}\log\Big(1+\frac{2^{2n_{22}}h_{22}^2}
            {1+2^{2n_{12}}h_{12}^2}\Big). 
        \end{align}
    \end{subequations}
\end{lemma}

The proof of Lemma~\ref{thm:upper-G} is reported in
Appendix~\ref{sec:appendix_upper-G}. Inequalities
\eqref{eq:upper-eliminate-R21-G}--\eqref{eq:sum-rate-up-last-G} are from
\cite[Lemma~5.2, Theorem~5.3]{huang08}. Inequalities
\eqref{eq:upper-bound-IC1-G}--\eqref{eq:upper-bound-IC4-G} are novel.

We proceed with the proof of the upper bound in Theorem~\ref{thm:arbitrary_gaussian}
for the Gaussian X-channel.  Note that, for $n_{mk}\in\Zp$ and
$h_{mk}\in(1,2]$, 
\begin{align*}
    \tfrac{1}{2}\log\big(1+2^{2n_{11}}h_{11}^2+2^{2n_{12}}h_{12}^2\big)
    & \leq \tfrac{1}{2}\log\big(1+4\cdot 2^{2n_{11}}+4\cdot 2^{2n_{12}}\big) \\
    & \leq \tfrac{1}{2}\log\big(9\max\{1,2^{2n_{11}},2^{2n_{12}}\}\big) \\
    & = \max\{n_{11},n_{12}\}+\tfrac{1}{2}\log(9) \\
    \shortintertext{and}\\
    \tfrac{1}{2}\log\Big(1+\frac{2^{2n_{22}}h_{22}^2}{1+2^{2n_{12}}h_{12}^2}\Big)
    & \leq \tfrac{1}{2}\log\big(1+ 2^{2n_{22}-2n_{12}}h_{22}^2\big) \\
    & \leq \tfrac{1}{2}\log\big(5\max\{1, 2^{2n_{22}-2n_{12}}\} \big) \\
    & = (n_{22}-n_{12})^+ + \tfrac{1}{2}\log(5).
\end{align*}
Hence, \eqref{eq:upper-eliminate-R21-G} yields
\begin{equation*}
    R_{11}+R_{12}+R_{22} 
    \leq \max\{n_{11},n_{12}\}+(n_{22}-n_{12})^+ + \tfrac{1}{2}\log(5\cdot 9).
\end{equation*}
In a similar manner, we can upper bound the right-hand sides of all terms
in Lemma~\ref{thm:upper-G} by quantities depending only on $\bm{N}$. For
example, \eqref{eq:upper-bound-X2-G} yields
\begin{equation*}
    R_{11}+R_{12}+R_{21}+R_{22} 
    \leq \max\{ n_{12}, n_{11}-n_{21} \} +  \max\{ n_{21}, n_{22}-n_{12} \} 
    + \tfrac{1}{2}\log(9^2),
\end{equation*}
and \eqref{eq:upper-bound-IC1-G} yields
\begin{equation*}
    2R_{11}+R_{12}+R_{21}+R_{22}
    \leq \max\{n_{11},n_{12}\}+\max\{n_{21}, n_{22}-n_{12} \} 
    + (n_{11}-n_{21})^+ + \tfrac{1}{2}\log(5\cdot 9^2).
\end{equation*}

Comparing this to the upper bounds in Lemma~\ref{thm:upper} in
Section~\ref{sec:proofs_det_upper} for the lower-triangular
deterministic X-channel, we see that Lemma~\ref{thm:upper-G} for the
Gaussian X-channel is identical up to a constant gap. This highlights
again the close connection between the two models. Using the same
derivation as for the deterministic case, Lemma~\ref{thm:upper-G} can
thus be used to show that, under the assumption 
\begin{equation*}
    \min\{n_{11}, n_{22}\} \geq \max\{n_{12}, n_{21}\},
\end{equation*}
the sum capacity of the Gaussian X-channel satisfies
\begin{equation*}
    C(\bm{N}) \leq D(\bm{N}) + 4.
\end{equation*}
This concludes the proof of the upper bound.\hfill\IEEEQED

\section{Mathematical Foundations for Receiver Analysis}
\label{sec:foundations}

This section lays the mathematical groundwork for the analysis of the
decoders used in Sections~\ref{sec:proofs_det_lower} and
\ref{sec:proofs_gaussian_lower}. For the deterministic channel model,
decoding is successful if the various message subspaces are linearly
independent. Conditions for this linear independence to hold are
presented in Section~\ref{sec:foundations_det}. For the Gaussian case,
decoding is successful if the minimum distance between the different
messages as seen at the receivers is large. As we will see, this
problem can be reformulated as a number-theoretic problem. Conditions
for successful decoding in the Gaussian case are presented in
Section~\ref{sec:foundations_gaussian}.

\subsection{Decoding Conditions for the Deterministic Channel}
\label{sec:foundations_det}

We start by analyzing a ``generic'' receiver (i.e., the bit allocation
seen at either receiver one or two). To this end, we assume there are
two desired vectors $\bar{\bm{u}}_1$ and $\bar{\bm{u}}_2$ and one
interference vector $\bar{\bm{u}}_0$. The interference vector
$\bar{\bm{u}}_0$ consists of two signal vectors that are aligned and
can therefore be treated as a single vector. These three vectors are
multiplied by the lower-triangular channel matrices $\bar{\bm{G}}_1,
\bar{\bm{G}}_2,$ and $\bar{\bm{G}}_0$ created via the binary expansion
of the channel gains $g_1, g_2, g_0$ as before. 

\begin{figure}[htbp]
    \centering
    % Generated with LaTeXDraw 2.0.8
% Wed Nov 09 14:20:39 EST 2011
% \usepackage[usenames,dvipsnames]{pstricks}
% \usepackage{epsfig}
% \usepackage{pst-grad} % For gradients
% \usepackage{pst-plot} % For axes
\scalebox{1} % Change this value to rescale the drawing.
{
\begin{pspicture}(0,-1.9745313)(2.8421874,1.9445312)
\definecolor{color60b}{rgb}{0.8,0.8,0.8}
\psframe[linewidth=0.04,dimen=outer](1.323125,1.9345312)(0.683125,-1.4854687)
\usefont{T1}{ptm}{m}{n}
\rput(0.39171875,0.24921875){\footnotesize $n_1$}
\usefont{T1}{ptm}{m}{n}
\rput(0.96671873,-1.7704687){\footnotesize $\bar{\bm{u}}$}
\psline[linewidth=0.02cm,arrowsize=0.05291667cm 2.0,arrowlength=1.4,arrowinset=0.4]{<->}(0.603125,1.9345312)(0.603125,-1.4654688)
\psline[linewidth=0.02cm,arrowsize=0.05291667cm 2.0,arrowlength=1.4,arrowinset=0.4]{<->}(1.403125,1.9345312)(1.403125,-0.16546875)
\psline[linewidth=0.02cm,arrowsize=0.05291667cm 2.0,arrowlength=1.4,arrowinset=0.4]{<->}(1.403125,-0.86546874)(1.403125,-1.4654688)
\usefont{T1}{ptm}{m}{n}
\rput(1.7417188,-1.1507813){\footnotesize $n^+$}
\usefont{T1}{ptm}{m}{n}
\rput(2.1017187,0.89921874){\footnotesize $n_1-n^-$}
\usefont{T1}{ptm}{m}{n}
\rput(2.1317186,-0.50078124){\footnotesize $n^--n^+$}
\psline[linewidth=0.02cm,arrowsize=0.05291667cm 2.0,arrowlength=1.4,arrowinset=0.4]{<->}(1.403125,-0.16546875)(1.403125,-0.86546874)
\psframe[linewidth=0.04,dimen=outer,fillstyle=solid,fillcolor=color60b](1.323125,-0.14546876)(0.683125,-0.8854687)
\end{pspicture} 
}
    \caption{A vector $\bar{\bm{u}}$ in the set
    $\bar{\mc{U}}(n^-,n^+)$. White regions represent bits set to
    zero.}
    \label{fig:udef}
\end{figure}
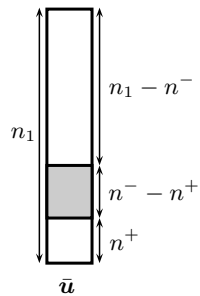

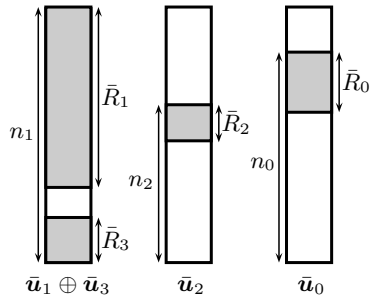
\begin{figure}[htbp]
    \centering
    % Generated with LaTeXDraw 2.0.8
% Tue Dec 20 17:04:08 EST 2011
% \usepackage[usenames,dvipsnames]{pstricks}
% \usepackage{epsfig}
% \usepackage{pst-grad} % For gradients
% \usepackage{pst-plot} % For axes
\scalebox{1} % Change this value to rescale the drawing.
{
\begin{pspicture}(0,-1.9845313)(7.3221874,1.9445312)
\definecolor{color151b}{rgb}{0.8,0.8,0.8}
\psframe[linewidth=0.04,dimen=outer](2.963125,1.9245312)(2.323125,-1.4954687)
\psframe[linewidth=0.04,dimen=outer,fillstyle=solid,fillcolor=color151b](2.963125,1.9445312)(2.323125,-0.49546874)
\psframe[linewidth=0.04,dimen=middle](4.543125,1.9245312)(3.943125,-1.4754688)
\psframe[linewidth=0.04,dimen=middle](6.143125,1.9245312)(5.543125,-1.4754688)
\psframe[linewidth=0.04,dimen=outer,fillstyle=solid,fillcolor=color151b](6.163125,1.3445313)(5.523125,0.50453126)
\usefont{T1}{ptm}{m}{n}
\rput(2.0317187,0.23921876){\footnotesize $n_1$}
\usefont{T1}{ptm}{m}{n}
\rput(3.6117187,-0.41078126){\footnotesize $n_2$}
\usefont{T1}{ptm}{m}{n}
\rput(2.6367188,-1.7804687){\footnotesize $\bar{\bm{u}}_{1}\oplus\bar{\bm{u}}_3$}
\usefont{T1}{ptm}{m}{n}
\rput(4.2767186,-1.7804687){\footnotesize $\bar{\bm{u}}_{2}$}
\usefont{T1}{ptm}{m}{n}
\rput(5.8767185,-1.7804687){\footnotesize $\bar{\bm{u}}_{0}$}
\psframe[linewidth=0.04,dimen=outer,fillstyle=solid,fillcolor=color151b](4.563125,0.64453125)(3.923125,0.12453125)
\psline[linewidth=0.02cm,arrowsize=0.05291667cm 2.0,arrowlength=1.4,arrowinset=0.4]{<->}(2.243125,1.9245312)(2.243125,-1.4754688)
\psline[linewidth=0.02cm,arrowsize=0.05291667cm 2.0,arrowlength=1.4,arrowinset=0.4]{<->}(3.843125,-1.4754688)(3.843125,0.62453127)
\psline[linewidth=0.02cm,arrowsize=0.05291667cm 2.0,arrowlength=1.4,arrowinset=0.4]{<->}(5.443125,-1.4754688)(5.443125,1.3245312)
\psline[linewidth=0.02cm,arrowsize=0.05291667cm 2.0,arrowlength=1.4,arrowinset=0.4]{<->}(3.043125,1.9245312)(3.043125,-0.47546875)
\psline[linewidth=0.02cm,arrowsize=0.05291667cm 2.0,arrowlength=1.4,arrowinset=0.4]{<->}(4.643125,0.62453127)(4.643125,0.12453125)
\psline[linewidth=0.02cm,arrowsize=0.05291667cm 2.0,arrowlength=1.4,arrowinset=0.4]{<->}(6.243125,1.3245312)(6.243125,0.52453125)
\psframe[linewidth=0.04,dimen=outer,fillstyle=solid,fillcolor=color151b](2.963125,-0.85546875)(2.323125,-1.4954687)
\psline[linewidth=0.02cm,arrowsize=0.05291667cm 2.0,arrowlength=1.4,arrowinset=0.4]{<->}(3.043125,-0.87546873)(3.043125,-1.4754688)
\usefont{T1}{ptm}{m}{n}
\rput(5.2317185,-0.16078125){\footnotesize $n_0$}
\usefont{T1}{ptm}{m}{n}
\rput(6.471719,0.93921876){\footnotesize $\bar{R}_0$}
\usefont{T1}{ptm}{m}{n}
\rput(3.2717187,-1.1607813){\footnotesize $\bar{R}_3$}
\usefont{T1}{ptm}{m}{n}
\rput(4.871719,0.38921875){\footnotesize $\bar{R}_2$}
\usefont{T1}{ptm}{m}{n}
\rput(3.2717187,0.7392188){\footnotesize $\bar{R}_1$}
\end{pspicture} 
}
    \caption{A generic receiver as analyzed in
    Lemma~\ref{thm:subspaces}. White regions correspond to zero
    bits; shaded regions carry information. Bits are labeled from 
    $1$ to $n_1$, starting from the top.}
    \label{fig:subspaces}
\end{figure}

We assume that certain components of the vectors $\bar{\bm{u}}_k$ are
set to zero. To formally capture this, we need to introduce some
notation. Let $n^-$ and $n^+$ be two nonnegative integers such that $n^-
\geq n^+$. Define
\begin{equation*}
    \bar{\mc{U}}(n^-, n^+)
    \triangleq \big\{\bar{\bm{u}}\in\{0,1\}^{n_1}:
    \bar{u}_i = 0 \ \forall 
    i\in\{1,\ldots, n_1-{n^-}\}\cup\{n_1-n^{+}+1,\ldots, n_1\}
    \big\}
\end{equation*}
as illustrated in Fig.~\ref{fig:udef}.
We consider vectors
$(\bar{\bm{u}}_0,\bar{\bm{u}}_1,\bar{\bm{u}}_2,\bar{\bm{u}}_3)$ in the
set 
\begin{align*}
    \bar{\mc{U}} & \triangleq
    \bar{\mc{U}}(n_0,n_0-\bar{R}_0)
    \times \bar{\mc{U}}(n_1,n_1-\bar{R}_1)
    \times \bar{\mc{U}}(n_2,n_2-\bar{R}_2)
    \times \bar{\mc{U}}(\bar{R}_3,0)
\end{align*}
with $n_1\geq n_0\geq n_2$, as is illustrated in Fig.~\ref{fig:subspaces}. 
Here, $\bar{\bm{u}}_1$ and $\bar{\bm{u}}_3$ are to be interpreted as the
common and private portions of the desired signal transmitted over the
direct link; $\bar{\bm{u}}_2$ is to be interpreted as the desired
signal transmitted over the cross link; and $\bar{\bm{u}}_0$ is to be
interpreted as the aligned interference.

The next lemma states that the subspaces spanned by the corresponding
columns of $\bar{\bm{G}}_k$ are linearly independent for most channel gains
$(g_0,g_1,g_2)$.

\begin{lemma}
    \label{thm:subspaces}
    Let $n_0, n_1, n_2\in\Zp$ such that $n_1 \geq n_0 \geq n_2$, and let
    $\bar{R}_0, \bar{R}_1, \bar{R}_2, \bar{R}_3\in\Zp$. Define the event
    \begin{equation*}
        B(\bar{\bm{u}}_0, \bar{\bm{u}}_1, \bar{\bm{u}}_2, \bar{\bm{u}}_3) 
        \triangleq \big\{(g_0, g_1, g_2)\in(1,2]^3:
        \bar{\bm{G}}_0\bar{\bm{u}}_0
        \oplus\bar{\bm{G}}_1(\bar{\bm{u}}_1\oplus\bar{\bm{u}}_3)
        \oplus\bar{\bm{G}}_2\bar{\bm{u}}_2 = \bm{0}
        \big\},
    \end{equation*}
    and set
    \begin{equation*}
        B \triangleq
        \bigcup_{
        (\bar{\bm{u}}_0, \bar{\bm{u}}_1, \bar{\bm{u}}_2, \bar{\bm{u}}_3)\in
        \bar{\mc{U}}\setminus \{(\bm{0}, \bm{0}, \bm{0}, \bm{0})\}
        } 
        B(\bar{\bm{u}}_0, \bar{\bm{u}}_1, \bar{\bm{u}}_2, \bar{\bm{u}}_3).
    \end{equation*}
    For any $\delta \in(0,1]$ satisfying
    \begin{align*}
        \bar{R}_1+\bar{R}_0+\bar{R}_2+\bar{R}_3 & \leq n_1-\log(16/\delta), \\
        \bar{R}_0+\bar{R}_2+\bar{R}_3 & \leq n_0-\log(16/\delta), \\
        \bar{R}_2+\bar{R}_3 & \leq n_2,
    \end{align*}
    we have
    \begin{equation*}
        \mu(B) \leq \delta.
    \end{equation*}
\end{lemma}

Observe that $B$ is the set of channel gains $g_0,g_1,g_2$ such that the
corresponding subspaces spanned by the selected columns of
$\bar{\bm{G}}_0, \bar{\bm{G}}_1, \bar{\bm{G}}_2$ are linearly dependent.
In other words, $B$ is the set of channel gains resulting in decoding
error. Thus the lemma states that if the rates $\bar{R}_k$ satisfy
certain conditions, then the subspaces under consideration are linearly
independent with high probability, and hence decoding is successful.

The condition on the rates in Lemma~\ref{thm:subspaces} can be
interpreted as follows. Let $n$ be some natural number. Since the
matrices $\bar{\bm{G}}_k$ are lower triangular, the subspaces spanned by
the last $n$ columns of $\bar{\bm{G}}_k$ are the same for all
$k\in\{0,1,2\}$.  Thus, a \emph{necessary} condition for the linear
independence of the three subspaces is that the total number of possible
nonzero components of $\bar{u}_{ki}$ with $i\geq n_1-n+1$ and
$k\in\{0,1,2,3\}$ is at most $n$ for every $n\in\{1,...,n_1\}$.  By the
structure of the set $\bar{\mc{U}}$, this condition can be verified by
considering only three values of $n$, namely $n\in\{n_0, n_1, n_2\}$
(see Fig.~\ref{fig:subspaces}). Thus, a necessary condition for the
linear independence of the subspaces is
\begin{align*}
    \bar{R}_1+\bar{R}_0+\bar{R}_2+\bar{R}_3 & \leq n_1, \\
    (\bar{R}_1-(n_1-n_0))^++\bar{R}_0+\bar{R}_2+\bar{R}_3 & \leq n_0, \\
    (\bar{R}_1-(n_1-n_2))^++(\bar{R}_0-(n_0-n_2))^++\bar{R}_2+\bar{R}_3 & \leq n_2.
\end{align*}
After some algebra these three conditions can be rewritten equivalently
as
\begin{align*}
    \bar{R}_1+\bar{R}_0+\bar{R}_2+\bar{R}_3 & \leq n_1, \\
    \bar{R}_0+\bar{R}_2+\bar{R}_3 & \leq n_0, \\
    \bar{R}_2+\bar{R}_3 & \leq n_2.
\end{align*}
Thus, Lemma~\ref{thm:subspaces} shows that, up to the constant
$\log(16/\delta)$ and for most channel gains $(g_0, g_1, g_2)$, these
\emph{necessary} conditions are also \emph{sufficient} for the linear
independence of the subspaces.

Before we provide the proof of Lemma~\ref{thm:subspaces}, we show how it
can be used to prove Lemma~\ref{thm:decoding_det} in
Section~\ref{sec:proofs_det_lower}. 

\begin{IEEEproof}[Proof of Lemma~\ref{thm:decoding_det}]
    We start by reformulating the conditions in Lemma~\ref{thm:subspaces}
    for each receiver. Consider first receiver one in
    Lemma~\ref{thm:decoding_det}. From Fig.~\ref{fig:Rx1-deterministic} in
    Section~\ref{sec:main_arbitrary}, we see that the corresponding message
    rates in Lemma~\ref{thm:subspaces} are given by
    \begin{align*}
        \bar{R}_0 & \triangleq \max\{\bar{R}_{21},\bar{R}_{22}^\sC\}, \\
        \bar{R}_1 & \triangleq \bar{R}_{11}^\sC, \\
        \bar{R}_2 & \triangleq \bar{R}_{12}, \\
        \bar{R}_3 & \triangleq \bar{R}_{11}^\sP.
    \end{align*}
    The choice of the bit levels $n_k$ in Lemma~\ref{thm:subspaces} depends
    on the values of $\bar{R}_0$ and $\bar{R}_2$. If $\bar{R}_0, \bar{R}_2 >
    0$, we need to set
    \begin{align*}
        n_0 & \triangleq n_{12}, \\
        n_1 & \triangleq n_{11}, \\
        n_2 & \triangleq n_{12}+n_{21}-n_{22},
    \end{align*}
    see again Fig.~\ref{fig:Rx1-deterministic}.  

    The conditions in Lemma~\ref{thm:subspaces} (with $\delta$ replaced by
    $\delta/2$ to guarantee that the outage event at each receiver has
    measure at most $\delta/2$) are then that
    \begin{subequations}
        \label{eq:decoding1_proof}
        \begin{align}
            \label{eq:decoding1a_proof}
            \bar{R}_{11}^{\sC} +\max\{\bar{R}_{21}, \bar{R}_{22}^{\sC}\} +\bar{R}_{12} +\bar{R}_{11}^{\sP}
            & \leq n_{11}-\log(32/\delta), \\
            \label{eq:decoding1b_proof}
            \max\{\bar{R}_{21}, \bar{R}_{22}^{\sC}\} +\bar{R}_{12} +\bar{R}_{11}^{\sP}
            & \leq n_{12}-\log(32/\delta), \\
            \label{eq:decoding1c_proof}
            \bar{R}_{12}+\bar{R}_{11}^{\sP}
            & \leq n_{12}+n_{21}-n_{22}.
        \end{align}
    \end{subequations}
    If $\bar{R}_2=0$, then the second column in
    Fig.~\ref{fig:subspaces} is empty, and hence the third condition in
    Lemma~\ref{thm:subspaces} does not need to be verified. Formally, note that in
    this case the value of $n_2$ is irrelevant to the decoding process. We
    may hence assume without loss of generality that $n_2$ is equal to $n_0$
    (thus still satisfying $n_0\geq n_2$). As a consequence, only conditions
    \eqref{eq:decoding1a_proof} and \eqref{eq:decoding1b_proof} need to be
    checked. If $\bar{R}_0=0$, then the value of $n_0$ is 
    irrelevant to the decoding process, and we can assume it to be equal
    to $n_1$ (thus still satisfying $n_1 \geq n_0$). As a consequence,
    only conditions \eqref{eq:decoding1a_proof} and
    \eqref{eq:decoding1c_proof} need to be checked.

    The decoding conditions for receiver two follow by symmetry.

    Denote by $B_1\subseteq\R^3$ the collection of triples $(g_{10}, g_{11},
    g_{12})$ such that decoding fails at receiver one.  Similarly, define
    $B_2\subseteq\R^3$ with respect to receiver two.  Finally, let $B\subset
    \R^6$ be the union of $B_1$ and $B_2$.  If the two sets of decoding
    conditions are satisfied, then Lemma~\ref{thm:subspaces} shows that
    \begin{equation*}
        \mu_3(B_m) \leq \delta/2
    \end{equation*}
    for $m\in\{1,2\}$, where here and in the following we use the notation
    $\mu_d$ to emphasize the Lebesgue measure is computed in $\R^d$. Then 
    \begin{align*}
        \mu_6(B) 
        & \leq \mu_3(B_1)\mu_3((1,2]^3)+\mu_3(B_2)\mu_3((1,2]^3) \\
        & \leq \delta,
    \end{align*}
    i.e., the collection of channel gains $(g_{mk})\in\R^{2\times 3}$ for
    which decoding fails is small. This concludes the proof of
    Lemma~\ref{thm:decoding_det}.
\end{IEEEproof}

It remains to prove Lemma~\ref{thm:subspaces}.

\begin{IEEEproof}[Proof of Lemma~\ref{thm:subspaces}]
    We start with a few preliminary observations. Note that, by the
    assumptions on $\bar{R}_k$, 
    \begin{align*}
        \bar{R}_1+\bar{R}_3 & \leq n_1, \\
        \bar{R}_0+\bar{R}_3 & \leq n_0, \\
        \bar{R}_2+\bar{R}_3 & \leq n_2,
    \end{align*}
    which implies that 
    \begin{equation}
        \label{eq:r3bound}
        \max\big\{\bar{R}_1, n_1-n_0+\bar{R}_0, n_1-n_2+\bar{R}_2\big\} \leq n_1-\bar{R}_3.
    \end{equation}
    From Fig.~\ref{fig:subspaces}, we see that this guarantees that if
    $\bar{\bm{u}}_k\neq 0$, then 
    \begin{equation}
        \label{eq:r3bound2}
        n(\bar{\bm{u}}_k) \leq n_1-\bar{R}_3
    \end{equation}
    for $k\in\{0,1,2\}$, where for a binary vector $\bar{\bm{u}}$ we
    use the notation $n(\bar{\bm{u}})$ to denote the smallest index
    $i$ such that $\bar{u}_i = 1$ with the convention that $n(\bm{0}) =
    +\infty$. Moreover, we see from the same figure that
    $n(\bar{\bm{u}_3}) > n_1-\bar{R}_3$.
    
    We now remove the dependence of $B$ on the private signal
    $\bar{\bm{u}}_3$. Since $\bar{\bm{G}}_k$ is lower triangular with
    unit diagonal (so that bits are only shifted downwards), we have
    $n(\bar{\bm{G}}_k\bar{\bm{u}}_k)=n(\bar{\bm{u}}_k)$.  
    Hence,
    \begin{equation*}
        \bar{\bm{G}}_0\bar{\bm{u}}_0
        \oplus\bar{\bm{G}}_1(\bar{\bm{u}}_1\oplus\bar{\bm{u}}_3)
        \oplus\bar{\bm{G}}_2\bar{\bm{u}}_2 = \bm{0} 
    \end{equation*}
    can hold only if
    \begin{align*}
        n\bigl(\bar{\bm{G}}_0\bar{\bm{u}}_0
        \oplus\bar{\bm{G}}_1\bar{\bm{u}}_1
        \oplus\bar{\bm{G}}_2\bar{\bm{u}}_2\bigr)
        & = n(\bar{\bm{G}_1}\bar{\bm{u}}_3) \\
        & = n(\bar{\bm{u}}_3) \\
        & > n_1-\bar{R}_3,
    \end{align*}
    where we have used that $n(\bar{\bm{u}}_3) > n_1-\bar{R}_3$.
    Furthermore, we have for $\bar{\bm{u}}_3\neq 0$ that
    \begin{equation*}
        \bar{\bm{G}}_0\bar{\bm{u}}_0
        \oplus\bar{\bm{G}}_1(\bar{\bm{u}}_1\oplus\bar{\bm{u}}_3)
        \oplus\bar{\bm{G}}_2\bar{\bm{u}}_2
        = \bm{0}
    \end{equation*}
    can hold only if $(\bar{\bm{u}}_0,\bar{\bm{u}}_1,\bar{\bm{u}}_2)\neq
    (\bm{0},\bm{0},\bm{0})$. 
    
    Defining the sets
    \begin{equation*}
        B^{\prime}(\bar{\bm{u}}_0, \bar{\bm{u}}_1, \bar{\bm{u}}_2) 
        \triangleq \big\{(g_0, g_1, g_2)\in(1,2]^3:
        n\big(
        \bar{\bm{G}}_0\bar{\bm{u}}_0
        \oplus\bar{\bm{G}}_1\bar{\bm{u}}_1
        \oplus\bar{\bm{G}}_2\bar{\bm{u}}_2
        \big) > n_1-\bar{R}_3
        \big\}
    \end{equation*}
    and
    \begin{align*}
        \bar{\mc{U}}^\prime 
        & \triangleq \bar{\mc{U}}(n_0,n_0-\bar{R}_0) 
        \times \bar{\mc{U}}(n_1,n_1-\bar{R}_1)
        \times \bar{\mc{U}}(n_2,n_2-\bar{R}_2),
    \end{align*}
    we hence have
    \begin{equation*}
        B \subseteq
        B^\prime 
        \triangleq \bigcup_{(\bar{\bm{u}}_0,\bar{\bm{u}}_1,\bar{\bm{u}}_2)
        \in\bar{\mc{U}}^\prime\setminus\{(\bm{0},\bm{0},\bm{0})\}}
        B^{\prime}(\bar{\bm{u}}_0, \bar{\bm{u}}_1, \bar{\bm{u}}_2).
    \end{equation*}
    We can then upper bound $\mu(B)$ using the union bound
    \begin{equation}
        \label{eq:mu3b_det}
        \mu(B)
        \leq
        \mu(B^\prime)
        \leq
        \sum_{
        \substack{
        (\bar{\bm{u}}_0, \bar{\bm{u}}_1, \bar{\bm{u}}_2)
        \in \bar{\mc{U}}^\prime\setminus\{(\bm{0}, \bm{0}, \bm{0})\}
        }
        } 
        \mu\big(B^\prime(\bar{\bm{u}}_0, \bar{\bm{u}}_1, \bar{\bm{u}}_2)\big).
    \end{equation}
    Observe that the right-hand side does not depend on the private
    signal $\bar{\bm{u}}_3$. We continue by analyzing each term in the
    summation on the right-hand side of~\eqref{eq:mu3b_det} separately. 

    Since we are integrating with respect to Lebesgue measure over $(g_0,
    g_1, g_2)\in(1,2]^3$, we can equivalently assume that $g_{0}, g_{1},
    g_{2}$ are independent and uniformly distributed over $(1,2]$. The bits
    in the binary expansion $([g_{k}]_i)_{i=-\infty}^{\infty}$ of these
    numbers are then binary random variables with the following properties.
    $[g_{k}]_i=0$ for $i \leq -1$, $[g_{k}]_0=1$, and
    $([g_{k}]_i)_{i=\infty}^{1}$ are \iid $\Bernoulli(1/2)$ (see, e.g.,
    \cite[Exercise~1.4.20]{durret04}). The lower-triangular Toeplitz matrix
    $\bar{\bm{G}}_k$ is then constructed from these binary random variables.
    Note that this implies that the three matrices $\bar{\bm{G}}_0,
    \bar{\bm{G}}_1, \bar{\bm{G}}_2$ are independent and identically
    distributed.

    Fix a binary vector $\bar{\bm{u}}$ and consider the product
    $\bar{\bm{G}}\bar{\bm{u}}$ for some $\bar{\bm{G}}=\bar{\bm{G}}_k$,
    $\bar{\bm{u}}=\bar{\bm{u}}_k$, and with addition again over $\Z_2$.
    We now describe the distribution of $\bar{\bm{G}}\bar{\bm{u}}$.
    Since $\bar{\bm{G}}$ is lower triangular with unit diagonal,
    $(\bar{\bm{G}}\bm{u})_i = 0$ whenever $1 \leq i < n(\bar{\bm{u}})$,
    and $(\bar{\bm{G}}\bar{\bm{u}})_{n(\bar{\bm{u}})}=1$.  Moreover, the
    components $(\bar{\bm{G}}\bar{\bm{u}})_i$ for $n(\bar{\bm{u}}) < i
    \leq n_1$ are \iid $\Bernoulli(1/2)$. 

    Assume first that 
    \begin{equation}
        \label{eq:nuordering}
        n(\bar{\bm{u}}_0) 
        \leq n(\bar{\bm{u}}_1) 
        \leq n(\bar{\bm{u}}_2) 
        < \infty.
    \end{equation}
    The summand in \eqref{eq:mu3b_det} can be written as
    \begin{align}
        \label{eq:summand}
        \mu\big(B^\prime(\bar{\bm{u}}_0, \bar{\bm{u}}_1,
        \bar{\bm{u}}_2)\big) = \sum_{\bm{b}_3: n(\bm{b}_3)>
        n_1-\bar{R}_3} \sum_{\bm{b}_1} \sum_{\bm{b}_2}
        \Pp(\bar{\bm{G}}_0\bar{\bm{u}}_0=
        \bm{b}_1\oplus\bm{b}_2\oplus\bm{b}_3)
        \Pp(\bar{\bm{G}}_1\bar{\bm{u}}_1=\bm{b}_1)
        \Pp(\bar{\bm{G}}_2\bar{\bm{u}}_2=\bm{b}_2),
    \end{align}
    where the probabilities are computed with respect to the random
    matrices $\bar{\bm{G}}_k$.  Using that
    $n(\bar{\bm{G}}_k\bar{\bm{u}}_k)=n(\bar{\bm{u}}_k)$, the three
    factors inside the summation are nonzero only if
    \begin{align*}
        n(\bm{b}_1\oplus\bm{b}_2\oplus\bm{b}_3) & = n(\bar{\bm{u}}_0), \\
        n(\bm{b}_1) & = n(\bar{\bm{u}}_1), \\
        n(\bm{b}_2) & = n(\bar{\bm{u}}_2).
    \end{align*}
    From this, we obtain that
    \begin{align*}
        n(\bar{\bm{u}}_0)
        & = n(\bm{b}_1\oplus\bm{b}_2\oplus\bm{b}_3) \\
        & \geq \min\{ n(\bm{b}_1), n(\bm{b}_2), n(\bm{b}_3) \} \\
        & \geq \min\{ n(\bar{\bm{u}}_1), n(\bar{\bm{u}}_2), n_1-\bar{R}_3 \} \\
        & = n(\bar{\bm{u}}_1),
    \end{align*}
    where for the last equality we have used \eqref{eq:nuordering} and
    that $n_1-\bar{R}_3$ is  larger than $n(\bar{\bm{u}}_1)$ 
    by \eqref{eq:r3bound2}. Since $n(\bar{\bm{u}}_0)\leq
    n(\bar{\bm{u}}_1)$ by \eqref{eq:nuordering}, this shows that
    $n(\bm{b}_1\oplus\bm{b}_2\oplus\bm{b}_3) = n(\bar{\bm{u}}_0)$ can
    hold only if $n(\bar{\bm{u}}_0) = n(\bar{\bm{u}}_1)$. 
    
    If these conditions on the $\bar{\bm{u}}_k$ and $\bm{b}_k$ are
    satisfied, then 
    \begin{equation*}
        \Pp(\bar{\bm{G}}_0\bar{\bm{u}}_0= \bm{b}_1\oplus\bm{b}_2\oplus\bm{b}_3)
        \Pp(\bar{\bm{G}}_1\bar{\bm{u}}_1=\bm{b}_1)
        \Pp(\bar{\bm{G}}_2\bar{\bm{u}}_2=\bm{b}_2)
        = 2^{-(n_1-n(\bar{\bm{u}}_0))-(n_1-n(\bar{\bm{u}}_1))-(n_1-n(\bar{\bm{u}}_2))}.
    \end{equation*}
    Substituting this into \eqref{eq:summand} shows that
    \begin{equation*}
        \mu\big(B^\prime(\bar{\bm{u}}_0, \bar{\bm{u}}_1, \bar{\bm{u}}_2)\big)
        \leq 2^{\bar{R}_3-n_1+n(\bar{\bm{u}}_0)}
    \end{equation*}
    whenever $n(\bar{\bm{u}}_0) = n(\bar{\bm{u}}_1)$, and 
    \begin{equation*}
        \mu\big(B^\prime(\bar{\bm{u}}_0, \bar{\bm{u}}_1, \bar{\bm{u}}_2)\big)
        = 0
    \end{equation*}
    otherwise.

    Assume more generally that $(\bar{\bm{u}}_1, \bar{\bm{u}}_2,
    \bar{\bm{u}}_3) \neq (\bm{0}, \bm{0}, \bm{0})$.
    Then a similar argument shows that
    \begin{equation}
        \label{eq:mu3b_det2}
        \mu\big(B^\prime(\bar{\bm{u}}_0, \bar{\bm{u}}_1, \bar{\bm{u}}_2)\big)
        \leq 2^{\bar{R}_3-n_1+\min_{k}n(\bar{\bm{u}}_k)}
    \end{equation}
    whenever there are two distinct indices $k, k'$ achieving the
    minimum $\min_k n(\bm{u}_k)$, and 
    \begin{equation}
        \label{eq:mu3b_det3}
        \mu\big(B^\prime(\bar{\bm{u}}_0, \bar{\bm{u}}_1, \bar{\bm{u}}_2)\big)
        = 0
    \end{equation}
    otherwise. In particular, the set 
    $B^\prime(\bar{\bm{u}}_0, \bar{\bm{u}}_1, \bar{\bm{u}}_2)$ has measure zero
    whenever at least two of the $\bar{\bm{u}}_k$ are equal to zero.

    Setting
    \begin{equation*}
        \bar{\mc{U}}^\prime(n^-, n^+)
        \triangleq \bar{\mc{U}}(n^-,n^+)\setminus\{\bm{0}\},
    \end{equation*}
    we can then rewrite \eqref{eq:mu3b_det} as
    \begin{align*}
        \mu(B^\prime)
        & \leq 
        \sum_{\bar{\bm{u}}_0\in\bar{\mc{U}}^\prime(n_0, n_0-\bar{R}_0)}
        \sum_{\bar{\bm{u}}_1\in\bar{\mc{U}}^\prime(n_1, n_1-\bar{R}_1)}
        \sum_{\bar{\bm{u}}_2\in\bar{\mc{U}}^\prime(n_2, n_2-\bar{R}_2)}
        \mu\big(B^\prime(\bar{\bm{u}}_0, \bar{\bm{u}}_1, \bar{\bm{u}}_2)\big) \\
        & \quad 
        + \sum_{\bar{\bm{u}}_1\in\bar{\mc{U}}^\prime(n_1, n_1-\bar{R}_1)}
        \sum_{\bar{\bm{u}}_2\in\bar{\mc{U}}^\prime(n_2, n_2-\bar{R}_2)}
        \mu\big(B^\prime(\bm{0}, \bar{\bm{u}}_1, \bar{\bm{u}}_2)\big) \\
        & \quad + 
        \sum_{\bar{\bm{u}}_0\in\bar{\mc{U}}^\prime(n_0, n_0-\bar{R}_0)}
        \sum_{\bar{\bm{u}}_2\in\bar{\mc{U}}^\prime(n_2, n_2-\bar{R}_2)}
        \mu\big(B^\prime(\bar{\bm{u}}_0, \bm{0}, \bar{\bm{u}}_2)\big) \\
        & \quad +
        \sum_{\bar{\bm{u}}_0\in\bar{\mc{U}}^\prime(n_0, n_0-\bar{R}_0)}
        \sum_{\bar{\bm{u}}_1\in\bar{\mc{U}}^\prime(n_1, n_1-\bar{R}_1)}
        \mu\big(B^\prime(\bar{\bm{u}}_0, \bar{\bm{u}}_1, \bm{0})\big).
    \end{align*}
    By \eqref{eq:mu3b_det3}, the set $B^\prime$ has measure zero
    whenever there is only a single minimizing $n(\bar{\bm{u}}_k)$. Together
    with the assumption $n_1\geq n_0\geq n_2$, this shows that we can
    restrict the lower boundaries of the sets $\bar{\mc{U}}^\prime$ in the
    various sums. For example
    \begin{align*}
        \sum_{\bar{\bm{u}}_0\in\bar{\mc{U}}^\prime(n_0, n_0-\bar{R}_0)} &
        \sum_{\bar{\bm{u}}_1\in\bar{\mc{U}}^\prime(n_1, n_1-\bar{R}_1)}
        \sum_{\bar{\bm{u}}_2\in\bar{\mc{U}}^\prime(n_2, n_2-\bar{R}_2)}
        \mu\big(B^\prime(\bar{\bm{u}}_0, \bar{\bm{u}}_1, \bar{\bm{u}}_2)\big) \\
        & = \sum_{\bar{\bm{u}}_0\in\bar{\mc{U}}^\prime(n_0, n_0-\bar{R}_0)}
        \sum_{\bar{\bm{u}}_1\in\bar{\mc{U}}^\prime(n_0, n_1-\bar{R}_1)}
        \sum_{\bar{\bm{u}}_2\in\bar{\mc{U}}^\prime(n_2, n_2-\bar{R}_2)}
        \mu\big(B^\prime(\bar{\bm{u}}_0, \bar{\bm{u}}_1, \bar{\bm{u}}_2)\big),
    \end{align*}
    where we have changed $\bar{\mc{U}}^\prime(n_1, n_1-\bar{R}_1)$ to
    $\bar{\mc{U}}^\prime(n_0, n_1-\bar{R}_1)$, and similarly for the other three
    summations.  Together with \eqref{eq:mu3b_det2} this yields that
    \begin{align*}
        \mu(B^\prime)
        & \leq 
        \sum_{\bar{\bm{u}}_0\in\bar{\mc{U}}^\prime(n_0, n_0-\bar{R}_0)}
        \sum_{\bar{\bm{u}}_1\in\bar{\mc{U}}^\prime(n_0, n_1-\bar{R}_1)}
        \sum_{\bar{\bm{u}}_2\in\bar{\mc{U}}^\prime(n_2, n_2-\bar{R}_2)}
        2^{\bar{R}_3-n_1+\min_{k}n(\bar{\bm{u}}_k)} \\
        & \quad 
        + \sum_{\bar{\bm{u}}_1\in\bar{\mc{U}}^\prime(n_2, n_1-\bar{R}_1)}
        \sum_{\bar{\bm{u}}_2\in\bar{\mc{U}}^\prime(n_2, n_2-\bar{R}_2)}
        2^{\bar{R}_3-n_1+\min_{k}n(\bar{\bm{u}}_k)} \\
        & \quad + 
        \sum_{\bar{\bm{u}}_0\in\bar{\mc{U}}^\prime(n_2, n_0-\bar{R}_0)}
        \sum_{\bar{\bm{u}}_2\in\bar{\mc{U}}^\prime(n_2, n_2-\bar{R}_2)}
        2^{\bar{R}_3-n_1+\min_{k}n(\bar{\bm{u}}_k)} \\
        & \quad +
        \sum_{\bar{\bm{u}}_0\in\bar{\mc{U}}^\prime(n_0, n_0-\bar{R}_0)}
        \sum_{\bar{\bm{u}}_1\in\bar{\mc{U}}^\prime(n_0, n_1-\bar{R}_1)}
        2^{\bar{R}_3-n_1+\min_{k}n(\bar{\bm{u}}_k)}.
    \end{align*}
    We consider each of the four terms in turn.

    For the first term, we have
    \begin{align*}
        \sum_{\bar{\bm{u}}_0\in\bar{\mc{U}}^\prime(n_0, n_0-\bar{R}_0)}
        &
        \sum_{\bar{\bm{u}}_1\in\bar{\mc{U}}^\prime(n_0, n_1-\bar{R}_1)}
        \sum_{\bar{\bm{u}}_2\in\bar{\mc{U}}^\prime(n_2, n_2-\bar{R}_2)}
        2^{\bar{R}_3-n_1+\min_{k}n(\bar{\bm{u}}_k)} \\
        & = \sum_{i=n_1-n_0+1}^{n_1}
        \sum_{\bar{\bm{u}}_0\in\bar{\mc{U}}^\prime(n_0, n_0-\bar{R}_0)}
        \sum_{\bar{\bm{u}}_1\in\bar{\mc{U}}^\prime(n_0, n_1-\bar{R}_1)}
        \sum_{\bar{\bm{u}}_2\in\bar{\mc{U}}^\prime(n_2, n_2-\bar{R}_2)}
        2^{\bar{R}_3-n_1+i}
        \ind_{\{\min_k n(\bar{\bm{u}}_k) = i\}} \\
        & \leq \sum_{i=n_1-n_0+1}^{n_1}
        \sum_{\bar{\bm{u}}_0\in\bar{\mc{U}}^\prime(n_1-i+1, n_0-\bar{R}_0)}
        \sum_{\bar{\bm{u}}_1\in\bar{\mc{U}}^\prime(n_1-i+1, n_1-\bar{R}_1)}
        \sum_{\bar{\bm{u}}_2\in\bar{\mc{U}}^\prime(n_2, n_2-\bar{R}_2)}
        2^{\bar{R}_3-n_1+i}.
    \end{align*}
    Using that
    \begin{equation*}
        \card{\bar{\mc{U}}^\prime(n^-, n^+)} 
        \leq 2^{n^--n^+},
    \end{equation*}
    the right-hand side can be further upper bounded by
    \begin{align*}
        \sum_{i=n_1-n_0+1}^{n_1} & 2^{n_1-n_0+\bar{R}_0-i+1}
        \cdot 2^{\bar{R}_1-i+1}
        \cdot 2^{\bar{R}_2} 
        \cdot 2^{\bar{R}_3-n_1+i} \\
        & = 2^{\bar{R}_0+\bar{R}_1+\bar{R}_2+\bar{R}_3-n_0+2} \sum_{i=n_1-n_0+1}^{n_1} 2^{-i} \\
        & \leq 2^{\bar{R}_0+\bar{R}_1+\bar{R}_2+\bar{R}_3-n_1+2}.
    \end{align*}

    We can upper bound the remaining three terms in a similar fashion,
    yielding
    \begin{align*}
        \mu(B^\prime)
        & \leq
        2^{\bar{R}_3+2}\big(
        2^{\bar{R}_0+ \bar{R}_1+\bar{R}_2-n_1}
        + 2^{\bar{R}_1+\bar{R}_2-n_1}
        + 2^{\bar{R}_0+\bar{R}_2-n_0 }
        + 2^{\bar{R}_0+\bar{R}_1-n_1}\big) \\
        & \leq
        16\cdot
        2^{\bar{R}_3+\max\{\bar{R}_0+\bar{R}_1+\bar{R}_2-n_1, 
        \bar{R}_0+\bar{R}_2-n_0\}}.
    \end{align*}
    This shows that if
    \begin{align*}
        \bar{R}_1+\bar{R}_0+\bar{R}_2+\bar{R}_3 & \leq n_1-\log(16/\delta), \\
        \bar{R}_0+\bar{R}_2+\bar{R}_3 & \leq n_0-\log(16/\delta), \\
        \shortintertext{and (in order to guarantee \eqref{eq:r3bound}) if } \\
        \bar{R}_2+\bar{R}_3 & \leq n_2,
    \end{align*}
    then 
    \begin{equation*}
        \mu(B) \leq \mu(B^\prime) \leq \delta,
    \end{equation*}
    completing the proof of the lemma.
\end{IEEEproof}

\subsection{Decoding Conditions for the Gaussian Channel}
\label{sec:foundations_gaussian}

In this section, we analyze a ``generic'' receiver for the Gaussian
case. To this end, we prove a variation of a well-known result from
Diophantine approximation called Groshev's theorem (see, e.g.,
\cite[Theorem 1.12]{sprindzuk79}).

Define
\begin{equation*}
    \mc{U}(n^-, n^+)
    \triangleq \big\{u\in[-1,1]:
    [u]_i = 0 \ \forall 
    i\in\{1,\ldots, n_1-{n^-}\}\cup\{n_1-n^{+}+1,\ldots\}
    \big\},
\end{equation*}
where we assume that the binary expansion of $u$ and $-u$ is identical.
Set
\begin{align*}
    \mc{U} & \triangleq
    \mc{U}(n_0,n_0-\bar{R}_0)
    \times \mc{U}(n_1,n_1-\bar{R}_1)
    \times \mc{U}(n_2,n_2-\bar{R}_2)
    \times \mc{U}(\bar{R}_3,0).
\end{align*}
$\mc{U}$ is the set of real numbers such that their binary expansions,
when viewed as vectors of length $n_1$, are in the set $\bar{\mc{U}}$ as
illustrated in Fig.~\ref{fig:subspaces} in
Section~\ref{sec:foundations_det}. Thus, $\mc{U}$ is the direct
translation of the set $\bar{\mc{U}}$ of possible channel inputs for the
deterministic setting to the Gaussian setting. The next lemma states
that if the channel inputs are chosen from $\mc{U}$, then the
resulting minimum constellation distance as observed at the receivers is
large for most channel gains $(g_0,g_1,g_2)$.
\begin{lemma}
    \label{thm:groshev3}
    Let $n_0, n_1, n_2\in\Zp$ such that $n_1 \geq n_0 \geq n_2$, and let
    $\bar{R}_0, \bar{R}_1, \bar{R}_2, \bar{R}_3 \in \Zp$. Define the event
    \begin{equation*}
        B(u_0, u_1, u_2, u_3) 
        \triangleq \big\{(g_0, g_1, g_2)\in(1,4]^3: 
        \abs{g_0u_0+g_1(u_1+u_3)+g_2u_2}\leq 2^{5-n_1}\}, 
    \end{equation*}
    and set
    \begin{equation*}
        B \triangleq
        \bigcup_{(u_0,u_1,u_2,u_3)\in\mc{U}\setminus\{(0,0,0,0)\}}
        B(u_0,u_1,u_2,u_3).
    \end{equation*}
    For any $\delta\in(0,1]$ satisfying
    \begin{align*}
        \bar{R}_1+\bar{R}_0+\bar{R}_2+\bar{R}_3 & \leq n_1-6-\log(6552/\delta), \\
        \bar{R}_0+\bar{R}_2+\bar{R}_3 & \leq n_0-6-\log(6552/\delta), \\
        \bar{R}_2+\bar{R}_3 & \leq n_2-6,
    \end{align*}
    we have
    \begin{equation*}
        \mu(B) \leq \delta.
    \end{equation*}
\end{lemma}

Lemma~\ref{thm:groshev3} is the equivalent for the Gaussian channel of
Lemma~\ref{thm:subspaces} for the deterministic channel. Note that,
except for the constants, the conditions on the rates
in the two lemmas are identical. 

We now prove Lemma~\ref{thm:decoding_gauss} in
Section~\ref{sec:proofs_gaussian_lower} using Lemma~\ref{thm:groshev3}.
  
\begin{IEEEproof}[Proof of Lemma~\ref{thm:decoding_gauss}]
    We will use Lemma~\ref{thm:groshev3} with $\delta/2$ instead of $\delta$ and
    the same rate allocations as in the deterministic case, see
    Figs.~\ref{fig:Rx1-deterministic} and \ref{fig:Rx2-deterministic} in
    Section~\ref{sec:main_arbitrary}. Let $\tilde{B}_m\subset(1,4]^3$ be the
    collection of triples $(g_{m0},g_{m1},g_{m2})$ such that decoding is
    successful at receiver $m$. Define $B_m$ as the collection of channel
    gains $(h_{mk})\subset(1,2]^{2\times 2}$ such that the corresponding
    $(g_{mk})$ are in $\tilde{B}_m$. Finally, let $B$ denote the union of
    $B_1$ and $B_2$. Following the same arguments as in the proof of
    Lemma~\ref{thm:decoding_det} from Lemma~\ref{thm:subspaces} presented in
    Section~\ref{sec:foundations_det}, it can be shown that if the decoding
    conditions in Lemma~\ref{thm:decoding_gauss} are satisfied, then
    Lemma~\ref{thm:groshev3} guarantees that
    \begin{equation*}
        \mu_3(\tilde{B}_m) \leq \delta/2
    \end{equation*}
    for $m\in\{1,2\}$.

    The next lemma allows us to transfer this statement about the products
    $g_{mk}$ of channel gains to the corresponding statement about the
    original channel gains $h_{mk}$. For ease of notation, the statement
    of the lemma uses $g_k$ as a shorthand for $g_{mk}$ as defined
    in~\eqref{eq:gdef} for some fixed value of $m\in\{1,2\}$. 
    \begin{lemma}
        \label{thm:measure}
        Let $\tilde{B}\subseteq(1,4]^3$ be a subset of channel gains
        $(g_0,g_1,g_2)$ such that $\mu_3(\tilde{B})\leq \delta$. 
        Define
        \begin{equation*}
            B \triangleq \big\{
            (h_{mk})\in(1,2]^{2\times 2}: (g_0, g_1, g_2)\in\tilde{B}
            \big\}.
        \end{equation*}
        Then $\mu_4(B) \leq \delta$.
    \end{lemma}

    The proof of Lemma~\ref{thm:measure} is reported in
    Appendix~\ref{sec:appendix_measure}. Applying
    Lemma~\ref{thm:measure} to the sets $\tilde{B}_1$ and $\tilde{B}_2$
    corresponding to the outage events defined above, this implies that
    \begin{equation*}
        \mu_4(B_m) \leq \delta/2.
    \end{equation*}
    Hence,
    \begin{equation*}
        \mu_4(B)\leq \mu_4(B_1)+\mu_4(B_2)\leq \delta,
    \end{equation*}
    proving Lemma~\ref{thm:decoding_gauss}.
\end{IEEEproof}

We continue with the proof of Lemma~\ref{thm:groshev3}.  Instead of directly
analyzing the set $B$ in the statement of Lemma~\ref{thm:groshev3}, it
will be convenient to work with an equivalent set. Note that
$B(u_0,u_1,u_2,u_3)$ can be written as
\begin{equation*}
    B(u_0,u_1,u_2,u_3) = \bigl\{ (g_0,g_1,g_2)\in(1,4]^3:
    \abs{g_02^{n_1}u_0+g_12^{n_1}(u_1+u_3)+g_22^{n_1}u_2} \leq 2^5
    \bigr\}.
\end{equation*}
By the definition of $\mc{U}$ (see also Fig.~\ref{fig:subspaces} in
Section~\ref{sec:foundations_det}), we can decompose
\begin{align*}
    2^{n_1}u_0 & = A_0^\prime q_0, \\
    2^{n_1}u_1 & = A_1^\prime q_1, \\
    2^{n_1}u_2 & = A_2^\prime q_2, \\
    2^{n_1}u_3 & = q_3, 
\end{align*}
with
\begin{equation*}
    A_k^\prime \triangleq 2^{n_k-\bar{R}_k}
\end{equation*}
for $k\in\{0,1,2\}$ and
\begin{equation*}
    q_k 
    \in \{-Q_k, -Q_k+1, \ldots, Q_k-1, Q_k\}
\end{equation*}
for $k\in\{0,1,2,3\}$, where
\begin{equation*}
    Q_k \triangleq 2^{\bar{R}_k}.
\end{equation*}

We now remove the dependence of $B$ on $u_3$. We can further rewrite $B$
using the triangle inequality as
\begin{align*}
    B(u_0, u_1, u_2, u_3) 
    & = \{\abs{A_0^\prime g_0q_0+A_1^\prime g_1q_1+g_1q_3+A_2^\prime g_2q_2} 
    \leq 2^5\} \\
    & \subseteq \{\abs{A_0^\prime g_0q_0+A_1^\prime g_1q_1+A_2^\prime g_2q_2} 
    \leq 2^5+2^{\bar{R}_3+2}\} \\
    & \subseteq \{\abs{A_0^\prime g_0q_0+A_1^\prime g_1q_1+A_2^\prime g_2q_2} 
    \leq \beta^\prime \} \\
    & \triangleq B^\prime(q_0, q_1, q_2),
\end{align*}
where all sets are defined over $(g_0,g_1,g_2)\in(1,4]^3$, and where we
have defined
\begin{align*}
    \beta^\prime & \triangleq 2^{\bar{R}_3+6}.
\end{align*}
Setting
\begin{equation*}
    B^\prime
    \triangleq \bigcup_{
    \substack{q_0, q_1, q_2\in \Z:\\ 
    (q_0, q_1, q_2) \neq \bm{0}, \\
    \abs{q_k}\leq Q_k \forall k} 
    } 
    B^\prime(q_0, q_1, q_2),
\end{equation*}
we then have
\begin{equation*}
    \mu_3(B) \leq \mu_3(B^\prime).
\end{equation*}
The next lemma analyzes the set $B^\prime$ with $A_0^\prime=1$.

\begin{lemma}
    \label{thm:groshev2}
    Let $\beta \in (0,1]$, $A_1, A_2\in\N$, and
    $Q_0, Q_1, Q_2 \in\N$.  Define the event
    \begin{equation*}
        B^\prime(q_0, q_1, q_2) 
        \triangleq 
        \big\{(g_0, g_1, g_2)\in(1,4]^3: 
        \abs{g_0q_0+A_1g_1 q_1+ A_2g_2 q_2} < \beta \big\},
    \end{equation*}
    and set
    \begin{equation*}
        B^\prime
        \triangleq \bigcup_{
        \substack{q_0, q_1, q_2\in \Z:\\ 
        (q_0, q_1, q_2) \neq \bm{0}, \\
        \abs{q_k}\leq Q_k \forall k} 
        } 
        B^\prime(q_0, q_1, q_2).
    \end{equation*}
    Then
    \begin{align*}
        \mu(B^\prime) 
        \leq 
        504\beta \bigg(&
        2\min\Big\{Q_2, \frac{Q_0}{A_2}\Big\}
        +\min\Big\{Q_1\tilde{Q}_2,  
        \frac{Q_0\tilde{Q}_2}{A_1}, 
        \frac{A_2\tilde{Q}_2^2}{A_1}\Big\} \\
        & +2\min\Big\{Q_1, \frac{Q_0}{A_1}\Big\}
        +\min\Big\{Q_2\tilde{Q}_1,  
        \frac{Q_0\tilde{Q}_1}{A_2}, 
        \frac{A_1\tilde{Q}_1^2}{A_2}\Big\}
        \bigg)
    \end{align*}
    with
    \begin{align*}
        \tilde{Q}_1 
        & \triangleq \min\Big\{Q_1, 8\frac{\max\{Q_0, A_2Q_2\}}{A_1}\Big\}, \\
        \tilde{Q}_2 
        & \triangleq \min\Big\{Q_2, 8\frac{\max\{Q_0, A_1Q_1\}}{A_2}\Big\}.
    \end{align*}
\end{lemma}

\begin{remark}
    The special case of Lemma~\ref{thm:groshev2} with $A_1=A_2=1$,
    $Q_0=Q_1=Q_2=Q$, and $Q\to\infty$ corresponds to the (converse part
    of) Groshev's theorem, see, e.g., \cite[Theorem 1.12]{sprindzuk79}.
    Hence, Lemma~\ref{thm:groshev2} extends Groshev's theorem to
    asymmetric and non-asymptotic settings. 
\end{remark}

Before we present the proof of
Lemma~\ref{thm:groshev2}, we show how to prove Lemma~\ref{thm:groshev3}
with the help of Lemma~\ref{thm:groshev2}. 

\begin{IEEEproof}[Proof of Lemma~\ref{thm:groshev3}]
    We consider the three cases $A_0^\prime \leq
    \min\{A_1^\prime, A_2^\prime\}$, $A_1^\prime \leq \min\{A_0^\prime,
    A_2^\prime\}$, and $A_2^\prime \leq \min\{A_0^\prime, A_1^\prime\}$
    separately.

    Assume first that
    $A_0^\prime \leq \min\{A_1^\prime, A_2^\prime\}$. Define
    \begin{align*}
        A_0 & \triangleq 1, \\
        A_1 & \triangleq A_1^\prime/A_0^\prime = 2^{\bar{R}_0-\bar{R}_1-n_0+n_1}, \\
        A_2 & \triangleq A_2^\prime/A_0^\prime = 2^{\bar{R}_0-\bar{R}_2-n_0+n_2}, \\
        \beta & \triangleq \beta^\prime/A_0^\prime = 2^{\bar{R}_0+\bar{R}_3-n_0+6}.
    \end{align*}
    Note that $A_1, A_2\in \N$, and that $\beta\in(0,1]$ if
    \begin{equation}
        \label{eq:betabound1}
        \bar{R}_0+\bar{R}_3\leq n_0-6,
    \end{equation}
    as required by Lemma~\ref{thm:groshev2}.  The quantities
    $\tilde{Q}_1$ and $\tilde{Q}_2$ in Lemma~\ref{thm:groshev2} can be
    upper bounded as
    \begin{equation*}
        \tilde{Q}_1
        \leq 8\max\{Q_0,A_2 Q_2\}/A_1 
        = 8Q_0/A_1,
    \end{equation*}
    since $n_0 \geq n_2$ implies that $Q_0 \geq A_2Q_2$, and as
    \begin{align*}
        \tilde{Q}_2 & \leq Q_2.
    \end{align*}

    Applying Lemma~\ref{thm:groshev2} yields then 
    \begin{align*}
        \mu(B) 
        & \leq \mu(B')  \\
        & \leq 504\beta 
        \Big(2Q_2+\frac{A_2\tilde{Q}_2^2}{A_1}+2\frac{Q_0}{A_1}+Q_2\tilde{Q}_1\Big) \\
        & \leq 504\beta 
        \Big(2Q_2+\frac{A_2Q_2^2}{A_1}+2\frac{Q_0}{A_1}+8\frac{Q_0Q_2}{A_1}\Big) \\
        & \leq 6652\beta\max
        \Big\{Q_2,\frac{A_2Q_2^2}{A_1},\frac{Q_0}{A_1},\frac{Q_0Q_2}{A_1}\Big\} \\
        & = 6652\beta \max\Big\{Q_2,\frac{Q_0Q_2}{A_1}\Big\},
    \end{align*}
    where we have used that $Q_2\geq 1$ and that $A_2Q_2 \leq Q_0$ implying
    \begin{equation*}
        \frac{A_2Q_2^2}{A_1}\leq \frac{Q_0Q_2}{A_1}.
    \end{equation*}
    Substituting the definitions of $\beta$, $A_k$, and $Q_k$, yields that
    \begin{equation*}
        \mu(B) 
        \leq 6652\cdot 2^{\bar{R}_0+\bar{R}_3-n_0+6}
        \max\big\{ 2^{\bar{R}_2}, 2^{\bar{R}_1+\bar{R}_2+n_0-n_1} \big\}.
    \end{equation*}
    Together with \eqref{eq:betabound1}, this shows that if
    \begin{align*}
        \bar{R}_1+\bar{R}_0+\bar{R}_2+\bar{R}_3 & \leq n_1-6-\log(6652/\delta), \\
        \bar{R}_0+\bar{R}_2+\bar{R}_3 & \leq n_0-6-\log(6652/\delta), \\
        \bar{R}_0+\bar{R}_3 & \leq n_0-6,
    \end{align*}
    then
    \begin{equation*}
        \mu(B) \leq \delta.
    \end{equation*}
    Since $\delta\in(0,1]$ and $\bar{R}_2\geq 0$, the third condition is redundant
    and can be removed, showing the result in Lemma~\ref{thm:groshev3}. We
    point out that the third condition in Lemma~\ref{thm:groshev3} is not
    active if $A_0^\prime \leq \min\{A_1^\prime, A_2^\prime\}$. This is
    consistent with it not appearing in the derivation here.

    Assume next that
    $A_1^\prime \leq \min\{A_0^\prime, A_2^\prime\}$. Define
    \begin{align*}
        A_0 & \triangleq A_0^\prime/A_1^\prime = 2^{-\bar{R}_0+\bar{R}_1+n_0-n_1}, \\
        A_1 & \triangleq 1, \\
        A_2 & \triangleq A_2^\prime/A_1^\prime = 2^{\bar{R}_1-\bar{R}_2-n_1+n_2}, \\
        \beta & \triangleq \beta^\prime/A_1^\prime = 2^{\bar{R}_1+\bar{R}_3-n_1+6}.
    \end{align*}
    Note that $A_0, A_2\in \N$, and that $\beta\in(0,1]$ if
    \begin{equation}
        \label{eq:betabound2}
        \bar{R}_1+\bar{R}_3\leq n_1-6.
    \end{equation}
    We can hence apply Lemma~\ref{thm:groshev2} by appropriately relabeling
    indices (i.e., by swapping indices $0$ and $1$).  The quantities
    $\tilde{Q}_0$ and $\tilde{Q}_2$ can be upper bounded as
    \begin{equation*}
        \tilde{Q}_0
        = \min\Big\{Q_0, 8\frac{\max\{Q_1, A_2Q_2\}}{A_0}\Big\}
        \leq Q_0,
    \end{equation*}
    and 
    \begin{align*}
        \tilde{Q}_2 
        = \min\Big\{Q_2, 8\frac{\max\{Q_1, A_0Q_0\}}{A_2}\Big\}
        \leq Q_2.
    \end{align*}

    Applying Lemma~\ref{thm:groshev2} yields then that
    \begin{align*}
        \mu(B) 
        & \leq \mu(B')  \\
        & \leq 504\beta 
        \Big(2Q_2+Q_0\tilde{Q}_2+2Q_0+Q_2\tilde{Q}_0\Big) \\
        & \leq 504\beta 
        \Big(2Q_2+Q_0Q_2+2Q_0+Q_2Q_0\Big) \\
        & \leq 3024\beta Q_0Q_2.
    \end{align*}
    Substituting the definitions of $\beta$ and $Q_k$, yields that
    \begin{equation*}
        \mu(B) 
        \leq 3024\cdot 2^{\bar{R}_0+\bar{R}_1+\bar{R}_2+\bar{R}_3+6-n_1}.
    \end{equation*}
    Together with \eqref{eq:betabound2}, this shows that if
    \begin{align*}
        \bar{R}_1+\bar{R}_0+\bar{R}_2+\bar{R}_3 & \leq n_1-6-\log(3024/\delta),\\
        \bar{R}_1+\bar{R}_3 & \leq n_1-6,
    \end{align*}
    then
    \begin{equation*}
        \mu(B) \leq \delta.
    \end{equation*}
    Since $\delta\in(0,1]$ and $\bar{R}_0,\bar{R}_2\geq 0$, the second condition is
    redundant and can be removed, showing the result in
    Lemma~\ref{thm:groshev3}. As can be verified, the second and third
    conditions in Lemma~\ref{thm:groshev3} are not active when $A_1^\prime
    \leq \min\{A_0^\prime, A_2^\prime\}$, consistent with them not appearing
    in the derivation here.

    Finally, assume that $A_2^\prime \leq \min\{A_0^\prime, A_1^\prime\}$.
    Define
    \begin{align*}
        A_0 & \triangleq A_0^\prime/A_2^\prime=2^{-\bar{R}_0+\bar{R}_2+n_0-n_2}, \\
        A_1 & \triangleq A_1^\prime/A_2^\prime=2^{-\bar{R}_1+\bar{R}_2+n_1-n_2}, \\
        A_2 & \triangleq 1, \\
        \beta & \triangleq \beta^\prime/A_2^\prime=2^{\bar{R}_2+\bar{R}_3-n_2+6}.
    \end{align*}
    Note that $A_0, A_1\in \N$, and that $\beta\in(0,1]$ if
    \begin{equation}
        \label{eq:betabound3}
        \bar{R}_2+\bar{R}_3\leq n_2-6.
    \end{equation}
    We can hence apply Lemma~\ref{thm:groshev2} by relabeling indices as
    before (this time by swapping indices $0$ and $2$). The quantities
    $\tilde{Q}_0$ and $\tilde{Q}_1$ can be upper bounded as
    \begin{equation*}
        \tilde{Q}_0 
        = \min\Big\{Q_0, 8\frac{\max\{Q_2, A_1Q_1\}}{A_0}\Big\}
        \leq Q_0,
    \end{equation*}
    and
    \begin{align*}
        \tilde{Q}_1
        & = \min\Big\{Q_1, 8\frac{\max\{Q_2, A_0Q_0\}}{A_1}\Big\} \\
        & \leq 8\frac{\max\{Q_2, A_0Q_0\}}{A_1} \\
        & = 8\frac{A_0Q_0}{A_1}
    \end{align*}
    since $n_0 \geq n_2$ implies $A_0Q_0\geq Q_2$. 

    Applying Lemma~\ref{thm:groshev2} yields then that
    \begin{align*}
        \mu(B) 
        & \leq \mu(B')  \\
        & \leq 504\beta 
        \Big(2\frac{Q_2}{A_0}+\frac{Q_2\tilde{Q}_0}{A_1}
        +2\frac{Q_2}{A_1}+\frac{Q_2\tilde{Q}_1}{A_0}\Big) \\
        & \leq 504\beta 
        \Big(2\frac{Q_2}{A_0}+\frac{Q_2Q_0}{A_1}
        +2\frac{Q_2}{A_1}+8\frac{Q_2Q_0}{A_1}\Big) \\
        & \leq 6552\beta \max
        \Big\{\frac{Q_2}{A_0},\frac{Q_2Q_0}{A_1}\Big\}.
    \end{align*}
    Substituting the definitions of $\beta$, $A_k$, and $Q_k$, yields that
    \begin{equation*}
        \mu(B) 
        \leq 6652\cdot 2^{\bar{R}_2+\bar{R}_3+6-n_2}\max\big\{
        2^{\bar{R}_0+n_2-n_0},
        2^{\bar{R}_0+\bar{R}_1+n_2-n_1}
        \big\}.
    \end{equation*}
    Together with \eqref{eq:betabound3}, this shows that if
    \begin{align*}
        \bar{R}_1+\bar{R}_0+\bar{R}_2+\bar{R}_3 & \leq n_1-6-\log(6652/\delta),\\
        \bar{R}_0+\bar{R}_2+\bar{R}_3 & \leq n_0-6-\log(6652/\delta),\\
        \bar{R}_2+\bar{R}_3 & \leq n_2-6,
    \end{align*}
    then
    \begin{equation*}
        \mu(B) \leq \delta,
    \end{equation*}
    showing the result in Lemma~\ref{thm:groshev3}. It can be verified that,
    unlike in the other two cases, all three conditions in
    Lemma~\ref{thm:groshev3} can be active when $A_2^\prime \leq
    \min\{A_0^\prime, A_1^\prime\}$. This is again consistent with the
    derivation here. This proves Lemma~\ref{thm:groshev3}.
\end{IEEEproof}

It remains to prove Lemma~\ref{thm:groshev2}. The proof builds on an
argument in \cite{dodson09}.

\begin{IEEEproof}[Proof of Lemma~\ref{thm:groshev2}]
    Define
    \begin{equation*}
        B^\prime(q_1, q_2) 
        \triangleq \bigcup_{\substack{q_0\in\Z: \\ \abs{q_0}\leq Q_0}}
        B^\prime(q_0, q_1, q_2)
    \end{equation*}
    for $(q_1, q_2)\neq (0,0)$, and
    \begin{equation*}
        B^\prime(0, 0) 
        \triangleq \bigcup_{\substack{q_0\in\Z\setminus\{0\}: \\ \abs{q_0}\leq Q_0}}
        B^\prime(q_0, 0, 0).
    \end{equation*}
    For $g_0\in(1,4]$, set
    \begin{equation*}
        B^\prime_{g_0}(q_1, q_2) 
        \triangleq 
        \big\{(g_1, g_2)\in(1,4]^2: (g_0, g_1, g_2)\in B^\prime(q_1, q_2)\big\}.
    \end{equation*}
    Observe that $B^\prime_{g_0}(q_1, q_2)$ is a subset of $\R^2$ and that
    \begin{equation*}
        \mu_3(B^\prime(q_1, q_2)) 
        = \int_{g_0 = 1}^4 \mu_2(B^\prime_{g_0}(q_1, q_2)) dg_0.
    \end{equation*}

    We treat the cases $A_1\abs{q_1} \leq A_2\abs{q_2}$ and
    $A_1\abs{q_1} > A_2\abs{q_2}$ separately. Assume first
    $A_1\abs{q_1} \leq A_2\abs{q_2}$ and $q_2 \neq 0$. If 
    \begin{equation*}
        A_2\abs{q_2} \geq 8\max\{Q_0, A_1Q_1\}+1,
    \end{equation*}
    then 
    \begin{align*}
        \abs{g_0q_0+A_1g_1q_1+A_2g_2q_2} 
        & \geq A_2g_2\abs{q_2}-A_1g_1\abs{q_1}-g_0\abs{q_0} \\
        & \geq A_2\abs{q_2}-4A_1Q_1-4Q_0 \\
        & \geq 1 \\
        & \geq \beta, 
    \end{align*}
    where we have used that $\beta \leq 1$. Hence, $\mu_2(B^\prime_{g_0}(q_1,
    q_2)) = 0$. We can therefore assume without loss of generality that
    \begin{equation*}
        A_2\abs{q_2} \leq 8\max\{Q_0, A_1Q_1\}
    \end{equation*}
    for any value of $q_1$. By a similar argument, we can assume that
    \begin{equation*}
        A_2\abs{q_2} \leq 4Q_0
    \end{equation*}
    for $q_1=0$.

    The set $B^\prime_{g_0}(q_1, q_2)$ consists of at most 
    \begin{equation*}
        \min\{3Q_0, 7A_2\abs{q_2}\}
    \end{equation*}
    strips of slope $-A_1q_1/(A_2q_2)$ and width $2\beta/(A_2\abs{q_2})$ in
    the $g_2$ direction, including several partial strips (see
    Fig.~\ref{fig:strips}).  The area of this set is at most
    \begin{align}
        \label{eq:groshev2_1}
        \mu_2(B^\prime_{g_0}(q_1, q_2)) 
        & \leq 3\cdot\frac{2\beta}{A_2\abs{q_2}}\cdot\min\{3Q_0, 7A_2\abs{q_2}\} \nonumber\\
        & \leq 42\beta\min\Big\{\frac{Q_0}{A_2\abs{q_2}},1\Big\}.
    \end{align}
    \begin{figure}[htbp]
        \begin{center}
            \scalebox{1}{\input{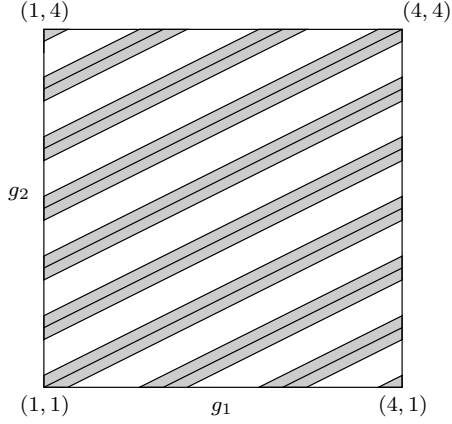}} 
        \end{center}

        \caption{Illustration of the set $B_{1}^\prime(1,-2)\subseteq
        (1,4]^2$ with $g_0=1, q_1=1, q_2=-2$, $A_1=A_2=1$, and
        $\beta=0.2$. In the figure, we assume that $Q_0 \gg
        A_2\abs{q_2}$. The set consists of $10\leq 7A_2\abs{q_2}=14$
        strips of slope $1/2=-A_1q_1/(A_2q_2)$.}

        \label{fig:strips}
    \end{figure}

    We now consider the case $A_1\abs{q_1}> A_2\abs{q_2}$
    and $q_1\neq 0$. As before, we can assume without loss of generality
    that
    \begin{equation*}
        A_1\abs{q_1} \leq 8\max\{Q_0, A_2Q_2\}
    \end{equation*}
    for any value of $q_2$, and that
    \begin{equation*}
        A_1\abs{q_1} \leq 4Q_0
    \end{equation*}
    for $q_2=0$. By the same analysis as in the last paragraph, we obtain that
    \begin{equation}
        \label{eq:groshev2_2}
        \mu_2(B^\prime_{g_0}(q_1, q_2))
        \leq 42\beta\min\Big\{\frac{Q_0}{A_1\abs{q_1}},1\Big\}.
    \end{equation}
    Finally, when $(q_1, q_2)=\bm{0}$ and $q_0 \neq 0$, then
    $g_0\abs{q_0} \geq 1 \geq \beta$, and hence
    \begin{equation}
        \label{eq:groshev2_3}
        \mu_2(B^\prime_{g_0}(0,0)) = 0.
    \end{equation}

    We can upper bound
    \begin{align*}
        \mu_3(B^\prime) 
        & = \mu_3\bigg(
        \cup_{\substack{q_1\in\Z: \\ \abs{q_1}\leq Q_1}}
        \cup_{\substack{q_2\in\Z: \\ \abs{q_2}\leq Q_2}}
        \cup_{\substack{q_0\in \Z: \abs{q_0}\leq Q_0\\ 
        (q_0, q_1, q_2) \neq \bm{0}}}
        B^\prime(q_0, q_1, q_2)\bigg) \nonumber\\
        & \leq \sum_{\substack{q_1\in\Z: \\ \abs{q_1}\leq Q_1}}
        \sum_{\substack{q_2\in\Z: \\ \abs{q_2}\leq Q_2}}
        \mu_3( B^\prime(q_1, q_2)) \nonumber\\
        & = \sum_{\substack{q_1\in\Z: \\ \abs{q_1}\leq Q_1}}
        \sum_{\substack{q_2\in\Z: \\ \abs{q_2}\leq Q_2}}
        \int_{g_0=1}^4 \mu_2(B^\prime_{g_0}(q_1, q_2))dg_0 \nonumber\\
        & = \sum_{\substack{q_2\in\Z\setminus\{0\}: \abs{q_2}\leq Q_2\\ 
        A_2\abs{q_2}\leq 4Q_0}}
        \int_{g_0=1}^4 \mu_2(B^\prime_{g_0}(0, q_2))dg_0 \nonumber\\
        & \quad {}+ \sum_{\substack{q_2\in\Z\setminus\{0\}: \abs{q_2}\leq Q_2\\ 
        A_2\abs{q_2}\leq 8\max\{Q_0, A_1Q_1\}}}
        \sum_{\substack{q_1\in\Z\setminus\{0\}: \abs{q_1}\leq Q_1\\ 
        A_1\abs{q_1}\leq A_2\abs{q_2} }} 
        \int_{g_0=1}^4 \mu_2(B^\prime_{g_0}(q_1, q_2))dg_0 \nonumber\\
        & \quad {}+\sum_{\substack{q_1\in\Z\setminus\{0\}: \abs{q_1}\leq Q_1\\ 
        A_1\abs{q_1}\leq 4Q_0}}
        \int_{g_0=1}^4 \mu_2(B^\prime_{g_0}(q_1, 0))dg_0 \nonumber\\
        & \quad {}+\sum_{\substack{q_1\in\Z\setminus\{0\}: \abs{q_1}\leq Q_1\\ 
        A_1\abs{q_1}\leq 8\max\{Q_0, A_2Q_2\}}}
        \sum_{\substack{q_2\in\Z\setminus\{0\}: \abs{q_2}\leq Q_2\\ 
        A_2\abs{q_2}\leq A_1\abs{q_1}}} 
        \int_{g_0=1}^4 \mu_2(B^\prime_{g_0}(q_1, q_2))dg_0 \\
        & \quad {}+\int_{g_0=1}^4 \mu_2(B^\prime_{g_0}(0,0))dg_0.
    \end{align*}
    Combined with \eqref{eq:groshev2_1}, \eqref{eq:groshev2_2}, and
    \eqref{eq:groshev2_3}, this yields
    \begin{align}
        \label{eq:groshev2_4}
        \mu_3(B^\prime) 
        & \leq 
        \sum_{\substack{q_2\in\Z\setminus\{0\}: \abs{q_2}\leq Q_2\\ 
        A_2\abs{q_2}\leq 4Q_0}}
        126\beta\min\Big\{\frac{Q_0}{A_2\abs{q_2}}, 1\Big\} \nonumber\\
        & \quad {} + \sum_{\substack{q_2\in\Z\setminus\{0\}: \abs{q_2}\leq Q_2\\ 
        A_2\abs{q_2}\leq 8\max\{Q_0, A_1Q_1\}}}
        \sum_{\substack{q_1\in\Z\setminus\{0\}: \abs{q_1}\leq Q_1\\ 
        A_1\abs{q_1}\leq A_2\abs{q_2} }} 
        126\beta\min\Big\{\frac{Q_0}{A_2\abs{q_2}}, 1\Big\} \nonumber\\
        & \quad {} +\sum_{\substack{q_1\in\Z\setminus\{0\}: \abs{q_1}\leq Q_1\\ 
        A_1\abs{q_1}\leq 4Q_0}}
        126\beta\min\Big\{\frac{Q_0}{A_1\abs{q_1}}, 1\Big\} \nonumber\\
        & \quad {} +\sum_{\substack{q_1\in\Z\setminus\{0\}: \abs{q_1}\leq Q_1\\ 
        A_1\abs{q_1}\leq 8\max\{Q_0, A_2Q_2\}}}
        \sum_{\substack{q_2\in\Z\setminus\{0\}: \abs{q_2}\leq Q_2\\ 
        A_2\abs{q_2}\leq A_1\abs{q_1}}} 
        126\beta\min\Big\{\frac{Q_0}{A_1\abs{q_1}}, 1\Big\}.
    \end{align}
    We now upper bound the four terms in the right-hand side of
    \eqref{eq:groshev2_4}. 

    For the first term in \eqref{eq:groshev2_4}, observe that
    \begin{align*}
        \big\lvert\big\{q_2\in\Z\setminus\{0\}: 
        \abs{q_2}\leq Q_2, A_2\abs{q_2}\leq 4Q_0\big\}\big\rvert
        & \leq 2\min\Big\{Q_2, 4\frac{Q_0}{A_2}\Big\} \\
        & \leq 8\min\Big\{Q_2, \frac{Q_0}{A_2}\Big\},
    \end{align*}
    so that
    \begin{align}
        \label{eq:groshev2_5a}
        \sum_{\substack{q_2\in\Z\setminus\{0\}: \abs{q_2}\leq Q_2\\ 
        A_2\abs{q_2}\leq 4Q_0}}
        126\beta\min\Big\{\frac{Q_0}{A_2\abs{q_2}}, 1\Big\} 
        \leq 1008\beta \min\Big\{Q_2, \frac{Q_0}{A_2}\Big\}.
    \end{align}

    For the second term in \eqref{eq:groshev2_4}, observe that
    \begin{equation*}
        \big\lvert\big\{ 
        q_1\in\Z\setminus\{0\}: \abs{q_1}\leq Q_1, A_1\abs{q_1}\leq A_2\abs{q_2}
        \big\}\big\rvert
        \leq 2\min\Big\{Q_1, \frac{A_2\abs{q_2}}{A_1}\Big\},
    \end{equation*}
    and hence
    \begin{align*}
        \sum_{\substack{q_1\in\Z\setminus\{0\}: \abs{q_1}\leq Q_1\\ 
        A_1\abs{q_1}\leq A_2\abs{q_2} }} 
        \min\Big\{\frac{Q_0}{A_2\abs{q_2}}, 1\Big\} 
        & \leq 2\min\Big\{Q_1, \frac{A_2\abs{q_2}}{A_1}\Big\} 
        \min\Big\{\frac{Q_0}{A_2\abs{q_2}}, 1\Big\} \\
        & \leq 2\min\Big\{\frac{Q_0Q_1}{A_2\abs{q_2}}, Q_1, 
        \frac{Q_0}{A_1}, \frac{A_2\abs{q_2}}{A_1}\Big\} \\
        & \leq 2\min\Big\{Q_1, \frac{Q_0}{A_1}, \frac{A_2\abs{q_2}}{A_1}\Big\}.
    \end{align*}
    Moreover, 
    \begin{align*}
        \big\{ 
        q_2\in\Z\setminus\{0\}: \abs{q_2}\leq Q_2, 
        A_2\abs{q_2}\leq 8\max\{Q_0, A_1Q_1\}
        \big\} 
        = \big\{ 
        q_2\in\Z\setminus\{0\}: \abs{q_2}\leq \tilde{Q}_2 \big\} 
    \end{align*}
    with
    \begin{equation*}
        \tilde{Q}_2
        \triangleq
        \min\Big\{Q_2, 8\frac{\max\{Q_0, A_1Q_1\}}{A_2}\Big\}.
    \end{equation*}
    Using these two facts, we can upper bound 
    \begin{align}
        \label{eq:groshev2_5b}
        \sum_{\substack{q_2\in\Z\setminus\{0\}: \abs{q_2}\leq Q_2\\ 
        A_2\abs{q_2}\leq 8\max\{Q_0, A_1Q_1\}}} &
        \sum_{\substack{q_1\in\Z\setminus\{0\}: \abs{q_1}\leq Q_1\\ 
        A_1\abs{q_1}\leq A_2\abs{q_2} }} 
        126\beta\min\Big\{\frac{Q_0}{A_2\abs{q_2}}, 1\Big\} \nonumber\\
        & \leq
        252\beta
        \sum_{q_2\in\Z\setminus\{0\}: \abs{q_2}\leq \tilde{Q}_2}
        \min\Big\{Q_1, \frac{Q_0}{A_1}, \frac{A_2\abs{q_2}}{A_1}\Big\} \nonumber\\
        & \leq
        252\beta\sum_{q_2\in\Z\setminus\{0\}: \abs{q_2}\leq \tilde{Q}_2}
        \min\Big\{Q_1, \frac{Q_0}{A_1}, \frac{A_2\tilde{Q_2}}{A_1}\Big\} \nonumber\\
        & \leq 504\beta
        \min\Big\{Q_1\tilde{Q}_2,  
        \frac{Q_0\tilde{Q}_2}{A_1}, 
        \frac{A_2\tilde{Q}_2^2}{A_1}\Big\}.
    \end{align}

    Similarly, for the third term in \eqref{eq:groshev2_4},
    \begin{align}
        \label{eq:groshev2_5c}
        \sum_{\substack{q_1\in\Z\setminus\{0\}: \abs{q_1}\leq Q_1\\ 
        A_1\abs{q_1}\leq 4Q_0}}
        126\beta\min\Big\{\frac{Q_0}{A_1\abs{q_1}}, 1\Big\} 
        \leq 1008\beta \min\Big\{Q_1, \frac{Q_0}{A_1}\Big\},
    \end{align}
    and for the fourth term
    \begin{align}
        \label{eq:groshev2_5d}
        \sum_{\substack{q_1\in\Z\setminus\{0\}: \abs{q_1}\leq Q_1\\ 
        A_1\abs{q_1}\leq 8\max\{Q_0, A_2Q_2\}}}
        \sum_{\substack{q_2\in\Z\setminus\{0\}: \abs{q_2}\leq Q_2\\ 
        A_2\abs{q_2}\leq A_1\abs{q_1} }} 
        126\beta\min\Big\{\frac{Q_0}{A_1\abs{q_1}}, 1\Big\}
        & \leq 504\beta
        \min\Big\{Q_2\tilde{Q}_1,  
        \frac{Q_0\tilde{Q}_1}{A_2}, 
        \frac{A_1\tilde{Q}_1^2}{A_2}\Big\}
    \end{align}
    with
    \begin{equation*}
        \tilde{Q}_1
        \triangleq \min\Big\{Q_1, 8\frac{\max\{Q_0, A_2Q_2\}}{A_1}\Big\}.
    \end{equation*}

    Substituting \eqref{eq:groshev2_5a}--\eqref{eq:groshev2_5d} 
    into \eqref{eq:groshev2_4} yields
    \begin{align*}
        \mu_3(B^\prime) 
        \leq 
        \beta \bigg( &
        1008\min\Big\{Q_2, \frac{Q_0}{A_2}\Big\}
        +504\min\Big\{Q_1\tilde{Q}_2,  
        \frac{Q_0\tilde{Q}_2}{A_1}, 
        \frac{A_2\tilde{Q}_2^2}{A_1}\Big\} \\
        & {}+1008\min\Big\{Q_1, \frac{Q_0}{A_1}\Big\}
        +504\min\Big\{Q_2\tilde{Q}_1,  
        \frac{Q_0\tilde{Q}_1}{A_2}, 
        \frac{A_1\tilde{Q}_1^2}{A_2}\Big\}
        \bigg),
    \end{align*}
    completing the proof.
\end{IEEEproof}

\section{Conclusion}
\label{sec:conclusion}

In this paper, we derived a constant-gap capacity approximation for the
Gaussian X-channel. This derivation was aided by a novel deterministic
channel model used to approximate the Gaussian channel. In the proposed
deterministic channel model, the actions of the channel are described by
a lower-triangular Toeplitz matrices with coefficients determined by the
bits in the binary expansion of the corresponding channel gains in the
original Gaussian problem. This is in contrast to traditional
deterministic models, in which the actions of the channel are only
dependent on the single most-significant bit of the channel gains in the
original Gaussian problem. Preserving this dependence on the fine
structure of the Gaussian channel gains turned out to be crucial to
successfully approximate the Gaussian X-channel by a deterministic
channel model.

Throughout this paper, we were only interested in obtaining a
constant-gap capacity approximation. Less emphasis was placed on the
actual value of that constant. For a meaningful capacity approximation
at smaller values of SNR, this constant needs to be optimized. More
sophisticated lattice codes (as opposed to the ones over the simple
integer lattice used in this paper) could be employed for this purpose,
see, e.g., \cite{ordentlich11}.  Furthermore, all the results in this
paper were derived for all channel gains outside an arbitrarily small
outage set. Analyzing the behavior of capacity for channel gains
that are inside this outage set is hence of interest. An approach
similar to the one in~\cite{wu11} could perhaps be utilized to this end.

Finally, the analysis in this paper focused on the Gaussian X-channel as
an example of a fully-connected communication network in which
interference alignment seems necessary. The hope is that the tools
developed in this paper can be used to help with the analysis of more
general networks requiring interference alignment. Ultimately, the goal
should be to move from degrees-of-freedom capacity approximations to
stronger constant-gap capacity approximations.

\appendices

\section{Verification of Decoding Conditions}
\label{sec:appendix_decoding}

This appendix verifies that the rate allocation in
Section~\ref{sec:proofs_det_lower} for the deterministic X-channel
satisfies the decoding conditions \eqref{eq:decoding1} and
\eqref{eq:decoding2} in Lemma~\ref{thm:decoding_det}.

Case I ($0\leq n_{12}+n_{21} \leq n_{11}$): Recall 
\begin{align*}
    \bar{R}_{11}^{\sP} & \triangleq n_{11}-n_{21}, \\
    \bar{R}_{22}^{\sP} & \triangleq n_{22}-n_{12}, \\
    \bar{R}_{22}^{\sC} & \triangleq \bar{R}_{11}^{\sC} \triangleq \bar{R}_{12} \triangleq \bar{R}_{21} \triangleq 0.
\end{align*}
This choice of rates satisfies \eqref{eq:decoding1a} and
\eqref{eq:decoding2a}. Since these are the only two relevant conditions
in this case, this shows that both receivers can recover the desired
messages.  

Case II ($n_{11} < n_{12}+n_{21} \leq n_{22}$): Recall 
\begin{align*}
    \bar{R}_{11}^{\sP} & \triangleq n_{11}-n_{21}, \\
    \bar{R}_{22}^{\sP} & \triangleq n_{22}-n_{12}, \\
    \bar{R}_{22}^{\sC} & \triangleq n_{12}-\bar{R}_{11}^{\sP}, \\
    \bar{R}_{11}^{\sC} & \triangleq \bar{R}_{12} \triangleq \bar{R}_{21} \triangleq 0.
\end{align*}
At receiver one, \eqref{eq:decoding1a} and \eqref{eq:decoding1b} are
satisfied since
\begin{equation*}
    \bar{R}_{22}^{\sC}+\bar{R}_{11}^{\sP} = n_{12} \leq n_{11}.
\end{equation*}
Condition \eqref{eq:decoding1c} does not need to be checked here.
At receiver two, \eqref{eq:decoding2a} is satisfied since
\begin{equation*}
    \bar{R}_{22}^{\sC}+\bar{R}_{22}^{\sP} = n_{22}+n_{21}-n_{11} \leq n_{22}.
\end{equation*}
Conditions \eqref{eq:decoding2b} and \eqref{eq:decoding2c} do not need
to be checked here.  Hence both receivers can decode successfully. 

Case III ($n_{22} < n_{12}+n_{21} \leq n_{11}+\tfrac{1}{2}n_{22}$):
Recall 
\begin{align*}
    \bar{R}_{11}^{\sP} & \triangleq n_{11}-n_{21}, \\
    \bar{R}_{22}^{\sP} & \triangleq n_{22}-n_{12}, \\
    \bar{R}_{12} & \triangleq (n_{12}+2n_{21}-n_{11}-n_{22})^+, \\
    \bar{R}_{21} & \triangleq (n_{21}+2n_{12}-n_{11}-n_{22})^+, \\
    \bar{R}_{11}^{\sC} & \triangleq n_{21}-\bar{R}_{22}^{\sP}-\bar{R}_{21}, \\
    \bar{R}_{22}^{\sC} & \triangleq n_{12}-\bar{R}_{11}^{\sP}-\bar{R}_{12}.
\end{align*}
To check the decoding conditions \eqref{eq:decoding1} and
\eqref{eq:decoding2}, we first argue that
\begin{subequations}
    \label{eq:max}
    \begin{align}
        \label{eq:maxa}
        \max\{\bar{R}_{21}, \bar{R}_{22}^{\sC}\} & =  \bar{R}_{22}^{\sC}, \\
        \label{eq:maxb}
        \max\{\bar{R}_{12}, \bar{R}_{11}^{\sC}\} & =  \bar{R}_{11}^{\sC}.
    \end{align}
\end{subequations}
The first equality trivially holds if $\bar{R}_{21} = 0$. Assuming then that
$\bar{R}_{21} > 0$, we have
\begin{equation*}
    \bar{R}_{22}^{\sC} - \bar{R}_{21}
    = n_{22}-n_{12}-\bar{R}_{12}.
\end{equation*}
If $\bar{R}_{12} = 0$, then this is nonnegative. Assuming then $\bar{R}_{12} > 0$,
we obtain
\begin{equation*}
    \bar{R}_{22}^{\sC} - \bar{R}_{21}
    = 2n_{22}+n_{11}-2(n_{12}+n_{21}) \geq 0,
\end{equation*}
where we have used that $n_{22}\geq n_{11}$ and that
$n_{11}+\tfrac{1}{2}n_{22} \geq n_{12}+n_{21}$. This
proves~\eqref{eq:maxa}. Using a similar argument, it can be shown that
$\bar{R}_{11}^{\sC} - \bar{R}_{12} \geq 0$, proving~\eqref{eq:maxb}.  To
check the decoding conditions \eqref{eq:decoding1} at receiver one,
observe now that
\begin{equation*}
    \bar{R}_{22}^{\sC}+\bar{R}_{12}+\bar{R}_{11}^{\sP} = n_{12},
\end{equation*}
satisfying \eqref{eq:decoding1b}. Moreover, 
\begin{align*}
    \bar{R}_{11}^\sC+\bar{R}_{22}^{\sC}+\bar{R}_{12}+\bar{R}_{11}^{\sP}
    & = \bar{R}_{11}^\sC+n_{12} \\
    & = 2n_{12}+n_{21}-n_{22}-\bar{R}_{21} \\
    & \leq n_{11}
\end{align*}
satisfying \eqref{eq:decoding1a}. Finally, if $\bar{R}_{12} > 0$, then 
\begin{equation*}
    \bar{R}_{12}+\bar{R}_{11}^{\sP} = n_{12}+n_{21}-n_{22},
\end{equation*}
satisfying \eqref{eq:decoding1c}; and if $\bar{R}_{12} = 0$, then
\eqref{eq:decoding1c} is irrelevant.
Using a similar argument, it can be shown that the decoding conditions 
\eqref{eq:decoding2} at receiver two hold. 

Case IV ($n_{11}+\tfrac{1}{2}n_{22} < n_{12}+n_{21} \leq \tfrac{3}{2}n_{22}$):
Recall 
\begin{align*}
    \bar{R}_{11}^{\sP} & \triangleq n_{11}-n_{21}, \\
    \bar{R}_{22}^{\sP} & \triangleq n_{22}-n_{12}, \\
    \bar{R}_{21} & \triangleq \big\lfloor n_{12}-\tfrac{1}{2}n_{22} \big\rfloor, \\
    \bar{R}_{12} & \triangleq \bar{R}_{11}^{\sC} \triangleq \big\lfloor n_{21}-\tfrac{1}{2}n_{22} \big\rfloor, \\
    \bar{R}_{22}^{\sC} & \triangleq n_{22}-n_{21}.
\end{align*}
To check the decoding conditions, note first that
\begin{align*}
    \max\{\bar{R}_{21}, \bar{R}_{22}^{\sC}\} = \bar{R}_{22}^{\sC}, \\
    \max\{\bar{R}_{12}, \bar{R}_{11}^{\sC}\} = \bar{R}_{11}^{\sC},
\end{align*}
since
\begin{equation*}
    \tfrac{3}{2}n_{22} \geq n_{12}+n_{21}
\end{equation*}
by assumption. For receiver one, we then have
\begin{equation*}
    \bar{R}_{11}^{\sC} +\bar{R}_{22}^{\sC} +\bar{R}_{12} +\bar{R}_{11}^{\sP} \leq n_{11},
\end{equation*}
satisfying \eqref{eq:decoding1a}. 
Moreover, 
\begin{equation*}
    \bar{R}_{22}^{\sC} +\bar{R}_{12} +\bar{R}_{11}^{\sP} 
    \leq n_{11}+\tfrac{1}{2}n_{22}-n_{21}
    \leq n_{12},
\end{equation*}
where we have used $n_{11}+\tfrac{1}{2}n_{22} < n_{12}+n_{21}$. Hence
\eqref{eq:decoding1b} is satisfied. Finally, 
\begin{equation*}
    \bar{R}_{12} +\bar{R}_{11}^{\sP}
    \leq n_{11}-\tfrac{1}{2}n_{22}
    \leq n_{12}+n_{21}-n_{22},
\end{equation*}
where we have again used $n_{11}+\tfrac{1}{2}n_{22} < n_{12}+n_{21}$.
Hence \eqref{eq:decoding1c} is satisfied. Together, this shows that
decoding is successful at receiver one. At receiver two, we have
\begin{equation*}
    \bar{R}_{22}^{\sC} +\bar{R}_{11}^{\sC} +\bar{R}_{21} +\bar{R}_{22}^{\sP} 
    \leq n_{22},
\end{equation*}
satisfying \eqref{eq:decoding2a}, and 
\begin{equation*}
    \bar{R}_{11}^{\sC} +\bar{R}_{21} +\bar{R}_{22}^{\sP} 
    \leq n_{21},
\end{equation*}
satisfying \eqref{eq:decoding2b}. Finally,
\begin{equation*}
    \bar{R}_{21} +\bar{R}_{22}^{\sP}
    \leq \tfrac{1}{2}n_{22}
    \leq n_{12}+n_{21}-n_{11},
\end{equation*}
where we have used $n_{11}+\tfrac{1}{2}n_{22} < n_{12}+n_{21}$. Hence
\eqref{eq:decoding2c} is satisfied. Together, this shows that decoding
is successful at receiver two. 

Case V ($\tfrac{3}{2}n_{22} < n_{12}+n_{21} \leq n_{11}+n_{22}$): Recall
\begin{align*}
    \bar{R}_{11}^{\sP} & \triangleq n_{11}-n_{21}, \\
    \bar{R}_{22}^{\sP} & \triangleq n_{22}-n_{12}, \\
    \bar{R}_{12} & \triangleq \bar{R}_{11}^{\sC} 
    \triangleq \big\lfloor \tfrac{2}{3}n_{21}-\tfrac{1}{3}n_{12} \big\rfloor, \\
    \bar{R}_{21} & \triangleq \bar{R}_{22}^{\sC} 
    \triangleq \big\lfloor \tfrac{2}{3}n_{12}-\tfrac{1}{3}n_{21} \big\rfloor.
\end{align*}
For decoding at receiver one, we need to verify the decoding conditions
\eqref{eq:decoding1}. We have
\begin{equation*}
    \bar{R}_{11}^{\sC} +\bar{R}_{22}^{\sC} +\bar{R}_{12} +\bar{R}_{11}^{\sP}
    \leq n_{11},
\end{equation*}
satisfying \eqref{eq:decoding1a}. Moreover, 
\begin{equation*}
    \bar{R}_{22}^{\sC} +\bar{R}_{12} +\bar{R}_{11}^{\sP}
    \leq n_{11} + n_{12} - \tfrac{2}{3}(n_{21}+n_{12})
    \leq n_{12},
\end{equation*}
where we have used $\tfrac{3}{2}n_{11} \leq \tfrac{3}{2}n_{22} < n_{12}+n_{21}$
This satisfies \eqref{eq:decoding1b}. Finally
\begin{align*}
    \bar{R}_{12}+\bar{R}_{11}^{\sP}
    & \leq n_{11} - \tfrac{1}{3}(n_{12}+n_{21}) \\
    & = n_{11}+n_{12}+n_{21} - \tfrac{4}{3}(n_{12}+n_{21}) \\
    & \leq n_{11}+n_{12}+n_{21} - 2n_{22}\\
    & \leq n_{12}+n_{21}-n_{22},
\end{align*}
where we have used that $n_{12}+n_{21} \geq \tfrac{3}{2}n_{22}$. Hence
\eqref{eq:decoding1c} is satisfied. A similar argument shows that the
decoding conditions \eqref{eq:decoding2} at receiver two hold. Hence
decoding is successful at both receivers.

\section{Proof of Lemma~\ref{thm:upper} in Section~\ref{sec:proofs_det_upper}}
\label{sec:appendix_upper}

Throughout this proof, we make use of the fact that, for
the (modulated) deterministic X-channel~\eqref{eq:xc_det_arb}, the
definition of capacity imposes that 
\begin{equation*}
    \bar{\bm{u}}_{mk}^{(T)} 
    \triangleq (\bar{\bm{u}}_{mk}[t])_{t=1}^T
\end{equation*}
is only a function of $w_{mk}$.

We start with~\eqref{eq:upper-eliminate-R21}. Define
$\bar{\bm{s}}_{12}$ as the contribution of the second transmitter
at the first receiver, i.e.,
\begin{align*}
    \bar{\bm{s}}_{12} 
    \triangleq
    \bar{\bm{G}}_{12}  
    \begin{pmatrix}
        \bm{0}  \\
        \bar{\bm{u}}_{12}^{\sC}
    \end{pmatrix}
    \oplus 
    \bar{\bm{G}}_{10} 
    \begin{pmatrix}
        \bm{0} \\
        \bar{\bm{u}}_{22}^{\sC}
    \end{pmatrix}.
\end{align*}
Let $\bar{\bm{s}}_{22}$ denote the contribution of the
second transmitter at the second receiver, i.e., 
\begin{align*}
    \bar{\bm{s}}_{22}
    & \triangleq \bar{\bm{G}}_{22} \bar{\bm{u}}_{22}
    \oplus \bar{\bm{G}}_{20} \bar{\bm{u}}_{12}.
\end{align*}
Similarly, we define $\bar{\bm{s}}_{11}$  and
$\bar{\bm{s}}_{21}$ as the contributions of the first transmitter
at the first and second receivers, respectively. With this, we can
rewrite the received vector at receiver $m$ as
\begin{equation*}
    \bar{\bm{y}}_m = \bar{\bm{s}}_{m1}\oplus\bar{\bm{s}}_{m2}.
\end{equation*}

For block length $T$, we have
\begin{align}
    T(\bar{R}_{22}-\varepsilon) 
    & \leq I \big(w_{22}; \bar{\bm{y}}_2^{(T)}\big) \nonumber\\
    & \leq I \big(w_{22}; \bar{\bm{y}}_2^{(T)} , \bar{\bm{s}}_{12}^{(T)},  
    \bar{\bm{u}}_{11}^{(T)}, \bar{\bm{u}}_{21}^{(T)}, w_{12}\big) \nonumber\\
    & = I \big(w_{22}; \bar{\bm{y}}_2^{(T)} , \bar{\bm{s}}_{12}^{(T)} 
    \bigm\vert  \bar{\bm{u}}_{11}^{(T)}, \bar{\bm{u}}_{21}^{(T)}, w_{12}\big) \nonumber\\
    & = I \big(w_{22};  \bar{\bm{s}}_{22}^{(T)}  , \bar{\bm{s}}_{12}^{(T)} 
    \bigm\vert   w_{12}\big) \nonumber\\
    & = I \big(w_{22}; \bar{\bm{s}}_{12}^{(T)} \bigm\vert w_{12}\big) 
    + I \big(w_{22};   \bar{\bm{s}}_{22}^{(T)} 
    \bigm\vert  \bar{\bm{s}}_{12}^{(T)}, w_{12}\big) \nonumber\\
    \label{eq:proof-outer-5}
    & \leq H \big(\bar{\bm{s}}_{12}^{(T)} \bigm\vert   w_{12}\big) 
    + H \big( \bar{\bm{s}}_{22}^{(T)} \bigm\vert  \bar{\bm{s}}_{12}^{(T)}, w_{12}\big),
\end{align}
where the first step follows from Fano's inequality.
In addition, using again Fano's inequality, 
\begin{align}
    T(\bar{R}_{11}+\bar{R}_{12}-\varepsilon) 
    & \leq I\big( w_{11}, w_{12} ; \bar{\bm{y}}_1^{(T)} \big) \nonumber\\
    & \leq I\big( w_{11}, w_{12} ,w_{21} ; \bar{\bm{y}}_1^{(T)} \big) \nonumber\\
    & = H\big(\bar{\bm{y}}_1^{(T)}\big)- H\big(\bar{\bm{y}}_1^{(T)} 
    \bigm\vert w_{11}, w_{12} ,w_{21} \big) \nonumber\\
    \label{eq:proof-outer-9} 
    & = H\big(\bar{\bm{y}}_1^{(T)}\big)
    - H\big(\bar{\bm{s}}_{12}^{(T)} \bigm\vert  w_{12} \big).
\end{align}
Adding~\eqref{eq:proof-outer-5} and~\eqref{eq:proof-outer-9} yields
\begin{align*}
    T(\bar{R}_{11}+\bar{R}_{12}+\bar{R}_{22}-2\varepsilon) 
    & \leq  H\big(\bar{\bm{y}}_1^{(T)}\big) 
    + H \big( \bar{\bm{s}}_{22}^{(T)} \bigm\vert \bar{\bm{s}}_{12}^{(T)}, w_{12}\big).
\end{align*}
For the first term on the right-hand side, we have
\begin{equation*}
    H\big(\bar{\bm{y}}_1^{(T)}\big) \leq T \max\{n_{11}, n_{12}\}.
\end{equation*}
For the second term, recall
that $\bar{\bm{u}}_{12}^{(T)}$ is a function of only $w_{12}$, and
hence
\begin{align*}
    H \big( \bar{\bm{s}}_{22}^{(T)} \bigm\vert \bar{\bm{s}}_{12}^{(T)}, w_{12}\big)
    & \leq H \big( \bar{\bm{s}}_{22}^{(T)} \bigm\vert \bar{\bm{s}}_{12}^{(T)},   
    \bar{\bm{u}}_{12}^{(T)} \big) \\
    & \leq H \bigg( \bar{\bm{G}}_{22} \bar{\bm{u}}_{22}^{(T)}
    \biggm\vert  \bar{\bm{G}}_{10}  
    \begin{pmatrix}
        \bm{0}^{(T)} \\
        (\bar{\bm{u}}_{22}^{\sC})^{(T)}
    \end{pmatrix}
    \bigg).
\end{align*}
Since $\bar{\bm{G}}_{mk}$ is lower triangular with nonzero diagonal,
it is invertible, implying that
\begin{align*}
    H \big( \bar{\bm{s}}_{22}^{(T)} \bigm\vert \bar{\bm{s}}_{12}^{(T)}, w_{12} \big)
    & \leq H \big( \bar{\bm{u}}_{22}^{(T)} \bigm\vert (\bar{\bm{u}}_{22}^{\sC})^{(T)} \big) \\
    & = H \big( (\bar{\bm{u}}_{22}^{\sP})^{(T)} 
    \bigm\vert (\bar{\bm{u}}_{22}^{\sC})^{(T)} \big) \\
    & \leq (n_{22}-n_{12})^+.
\end{align*}
Together, this shows that
\begin{align}
    \label{eq:proof-outer15}
    T(\bar{R}_{11}+\bar{R}_{12}+\bar{R}_{22}-2\varepsilon) 
    & \leq T\max\{n_{11}, n_{12}\}
    +H\big( (\bar{\bm{u}}_{22}^{\sP})^{(T)} 
    \bigm\vert (\bar{\bm{u}}_{22}^{\sC})^{(T)} \big) \\
    & \leq T ( \max\{n_{11}, n_{12}\}+(n_{22}-n_{12})^+ ) \nonumber.
\end{align}
Therefore, as $T \to \infty$ and $\varepsilon \to 0$, we
have~\eqref{eq:upper-eliminate-R21}. Similarly, we can
prove~\eqref{eq:upper-eliminate-R12},~\eqref{eq:upper-eliminate-R22},
and~\eqref{eq:upper-eliminate-R11}.

We now establish the upper bound~\eqref{eq:upper-bound-X2}. Starting with
Fano's inequality,
\begin{align}
    T(\bar{R}_{11}+\bar{R}_{12}-\varepsilon) 
    & \leq I\big(w_{11}, w_{12} ; \bar{\bm{y}}_{1}^{(T)}\big) \nonumber\\
    & \leq I\big(w_{11}, w_{12} ; \bar{\bm{y}}_{1}^{(T)}, \bar{\bm{s}}_{21}^{(T)},w_{21}\big) 
    \nonumber\\
    & = I\big(w_{11}, w_{12} ; \bar{\bm{y}}_{1}^{(T)}, \bar{\bm{s}}_{21}^{(T)} 
    \bigm\vert w_{21}\big) \nonumber\\
    &= I\big(w_{11}, w_{12} ;  \bar{\bm{s}}_{21}^{(T)} \bigm\vert w_{21}\big)
    +I\big(w_{11}, w_{12} ; \bar{\bm{y}}_{1}^{(T)} 
    \bigm\vert \bar{\bm{s}}_{21}^{(T)}, w_{21}\big) \nonumber\\
    & \leq H\big( \bar{\bm{s}}_{21}^{(T)} \bigm\vert w_{21}\big)
    +H\big( \bar{\bm{y}}_{1}^{(T)} \bigm\vert \bar{\bm{s}}_{21}^{(T)}, w_{21}\big) 
    -H\big( \bar{\bm{y}}_{1}^{(T)} 
    \bigm\vert \bar{\bm{s}}_{21}^{(T)}, w_{21}, w_{11}, w_{12} \big) \nonumber\\
    & = H\big( \bar{\bm{s}}_{21}^{(T)} \bigm\vert w_{21} \big)
    +H\big( \bar{\bm{y}}_{1}^{(T)} \bigm\vert \bar{\bm{s}}_{21}^{(T)}, w_{21}\big)
    -H\big( \bar{\bm{y}}_{1}^{(T)} 
    \bigm\vert \bar{\bm{s}}_{21}^{(T)}, w_{21}, w_{11}, w_{12}, 
    \bar{\bm{u}}_{11}^{(T)}, \bar{\bm{u}}_{21}^{(T)}\big) \nonumber\\
    \label{eq:proof-outer-10} 
    & = H\big( \bar{\bm{s}}_{21}^{(T)} \bigm\vert w_{21}\big)
    +H\big( \bar{\bm{y}}_{1}^{(T)} \bigm\vert \bar{\bm{s}}_{21}^{(T)}, w_{21}\big) 
    -H\big( \bar{\bm{s}}_{12}^{(T)} \bigm\vert w_{12}\big).
\end{align}
Similarly, we have
\begin{align}
    \label{eq:proof-outer-11} 
    T(\bar{R}_{21}+\bar{R}_{22}-\varepsilon) 
    & \leq  H\big( \bar{\bm{s}}_{12}^{(T)} \bigm\vert w_{12}\big)
    +H\big( \bar{\bm{y}}_{2}^{(T)} \bigm\vert \bar{\bm{s}}_{12}^{(T)}, w_{12}\big)
    -H\big( \bar{\bm{s}}_{21}^{(T)} \bigm\vert w_{21} \big).
\end{align}
Adding~\eqref{eq:proof-outer-10} and~\eqref{eq:proof-outer-11}, we obtain
\begin{align*}
    T(\bar{R}_{11}+\bar{R}_{12}+\bar{R}_{21}+\bar{R}_{22}-2\varepsilon) 
    & \leq H\big( \bar{\bm{y}}_{1}^{(T)} \bigm\vert \bar{\bm{s}}_{21}^{(T)}, w_{21}\big) 
    +H\big( \bar{\bm{y}}_{2}^{(T)} \bigm\vert \bar{\bm{s}}_{12}^{(T)}, w_{12}\big) \\
    & \leq H\big( \bar{\bm{y}}_{1}^{(T)} \bigm\vert \bar{\bm{s}}_{21}^{(T)}, 
    \bar{\bm{u}}_{21}^{(T)}\big)
    +H\big( \bar{\bm{y}}_{2}^{(T)} 
    \bigm\vert \bar{\bm{s}}_{12}^{(T)}, \bar{\bm{u}}_{12}^{(T)}\big),
\end{align*}
where in the last line we have used that $\bar{\bm{u}}_{mk}^{(T)}$
is only a function of $w_{mk}$. For the first term, we obtain using
invertibility of the matrices $\bar{\bm{G}}_{mk}$,
\begin{align*}
    H\big( \bar{\bm{y}}_{1}^{(T)} & \bigm\vert \bar{\bm{s}}_{21}^{(T)}, 
    \bar{\bm{u}}_{21}^{(T)}\big) \\
    & = H\big( \bar{\bm{y}}_{1}^{(T)} \bigm\vert (\bar{\bm{u}}_{11}^{\sC})^{(T)}, 
    \bar{\bm{u}}_{21}^{(T)}\big) \\
    & =
    H\left(
    \bar{\bm{G}}_{11} 
    \begin{pmatrix}
        (\bar{\bm{u}}_{11}^{\sC})^{(T)} \\
        (\bar{\bm{u}}_{11}^{\sP})^{(T)} 
    \end{pmatrix}
    \oplus 
    \bar{\bm{G}}_{12} 
    \begin{pmatrix}
        \bm{0}^{(T)}  \\
        (\bar{\bm{u}}_{12}^{\sC})^{(T)}
    \end{pmatrix}
    \oplus 
    \bar{\bm{G}}_{10} 
    \begin{pmatrix}
        \bm{0}^{(T)} \\
        (\bar{\bm{u}}_{22}^{\sC})^{(T)}
    \end{pmatrix}
    \biggm\vert (\bar{\bm{u}}_{11}^{\sC})^{(T)} 
    \right) \\
    & \leq
    H\left(
    \bar{\bm{G}}_{11} 
    \begin{pmatrix}
        \bm{0}^{(T)} \\
        (\bar{\bm{u}}_{11}^{\sP})^{(T)} \\
    \end{pmatrix}
    \oplus 
    \bar{\bm{G}}_{12} 
    \begin{pmatrix}
        \bm{0}^{(T)} \\
        (\bar{\bm{u}}_{12}^{\sC})^{(T)}
    \end{pmatrix}
    \oplus 
    \bar{\bm{G}}_{10} 
    \begin{pmatrix}
        \bm{0}^{(T)} \\
        (\bar{\bm{u}}_{22}^{\sC})^{(T)}
    \end{pmatrix}
    \right).
\end{align*}
Since the matrices $\bar{\bm{G}}_{mk}$ are lower triangular, this
last term is upper bounded by 
\begin{equation*}
    T \max \{ n_{12}, n_{11}-n_{21}\}.
\end{equation*}
By an analogous argument, 
\begin{equation*}
    H\big( \bar{\bm{y}}_{2}^{(T)} 
    \bigm\vert \bar{\bm{s}}_{12}^{(T)}, \bar{\bm{u}}_{12}^{(T)}\big)
    \leq T \max \{n_{21}, n_{22}-n_{12}\}.
\end{equation*}
Together, this shows that
\begin{align*}
    T(\bar{R}_{11}+\bar{R}_{12}+\bar{R}_{21}+\bar{R}_{22}-2\varepsilon) 
    & \leq  T \big( \max \{ n_{12}, n_{11}-n_{21}\} 
    + \max \{n_{21}, n_{22}-n_{12}\} \big),
\end{align*}
proving \eqref{eq:upper-bound-X2} as $T\to\infty$ and
$\varepsilon\to 0$. Similarly, we can
prove~\eqref{eq:sum-rate-up-last}.

We now establish the bound \eqref{eq:upper-bound-IC1}.  By Fano's
inequality, 
\begin{align}
    \label{eq:proof-outer12}
    T(\bar{R}_{21}+\bar{R}_{22}-\varepsilon)
    & \leq I\big(w_{21}, w_{22}; \bar{\bm{y}}_2^{(T)}\big) \nonumber\\
    & \leq I\big(w_{21}, w_{22}; \bar{\bm{y}}_2^{(T)},
    (\bar{\bm{u}}_{22}^{\sC})^{(T)}, w_{12} \big) \nonumber\\
    & = I\big(w_{21}, w_{22}; \bar{\bm{y}}_2^{(T)},
    (\bar{\bm{u}}_{22}^{\sC})^{(T)} \bigm\vert w_{12} \big) \nonumber\\
    & = I\big(w_{21}, w_{22}; (\bar{\bm{u}}_{22}^{\sC})^{(T)} \bigm\vert w_{12} \big) 
    + I\big(w_{21}, w_{22}; \bar{\bm{y}}_2^{(T)}
    \bigm\vert  w_{12}, (\bar{\bm{u}}_{22}^{\sC})^{(T)}\big) \nonumber\\
    & = H\big((\bar{\bm{u}}_{22}^{\sC})^{(T)}\big)
    + I\big(w_{21}, w_{22}, \bar{\bm{s}}_{21}^{(T)}; \bar{\bm{y}}_2^{(T)}
    \bigm\vert  w_{12}, (\bar{\bm{u}}_{22}^{\sC})^{(T)}\big) \nonumber\\
    & \quad {} - I\big(\bar{\bm{s}}_{21}^{(T)}; \bar{\bm{y}}_2^{(T)}
    \bigm\vert  w_{12}, w_{21}, w_{22} , (\bar{\bm{u}}_{22}^{\sC})^{(T)}\big) \nonumber\\
    & = H\big((\bar{\bm{u}}_{22}^{\sC})^{(T)}\big)
    + H\big(\bar{\bm{y}}_2^{(T)} \bigm\vert  w_{12}, (\bar{\bm{u}}_{22}^{\sC})^{(T)}\big) 
    -H\big((\bar{\bm{u}}_{11}^{\sC})^{(T)}\big).
\end{align}
Moreover, using again Fano's inequality, 
\begin{align}
    \label{eq:proof-outer13}
    T(\bar{R}_{11}-\varepsilon)
    & \leq I\big(w_{11}; \bar{\bm{y}}_1^{(T)}\big) \nonumber\\
    & \leq I\big(w_{11}; \bar{\bm{y}}_1^{(T)}, w_{12}, w_{21}, w_{22}\big) \nonumber\\
    & = I\big(w_{11}; \bar{\bm{y}}_1^{(T)} \bigm\vert w_{12}, w_{21}, w_{22}\big) \nonumber\\
    & = H\big(\bar{\bm{y}}_1^{(T)} \bigm\vert w_{12}, w_{21}, w_{22}\big) \nonumber\\
    & = H\big(\bar{\bm{u}}_{11}^{(T)}\big) \nonumber\\
    & = H\big((\bar{\bm{u}}_{11}^{\sC})^{(T)}\big) 
    + H\big((\bar{\bm{u}}_{11}^{\sP})^{(T)} \bigm\vert (\bar{\bm{u}}_{11}^{\sC})^{(T)} \big).
\end{align}
Adding \eqref{eq:proof-outer12} and \eqref{eq:proof-outer13} yields
\begin{align*}
    T(\bar{R}_{11} & +\bar{R}_{21}+\bar{R}_{22}-2\varepsilon) \nonumber\\
    & \leq H\big(\bar{\bm{y}}_2^{(T)} \bigm\vert  w_{12}, (\bar{\bm{u}}_{22}^{\sC})^{(T)}\big) 
    + H\big((\bar{\bm{u}}_{11}^{\sP})^{(T)} \bigm\vert (\bar{\bm{u}}_{11}^{\sC})^{(T)} \big)
    + H\big((\bar{\bm{u}}_{22}^{\sC})^{(T)}\big) \nonumber\\
    & \leq T\big(\max\{n_{21}, n_{22}-n_{12}\} 
    + (n_{11}-n_{21})^+ \big)
    + H\big((\bar{\bm{u}}_{22}^{\sC})^{(T)}\big).
\end{align*}
Combined with \eqref{eq:proof-outer15} derived earlier, we obtain
\begin{align*}
    T(2\bar{R}_{11} & +\bar{R}_{12}+\bar{R}_{21}+2\bar{R}_{22}-4\varepsilon) \nonumber\\
    & \leq T\big(\max\{n_{11},n_{12}\}+\max\{n_{21}, n_{22}-n_{12}\} 
    + (n_{11}-n_{21})^+ \big) \\
    & \quad {} + H\big((\bar{\bm{u}}_{22}^{\sC})^{(T)}\big)
    + H\big( (\bar{\bm{u}}_{22}^{\sP})^{(T)} \mid (\bar{\bm{u}}_{22}^{\sC})^{(T)} \big) \\
    & = T\big(\max\{n_{11},n_{12}\}+\max\{n_{21}, n_{22}-n_{12}\} 
    + (n_{11}-n_{21})^+ \big)
    + H\big((\bar{\bm{u}}_{22})^{(T)}\big).
\end{align*}
Since $(\bar{\bm{u}}_{22})^{(T)}$ is a deterministic function of
$w_{22}$, we have
\begin{equation}
    \label{eq:key}
    H\big((\bar{\bm{u}}_{22})^{(T)}\big)
    \leq H(w_{22})
    = T\bar{R}_{22}.
\end{equation}
From \eqref{eq:key}, we obtain 
\begin{equation*}
    T(2\bar{R}_{11}+\bar{R}_{12}+\bar{R}_{21}+\bar{R}_{22}-4\varepsilon) 
    \leq T\big(\max\{n_{11},n_{12}\}+\max\{n_{21}, n_{22}-n_{12}\} 
    + (n_{11}-n_{21})^+ \big).
\end{equation*}
Letting $T\to\infty$ and $\varepsilon\to 0$ yields the upper bound
\eqref{eq:upper-bound-IC1}. Similarly, we can prove
\eqref{eq:upper-bound-IC2}--\eqref{eq:upper-bound-IC4}. \hfill\IEEEQED

\begin{remark}
    Equation~\eqref{eq:key} is a key step in the derivation of the outer
    bound \eqref{eq:upper-bound-IC1}. If we had used the standard bound
    $H\big((\bar{\bm{u}}_{22})^{(T)}\big) \leq T\max\{n_{22},n_{21}\}$,
    we would have obtained a looser bound than
    \eqref{eq:upper-bound-IC1}.  
\end{remark}

\section{Analysis of Mismatched Encoders and Decoders}
\label{sec:appendix_mismatch}

The proof of Theorem~\ref{thm:arbitrary_gaussian} in
Section~\ref{sec:proofs_gaussian_lower} assumes that the precise channel
gains $h_{mk}$ are available at all encoders and decoders. Here we
assume instead that these channel gains are only known approximately at
any node in the network. As we will see, the only effect of this change
in available channel state information is to decrease the minimum
constellation distance seen at the receivers. 

Formally, assume both transmitters and receivers have only access to
estimates $\hat{h}_{mk}$ of $h_{mk}$ satisfying
\begin{equation}
    \label{eq:quant}
    \abs{h_{mk}-\hat{h}_{mk}} 
    \leq \varepsilon
    \triangleq 2^{-\max_{m,k} n_{mk}}.
\end{equation}
In other words, all transmitters and receivers have access to a
$\max_{m,k} n_{mk}$-bit quantization of the channel gains. Since we know
a priori that $h_{mk}\in(1,2]$, we can assume without loss of generality
that $\hat{h}_{mk} \in (1, 2]$ as well.

Each transmitter $k$ forms the modulated symbol $u_{mk}$ from the message
$w_{mk}$. From these modulated signals, the channel inputs
\begin{align*}
    x_1 & \triangleq \hat{h}_{22}u_{11}+\hat{h}_{12}u_{21}, \\
    x_2 & \triangleq \hat{h}_{21}u_{12}+\hat{h}_{11}u_{22}
\end{align*}
are formed. In other words, the transmitters treat the estimated channel
gains $\hat{h}_{mk}$ as if they were the correct ones; the encoders are
thus mismatched.  The modulation process from $w_{1k}$ to $u_{1k}$ is
the same as in the matched case analyzed in
Section~\ref{sec:proofs_gaussian_lower}.  Since $\abs{\hat{h}_{mk}}\leq 2$
and $\abs{u_{mk}}\leq 1/4$, the resulting channel input $x_k$ satisfies the
unit average power constraint at the transmitters. 

The channel output at receiver one is
\begin{align*}
    y_1 
    & = 2^{n_{11}}h_{11}x_1+2^{n_{12}}h_{12}x_2+z_1 \\
    & = 
    \big( h_{11}\hat{h}_{22} 2^{n_{11}}u_{11}+
    h_{12}\hat{h}_{21} 2^{n_{12}}u_{12} \big)
    +\big( \hat{h}_{12}h_{11} 2^{n_{11}}u_{21}
    +h_{12}\hat{h}_{11}2^{n_{12}}u_{22}^\sC \big)
    + \big(h_{12}\hat{h}_{11}2^{n_{12}}u_{22}^\sP 
    +z_1 \big),
\end{align*}
As in the matched case, the received signal consists of desired signals,
interference signals, and signals treated as noise.
For the third term treated as noise, we have
\begin{equation}
    \label{eq:noisebound_mm}
    \abs{h_{12}\hat{h}_{11}2^{n_{12}}u_{22}^\sP} 
    \leq 1,
\end{equation}
since
\begin{equation*}
    2^{n_{12}}u_{22}^\sP \in [0,1/4).
\end{equation*}

The demodulator at receiver one searches for $(\hat{s}_{11},
\hat{s}_{12}, \hat{s}_{10})$ minimizing
\begin{equation*}
    \abs{y_1-\hat{h}_{11}\hat{h}_{22}\hat{s}_{11}
    -\hat{h}_{12}\hat{h}_{21}\hat{s}_{12}
    -\hat{h}_{12}\hat{h}_{11}\hat{s}_{10}}.
\end{equation*}
Note that the entire demodulation process depends solely on the
estimated channel gains $\hat{h}_{mk}$ and not on the actual channel
gains $h_{mk}$.  Furthermore, the demodulator is the maximum-likelihood
detector only if the estimated channel gains coincide with the actual
channel gains.  Thus, the demodulator is mismatched. 

We now analyze the probability of error of this mismatched demodulator.
There are two contributions to this probability of error. One is due to
noise, the other one due to mismatched detection. 
Set
\begin{equation*}
    v \triangleq \hat{h}_{11}\hat{h}_{22}s_{11}
    -\hat{h}_{12}\hat{h}_{21}s_{12}
    -\hat{h}_{12}\hat{h}_{11}s_{10},
\end{equation*}
and define $\hat{v}$ similarly, but with respect to
$(\hat{s}_{11},\hat{s}_{12},\hat{s}_{12})$. 

We need to upper bound
\begin{equation*}
    \Pp(\hat{v}=\hat{q} \mid v = q(\hat{q},\ell) )
\end{equation*}
with $q(\hat{q},\ell)$ as defined in in
Section~\ref{sec:proofs_gaussian_lower}. Let $d^\prime$ be the minimum
distance between any two noiseless estimated received signals (as
assumed by the mismatched demodulator using $\hat{h}_{mk}$), i.e.,
between any two possible values of $v$. Let
$\hat{d}$ be the maximum distance between the noiseless received signal
$y_1-z_1-h_{12}\hat{h}_{11}2^{n_{12}}u_{22}^\sP$ and the estimated
received $v$ signal with the same channel inputs. Then
\begin{align}
    \label{eq:xc_error_mm}
    \Pp(\hat{v}=\hat{q} \mid v = q(\hat{q},\ell) )
    & \leq \Pp\bigl( z_1+\abs{h_{12}\hat{h}_{11}2^{n_{12}}u_{22}^\sP}+\hat{d} 
    \geq \abs{\ell} d^\prime/2 \bigr) \notag\\
    & \leq \Pp\bigl( z_1 \geq \abs{\ell} d^\prime/2-\hat{d}-1 \bigr),
\end{align}
where we have used \eqref{eq:noisebound_mm}.

We start by upper bounding the mismatch distance $\hat{d}$. We have
\begin{align}
    \label{eq:xc_dhat_mm}
    \hat{d}
    & \triangleq \max_{(u_{mk})}
    \big\lvert \hat{h}_{22}2^{n_{11}}u_{11}(h_{11}-\hat{h}_{11})
    +\hat{h}_{21}2^{n_{12}}u_{12}(h_{12}-\hat{h}_{12}) \nonumber\\
    & \quad \phantom{ \max_{(u_{mk})} \big\lvert }
    \ +\hat{h}_{12}2^{n_{11}}u_{21}(h_{11}-\hat{h}_{11})
    +\hat{h}_{11}2^{n_{12}}u_{22}^\sC(h_{12}-\hat{h}_{12}) \big\rvert \nonumber\\
    & \leq 4\cdot2\cdot2^{n_{11}}\cdot\tfrac{1}{4}\cdot\varepsilon \nonumber\\
    & \leq 2,
\end{align}
where we have used \eqref{eq:quant}, that $\abs{u_{mk}}\leq 1/4$, and
that $\abs{\hat{h}_{mk}}\leq 2$.

We continue by lower bounding the distance $d^\prime$ between the
estimated received signal (i.e., as assumed by the mismatched detector)
generated by the correct $(s_{11}, s_{12}, s_{10})$ and by any other
triple $(s_{11}', s_{12}', s_{10}')$. By the triangle inequality, 
\begin{align}
    \label{eq:xc_d2_mm}
    d^\prime
    & \triangleq \min_{\substack{(s_{11}, s_{12}, s_{10}) \\ \neq (s_{11}', s_{12}', s_{10}')}}
    \big\lvert
    \hat{h}_{11}\hat{h}_{22}(s_{11}-s_{11}')
    +\hat{h}_{12}\hat{h}_{21}(s_{12}-s_{12}')
    +\hat{h}_{12}\hat{h}_{11}(s_{10}-s_{10}') \big\rvert \nonumber\\
    & \geq \min_{\substack{(s_{11}, s_{12}, s_{10}) \\ \neq(s_{11}', s_{12}', s_{10}')}}
    \big\lvert
    h_{11}h_{22}(s_{11}-s_{11}')
    +h_{12}h_{21}(s_{12}-s_{12}')
    +h_{12}h_{11}(s_{10}-s_{10}') \big\rvert 
    - 3\cdot 5/2 \nonumber\\
    & \geq d-8,
\end{align}
where $d$ denotes the minimum distance \eqref{eq:xc_d} in the matched
case as analyzed in Section~\ref{sec:proofs_gaussian_lower}.  Here we
have used that
\begin{align*}
    \abs{s_{11}-s_{11}'}\abs{ \hat{h}_{11}\hat{h}_{22}-h_{11}h_{22} }
    & \leq 2^{n_{11}-1}\abs{(\hat{h}_{11}-h_{11})(\hat{h}_{22}-h_{22})
    +h_{22}(\hat{h}_{11}-h_{11})+h_{11}(\hat{h}_{22}-h_{22})} \\
    & \leq 2^{n_{11}-1}\cdot 5 \cdot\varepsilon \\
    & \leq 5/2
\end{align*}
by \eqref{eq:quant}, and similarly for the other two terms.

Combining \eqref{eq:xc_dhat_mm} and \eqref{eq:xc_d2_mm} shows that
\begin{equation*}
    \abs{\ell} d^\prime/2-\hat{d}-1 
    \geq \abs{\ell} (d-8)/2-3.
\end{equation*}
By \eqref{eq:xc_error_mm}, this implies
\begin{equation}
    \label{eq:xc_error3_mm}
    \Pp\bigl(\hat{v}=\hat{q} \mid v = q(\hat{q},\ell)\bigr)
    \leq \Pp\bigl( z_1 \geq \abs{\ell} (d-8)/2-3 \bigr).
\end{equation}

\section{Proof of Lemma~\ref{thm:upper-G} in Section~\ref{sec:proofs_gaussian_upper}}
\label{sec:appendix_upper-G}

The inequalities
\eqref{eq:upper-eliminate-R21-G}--\eqref{eq:sum-rate-up-last-G} have
been already proved in \cite[Lemma~5.2, Theorem~5.3]{huang08}. Here 
we present the proof for inequalities
\eqref{eq:upper-bound-IC1-G}--\eqref{eq:upper-bound-IC4-G}. First,
we establish the bound \eqref{eq:upper-bound-IC1-G}.

Define $s_{mk}[t]$ as the contribution of transmitter $k$ at
receiver $m$ corrupted by receiver noise $z_m[t]$, i.e.,
\begin{equation*} 
    s_{mk}[t] \triangleq 2^{n_{mk}}h_{mk}x_{k}[t]+z_m[t].  
\end{equation*}
For block length $T$, we have
\begin{align}
    T(R_{22}-\varepsilon) 
    & \leq I\big(w_{22};y_2^{(T)}\big) \nonumber\\
    & \leq I\big(w_{22};y_2^{(T)} ,s_{12}^{(T)},    x_{1}^{(T)}, w_{12}\big) \nonumber\\
    & = I\big(w_{22};y_2^{(T)} ,s_{12}^{(T)} 
    \bigm\vert  x_{1}^{(T)}, w_{12}\big) \nonumber\\
    & = I\big(w_{22}; s_{22}^{(T)}  ,s_{12}^{(T)} 
    \bigm\vert w_{12}\big) \nonumber\\
    & = I\big(w_{22};s_{12}^{(T)} \bigm\vert w_{12}\big) 
    + I\big(w_{22};  s_{22}^{(T)} 
    \bigm\vert s_{12}^{(T)}, w_{12}\big) \nonumber\\
    \label{eq:proof-outer-5-G}
    & = h\big(s_{12}^{(T)} \bigm\vert w_{12}\big)-h\big(z_1^{(T)}\big) 
    + h\big( s_{22}^{(T)} \bigm\vert s_{12}^{(T)}, w_{12}\big)-h\big(z_2^{(T)}\big),
\end{align}
where the first step follows from Fano's inequality.
Again from  Fano's inequality, we have
\begin{align}
    T(R_{11}+R_{12}-\varepsilon) 
    & \leq I\big(w_{11}, w_{12} ;y_1^{(T)} \big) \nonumber\\
    & \leq I\big(w_{11}, w_{12} ,w_{21} ;y_1^{(T)} \big) \nonumber\\
    & = h\big(y_1^{(T)}\big) 
    -  h\big(y_1^{(T)} \bigm\vert w_{11}, w_{12} ,w_{21} \big) \nonumber\\
    \label{eq:proof-outer-9-G} 
    & = h\big(y_1^{(T)}\big) -  h\big(s_{12}^{(T)} \bigm\vert w_{12}\big).
\end{align}
Adding~\eqref{eq:proof-outer-5-G} and~\eqref{eq:proof-outer-9-G} yields
\begin{align} \label{eq:proof-outer15-G}
    T(R_{11}+R_{12}+R_{22}-2\varepsilon) 
    & \leq  h\big(y_1^{(T)}\big)-h\big(z_1^{(T)}\big) +
    h\big(s_{22}^{(T)} \bigm\vert s_{12}^{(T)}, w_{12}\big)
    -h\big(z_2^{(T)}\big).
\end{align}

Using  Fano's inequality at receiver two, we have
\begin{align}
    \label{eq:proof-outer12-G} 
    T(R_{21}+R_{22}-\varepsilon)
    & \leq I\big(w_{21}, w_{22};y_2^{(T)}\big) \nonumber\\
    & \leq I\big(w_{21}, w_{22};y_2^{(T)},
    s_{12}^{(T)}, w_{12}\big) \nonumber\\
    & = I\big(w_{21}, w_{22};y_2^{(T)},
    s_{12}^{(T)} \bigm\vert w_{12}\big) \nonumber\\
    & = I\big(w_{21}, w_{22};s_{12}^{(T)} \bigm\vert w_{12}\big) 
    + I\big(w_{21}, w_{22};y_2^{(T)}
    \bigm\vert  s_{12}^{(T)}, w_{12}\big) \nonumber\\
    & = h\big(s_{12}^{(T)} \bigm\vert w_{12}\big)
    - h\big(z_1^{(T)}\big)
    + h\big(y_2^{(T)} \bigm\vert s_{12}^{(T)}, w_{12}\big) 
    - h\big(s_{21}^{(T)} \bigm\vert w_{21}\big).
\end{align}
Moreover, Fano's inequality at receiver one yields
\begin{align}
    \label{eq:proof-outer13-G}
    T(R_{11}-\varepsilon)
    & \leq I\big(w_{11};y_{1}^{(T)}\big) \nonumber\\
    & \leq I\big(w_{11};y_{1}^{(T)},s_{21}^{(T)},  w_{12}, w_{21}, w_{22}\big) \nonumber\\
    & = I\big(w_{11};y_{1}^{(T)},s_{21}^{(T)} 
    \bigm\vert w_{12}, w_{21}, w_{22}\big) \nonumber\\
    & = I\big(w_{11};s_{11}^{(T)},s_{21}^{(T)} 
    \bigm\vert w_{12}, w_{21}, w_{22}\big) \nonumber\\
    & = h\big(s_{11}^{(T)},s_{21}^{(T)} \bigm\vert w_{12}, w_{21}, w_{22}\big)
    - h\big(s_{11}^{(T)},s_{21}^{(T)} 
    \bigm\vert w_{11}, w_{12}, w_{21}, w_{22}\big) \nonumber\\
    & = h\big( s_{21}^{(T)} \bigm\vert w_{21}\big) 
    + h\big(s_{11}^{(T)} \bigm\vert s_{21}^{(T)}, w_{21}\big)
    - h\big( z_{1}^{(T)}, z_{2}^{(T)}\big).
\end{align}
Adding \eqref{eq:proof-outer12-G} and \eqref{eq:proof-outer13-G} yields
\begin{align} 
    \label{eq:proof-outer14-G}
    T(R_{11} + & R_{21} +R_{22}  -2\varepsilon) \nonumber\\
    & = h\big(s_{12}^{(T)} \bigm\vert w_{12}\big)
    - h\big(z_1^{(T)}\big) 
    +  h\big(y_2^{(T)} \bigm\vert s_{12}^{(T)}, w_{12}\big) 
    +  h\big(s_{11}^{(T)} \bigm\vert s_{21}^{(T)}, w_{21}\big)
    -  h\big( z_{1}^{(T)} , z_{2}^{(T)}\big).
\end{align}

Adding \eqref{eq:proof-outer14-G} and \eqref{eq:proof-outer15-G}
derived earlier, we obtain
\begin{align}
    \label{eq:proof-outer20-G}
    T(2R_{11} & +R_{12}+ R_{21}+2R_{22}-4\varepsilon) \nonumber\\
    & \leq h\big(y_1^{(T)}\big)
    - 2h\big(z_1^{(T)}\big) +
    h\big( s_{22}^{(T)} \bigm\vert s_{12}^{(T)},  w_{12}\big)
    - h\big(z_2^{(T)}\big)
    + h\big(s_{12}^{(T)} \bigm\vert w_{12} \big) \nonumber \\
    &  \quad {} + h\big(y_2^{(T)} \bigm\vert s_{12}^{(T)}, w_{12}\big) 
    + h\big(s_{11}^{(T)} \bigm\vert  s_{21}^{(T)}, w_{21}\big)
    - h\big( z_{1}^{(T)}, z_{2}^{(T)}\big).
\end{align}
Since
\begin{align}
    \label{eq:proof-outer21-G}
    h\big( s_{12}^{(T)} \bigm\vert w_{12}\big)
    +h\big( s_{22}^{(T)} \bigm\vert s_{12}^{(T)}, w_{12}\big)
    -h\big(z_1^{(T)},  z_2^{(T)}\big) 
    & = I\big(w_{22} ; s_{22}^{(T)} , s_{12}^{(T)} \bigm\vert w_{12}\big) \nonumber\\
    & \leq H(w_{22}) \nonumber\\
    & = T R_{22},
\end{align}
we obtain from \eqref{eq:proof-outer20-G} that
\begin{align*}
    T(2R_{11}+R_{12} & + R_{21}+R_{22}-4\varepsilon) \nonumber\\
    & \leq h\big(y_1^{(T)}\big)
    - 2h\big(z_1^{(T)}\big)
    - h\big(z_2^{(T)}\big)
    + h\big(y_2^{(T)} \bigm\vert s_{12}^{(T)}, w_{12}\big) 
    + h\big(s_{11}^{(T)} \bigm\vert  s_{21}^{(T)}, w_{21}\big) \nonumber \\
    & \leq h\big(y_1^{(T)}\big)
    - h\big(z_1^{(T)}\big) 
    + h\big(y_2^{(T)} \bigm\vert s_{12}^{(T)}\big)
    - h\big(z_2^{(T)}\big) 
    + h\big(s_{11}^{(T)} \bigm\vert s_{21}^{(T)}\big) 
    - h\big(z_1^{(T)}\big) \nonumber \\
    & \leq \frac{T}{2}\log\big(1+2^{2n_{11}}h_{11}^2+2^{2n_{12}}h_{12}^2\big)
    + \frac{T}{2}\log\Big( 1+2^{2n_{21}}h_{21}^2
    +\frac{2^{2n_{22}}h_{22}^2}{1+2^{2n_{12}}h_{12}^2} \Big) \\
    & \quad {} + \frac{T}{2}
    \log\Big(1+\frac{2^{2n_{11}}h_{11}^2}{1+2^{2n_{21}}h_{21}^2} \Big),
\end{align*}
where the last inequality follows from the fact that \iid Gaussian
random variables maximize conditional differential entropy.  Letting
$T\to\infty$ and $\varepsilon\to 0$ proves \eqref{eq:upper-bound-IC1-G}.
Inequalities \eqref{eq:upper-bound-IC2-G}--\eqref{eq:upper-bound-IC4-G}
can be proved similarly.  \hfill\IEEEQED

\begin{remark}
    We point out that, as in the deterministic case,
    \eqref{eq:proof-outer21-G} is a key step to the derivation of the
    outer bound for the Gaussian X-channel.
\end{remark}

\section{Proof of Lemma~\ref{thm:measure} in Section~\ref{sec:foundations_gaussian}}
\label{sec:appendix_measure}

By Fubini's theorem, we have for $m=1$
\begin{align*}
    \mu_4(B)
    & = \int_{h_{11}=1}^{2} \int_{h_{12}=1}^{2} 
    \int_{h_{21}=1}^{2}\int_{h_{22}=1}^{2}
    \ind_B(h_{11}, h_{12}, h_{21}, h_{22})
    dh_{22} dh_{21} dh_{12} dh_{11} \\
    & = \int_{h_{11}=1}^{2} \int_{h_{12}=1}^{2} 
    \int_{h_{21}=1}^{2}\int_{h_{22}=1}^{2}
    \ind_{\tilde{B}}(h_{11}h_{12}, h_{11}h_{22}, h_{12}h_{21})
    dh_{22} dh_{21} dh_{12} dh_{11} \\
    & \leq \int_{h_{11}=1}^{2}\int_{g_{0}=1}^4
    \int_{g_{2}=1}^4\int_{g_{1}=1}^4
    \ind_{\tilde{B}}(g_{0}, g_{1}, g_{2})
    g_{0}^{-1}h_{11}^{-1}
    dg_{1} dg_{2} dg_{0} dh_{11} \\
    & \leq \int_{h_{11}=1}^{2} 
    \mu_3(\tilde{B}) dh_{11} \\
    & \leq \delta.
\end{align*}
The situation is analogous for $m=2$. \hfill\IEEEQED

\section*{Acknowledgment}

The authors would like to thank G.~Kramer for pointing them to Slepian's
1974 Shannon Lecture~\cite{slepian76} and the reviewers for their
careful reading of the manuscript and their thoughtful comments.

\end{document}